\documentclass[a4paper, 11pt]{article}
\usepackage{a4wide, amsfonts,amsmath,amsthm,amssymb,bbm}
\numberwithin{equation}{section}
\begin{document}

\title{Topological charges in $SL(2,\bf{R})$ covariant massive 11-dimensional and Type IIB SUGRA}
%\author{Andrew K. Callister\footnote{E-mail: a.k.callister@durham.ac.uk} and Douglas J. Smith\footnote{E-mail: douglas.smith@durham.ac.uk}\\
%\textit{Centre for Particle Theory \\
%Department of Mathematical Sciences \\
%University of Durham, Durham, DH1 3LE, U.K.}
%}
\author{Andrew K Callister\footnote{E-mail: a.k.callister@durham.ac.uk} and Douglas J Smith\footnote{E-mail: douglas.smith@durham.ac.uk} \\ \\
Department of Mathematical Sciences,
University of Durham\\
Science Laboratories,
South Road,
Durham\\
DH1 3LE.
UK \\
%Email:
%\email{a.k.callister@durham.ac.uk},
%\email{douglas.smith@durham.ac.uk}
}

\maketitle

\begin{abstract}
In this paper we construct closed expressions that correspond to the topological charges of the various 1/2-BPS states of the maximal 10 and 11-dimensional supergravity theories. These expressions are related to the structure of the SUSY algebras in curved spacetimes. We mainly focus on IIB SUGRA and 11-dimensional SUGRA in a double M9-brane background, with an emphasis on the $SL(2,\mathbb{R})$ multiplet structure of the charges and how these map between theories. This includes the charges corresponding to the multiplets of 7- and 9-branes in IIB. We find that examining the possible multiplet structures of the charges provides another tool for exploring the spectrum of BPS states that appear in these theories. As a prerequisite to constructing the charges we determine the field equations and multiplet structure of the 11-dimensional gauge potentials, extending previous results on the subject. The massive gauge transformations of the fields are also discussed. We also demonstrate how these massive gauge transformations are compatible with the construction of an $SL(2,\mathbb{R})$ covariant kinetic term in the 11-dimensional Kaluza-Klein monopole worldvolume action.
%\end{abstract}

\vspace*{-16cm}
\begin{flushright}
{\bf DCPT-09/37}
\end{flushright}
\vspace*{15.5cm}

\end{abstract}

%\preprint{DCPT-09/??}
%\keywords{Supersymmetry, Massive supergravity}

%\begin{document}

%\maketitle

%\begin{center}
%
%{\em Department of Mathematical Sciences,
%University of Durham,
%Science Laboratories,
%South Road,
%Durham. DH1 3LE.
%UK}
%
%\end{center}
%
%\vspace{1.4cm}
%
%\begin{abstract}
%In this paper we construct closed expressions that correspond to the topological charges of the various 1/2-BPS states of the maximal 10 and 11-dimensional supergravity theories. These expressions are related to the structure of the SUSY algebras in curved spacetimes. We mainly focus on IIB SUGRA and 11-dimensional SUGRA in a double M9-brane background, with an emphasis on the $SL(2,\mathbb{R})$ multiplet structure of the charges and how these map between theories. This includes the charges corresponding to the multiplets of 7- and 9-branes in IIB. We find that examining the possible multiplet structures of the charges provides another tool for exploring the spectrum of BPS states that appear in these theories. As a prerequisite to constructing the charges we determine the field equations and multiplet structure of the 11-dimensional gauge potentials, extending previous results on the subject. The massive gauge transformations of the fields are also discussed. We also demonstrate how these massive gauge transformations are compatible with the construction of an $SL(2,\mathbb{R})$ covariant kinetic term in the 11-dimensional Kaluza-Klein monopole worldvolume action.
%\end{abstract}

\section{Introduction}

It is well known that the flatspace SUSY algebras of the various supergravity theories contain terms that correspond to the topological charges of the 1/2-BPS states or branes that couple to the theories \cite{de Azcarraga:1989gm}. Such terms have even been used to infer the existence of previously unknown BPS states \cite{Hull:1997kt,Townsend:1997wg}. It is apparent however that in certain instances the one-to-one relation between the BPS states and the flatspace SUSY algebra breaks down. Perhaps the most well known example of this is the triplet of 7-branes in IIB where it is often stated they each correspond to the same SUSY charge which is invariant under the classical $SL(2,\mathbb{R})$ symmetry group\footnote{In this paper we only consider the classical solutions and therefore do not consider the quantisation of this group to $SL(2,\mathbb{Z})$.} \cite{Meessen:1998qm,Eyras:1999at}. Another example of this also occurs with the 9-brane multiplets in IIB \cite{Bergshoeff:2005ac,Bergshoeff:2006ic}. There exist six 9-branes in total, four of which transform as a quadruplet and two as a doublet. The IIB SUSY algebra however only contains two 9-form charges and so once again the relationship between the SUSY charges and BPS states contains discrepancies. 

Another short coming in reading off the BPS states from the flatspace SUSY algebra occurs for states whose spacetime solution contains a Killing vector, or equivalently states whose worldvolume actions require a gauged isometry. The most notable examples are the Kaluza-Klein (KK) monopoles \cite{Sorkin:1983ns,Gross:1983hb,Bergshoeff:1997gy,Bergshoeff:1998ef,Eyras:1998kx} and M9-branes \cite{Bergshoeff:1998bs} in 11-dimensional SUGRA, and many of their various dimensional reductions to the lower dimensional theories. Although such states have associated charges in the flatspace SUSY algebra there is no way to infer from the SUSY algebra alone that these states require Killing vectors, and whether the Killing vectors lie transverse or parallel to the brane worldvolume. Such problems led \cite{LozanoTellechea:2000mc} to conclude that the usefulness of the flatspace SUSY algebra in determining the BPS states had been overstated, and they proposed a modification to the interpretation of the charges which improved the situation. These problematic states on the 11-dimensional/IIA side are related to the IIB 7 and 9-branes via T-duality.   

In this paper we address these issues by following the work of \cite{Callister:2007jy} based on \cite{HackettJones:2003vz}. Here the notion of a `generalised' charge was introduced. These are spacetime expressions which generalise the charges found in the flatspace SUSY algebra to supersymmetric curved backgrounds. Each generalised charge is associated to a specific BPS state, the idea being that they appear in the SUSY algebra for backgrounds that are sourced by their respective BPS states. Therefore these expressions provide an elegant way of deducing the structure of the SUSY algebra in any background that satisfies the equations of motion. By construction the generalised charges are closed expressions\footnote{In this paper we will loosely use the terms generalised charges or simply charges when strictly speaking we are referring to the charge densities.} for any supersymmetric field configuration,
% that satisfies the equations of motion, 
and therefore provide the topological charge of a BPS state when pulled back to and integrated over its worldvolume. They are constructed out of combinations of Killing spinors and Dirac $\Gamma$ matrices, which combine into bilinear forms, as well as the gauge potentials to which the branes couple; and in this way implicitly depend on the supersymmetry and curvature of the background spacetime. 

In \cite{Callister:2007jy} generalised charges were constructed for the `standard' spectrum of states of the maximal 10 and 11-dimensional supergravities, namely the D-branes, M-branes (including the M9-brane), F-strings, NS5-branes and KK-monopoles. In this paper we extend these results by determining the generalised charges for the 7-branes and 9-branes in IIB that have not already been considered. Furthermore, we also consider those states on the IIA side related to these by T-duality, as well as their 11-dimensional upliftings. For the IIB states we show in Section \ref{sec:cov IIB charges} that they transform under $SL(2,\mathbb{R})$ in the same representation as the BPS states they are associated to. Their T-duals on the IIA side always contain a Killing vector as part of their solution, and this is found to be faithfully represented in the structures of their generalised charges, in fact such observations were already made for the KK-monopoles and M9-brane in \cite{Callister:2007jy}. In this way we see that the discrepancies between the SUSY algebra and spectrum of BPS states described in the opening paragraphs only exists if one considers the flatspace SUSY algebra, but does not exist generally. 

The IIB 7-branes have been studied many times in the literature  (see \cite{Bergshoeff:2007aa,Bergshoeff:2006jj,Bergshoeff:2002mb} for some recent work, and references therein). In \cite{Meessen:1998qm}, extending the earlier works of \cite{Bergshoeff:1996ui,Lavrinenko:1997qa}, it was shown that performing a Scherk-Schwarz \cite{Scherk:1979zr} dimensional reduction of the IIB theory in a background containing the 7-branes leads to a triplet of $SL(2,\mathbb{R})$ covariant massive 9-dimensional SUGRAs, with the 7-branes becoming domain walls. It was also shown that on the IIA side the same 9-dimensional theories could be produced from a massive version of 11-dimensional SUGRA containing two M9-branes and a symmetric $2\times 2$ mass matrix. The Killing isometries associated with the M9-branes make the theory inherently non-covariant (in the spacetime sense) and define a 2-torus, and dimensional reduction over this sub-manifold leads to the massive 9-dimensional theories. The $SL(2,\mathbb{R})\subset GL(2,\mathbb{R})$ symmetry associated with this reduction is mapped to the IIB symmetry group \cite{Bergshoeff:1995as,Bergshoeff:1995cg,Bergshoeff:1995cw,Aspinwall:1995fw}. Dimensional reduction over just an $S^1$ on the other hand yields a non-covariant massive IIA theory, which is mapped to the IIB theory using the `massive' T-duality rules. The subset of these rules which map the IIB theory with a D7-brane to Romans' massive IIA SUGRA \cite{Romans:1985tz} were first given in \cite{Bergshoeff:1996ui} and used a subgroup of the full $SL(2,\mathbb{R})$ group to perform the Scherk-Schwarz reduction.

Our approach in this paper is to initially focus on the states and charges in 11-dimensional SUGRA, then relate these to the charges in IIA and IIB by performing dimensional reductions and T-dualities. We find that the IIB charges of the 7-branes and 9-brane quadruplet map to certain variations of the 11-dimensional KK-monopole and M9-brane charges, which form the appropriate multiplets in 11-dimensions. For example, carrying out a double dimensional reduction on the standard M9-brane gives either a D8-brane or KK8-monopole depending upon whether the reduction is along the gauged isometry or not \cite{Bergshoeff:1998bs}. Therefore reducing 11-dimensional SUGRA in a background of two non-parallel M9-branes over a single isometry direction will produce a non-covariant massive IIA theory containing both a D8-brane and KK8-monopole, which are the T-duals of the D7-brane and its S-dual partner, the NS7-brane, respectively \cite{Meessen:1998qm}. (By `S-dual partner' we mean the branes related by the discrete S-duality transformation.) However the triplet of IIB 7-branes contains a third member which in this paper we refer to as the `r7'-brane. One would therefore expect there to be a corresponding third brane in 11-dimensions along with the two M9-branes, and that these should transform as a triplet. We find that the charge for such a brane does exist and it has a structure that is essentially the same as the M9-brane but explicitly requires two distinct Killing vectors. Furthermore, we also find that there exists a triplet of 10-form gauge potentials which these states couple to.

Alternatively, one can consider the 11-dimensional KK-monopole. Performing a direct dimensional reduction on the KK-monopole gives either the D6-brane or KK6-monopole \cite{Eyras:1999at} depending upon whether the reduction is along the Taub-NUT direction or not. These states also map to the D7- and NS7-branes in IIB respectively. As with the M9-branes we would therefore expect there to be a third type of brane present in 11-dimensions related to the KK-monopole, and which maps to the IIB r7-brane forming a triplet of 11-dimensional states. Once again we find the charge of such a state, which has a structure similar to that of the KK-monopole charge but depends explicitly on two Killing vectors. 

A similar story exists for the quadruplet of 9-branes in IIB whose charges are mapped to variations of the M9-brane charge. The IIB 9-brane doublet on the other hand seems to follow a separate path and is unrelated to the M9-brane charges. 

To determine the structure of the charges we demand that they are closed expressions, and therefore in order to construct them we must know the field equations of the potentials to which the corresponding branes couple. The required field equations in IIB are known \cite{Dall'Agata:1998va,Bergshoeff:2005ac}, however their 11-dimensional counterparts have not yet been given. The first half of this paper is devoted to finding these equations. It is a tricky matter to determine all the field equations in 11-dimensional SUGRA directly so we use an indirect method here.  Our method will be to initially consider the massive SUGRA presented in \cite{Bergshoeff:1997ak}, which has a single M9-brane background and therefore only one mass parameter. Dimensional reduction over the `massive' Killing isometry associated with the M9-brane yields Romans' IIA theory. We will however introduce a second `non-massive' Killing isometry into the theory, such as that which occurs in the KK-monopole solution. The reason for doing this is that we can determine the field equations that intrinsically depend upon this Killing vector, for example that associated with its 8-form dual, by considering the associated field equations in Romans' theory which must have standard gauge properties. Once we have these equations it is then a simple matter to `promote' this Killing vector to a massive Killing vector and determine the full $SL(2,\mathbb{R})$ covariant equations.

In the rest of this paper when we are discussing one of the theories specifically we will refer to the 11-dimensional SUGRA containing a single M9-brane as simply massive 11-dimensional SUGRA, and that containing two M9-branes as $SL(2,\mathbb{R})$ covariant 11-dimensional SUGRA. The IIA theories obtained by the dimensional reductions of these will be referred to as Romans' IIA and non-covariant IIA respectively.

The organisation of this paper is as follows: In Sections \ref{sec:11-d field eqns}-\ref{sec:11-d cov SUGRA} we present the field equations for the massive 11- and 10-dimensional SUGRAs considered in this paper. Specifically in Section \ref{sec:11-d field eqns} we review the details of massive 11-dimensional SUGRA required for our purposes and then determine the field equations by relating the 11-dimensional fields to those produced by dimensionally reducing to Romans' IIA theory. Then in Section \ref{sec:10-d field eqns} we consider the 10-dimensional equations which are obtained through dimensional reduction and T-duality. In Section \ref{sec:11-d cov field eqns} we use the results of the previous sections to determine the 11-dimensional $SL(2,\mathbb{R})$ covariant field equations.
In Sections \ref{sec:11-d cov charges}-\ref{sec:IIB charges} we present the generalised charges for these massive SUGRA theories. We begin in Section \ref{sec:11-d cov charges} by constructing the $SL(2,\mathbb{R})$ covariant generalised charges. Next, we map these charges to IIB by first dimensionally reducing them to non-covariant IIA in Section \ref{sec:IIA charges}, then T-dualising these to IIB in Section \ref{sec:non-cov IIB charges}. We re-express these in Section \ref{sec:cov IIB charges} in an $SL(2,\mathbb{R})$ covariant manner, before discussing the general nature of the massive T-duality rules between IIB and non-covariant IIA in Section \ref{sec:cov T-duality}. We then summarise our results in Section \ref{sec:discussion}. We have five appendices. We give our conventions in Appendix \ref{sec:conventions}, and list the reduction rules from 11-dimensions to IIA in Appendix \ref{sec:reducs}. Then in Appendix \ref{sec:KK action} we show how the massive gauge transformation rules are compatible with the construction of the $SL(2,\mathbb{R})$ covariant kinetic term of the worldvolume action of the KK-monopoles. We determine the T-duality rules which produce the quadruplet of IIB potentials in Appendix \ref{sec:mapping 10-forms} and then finally in Appendix \ref{sec:IIB massive fields} we give the IIB field equations which map to those in $SL(2,\mathbb{R})$ covariant 11-dimensional SUGRA.

\section{Field equations in massive 11-dimensional SUGRA}\label{sec:11-d field eqns}

As mentioned in the introduction, the first theory we will consider is the massive version of 11-dimensional SUGRA presented in \cite{Bergshoeff:1997ak}. We will begin by reviewing the key characteristics of this theory that are relevant for our purposes.

The theory contains a scalar mass parameter $\hat{m}$.\footnote{We denote 11-dimensional objects with a hat.} The mass parameter can be introduced via an auxiliary 10-form gauge potential $\hat{A}^{(10)}$ \cite{Sato:1999bu,Bergshoeff:1996ui} which minimally couples to the M9-brane and has an 11-form field strength $\hat{F}^{(11)}$ which is related to $\hat{m}$ via a Hodge duality relation. In this way the M9-brane sources the mass parameter. The equation of motion for $\hat{m}$ restricts it to being piece-wise constant, with a possible discontinuity across the M9-brane.

The M9-brane spacetime solution \cite{Bergshoeff:1998bs} contains a Killing vector which is parallel to the brane worldvolume. The presence of this Killing vector makes the theory non-covariant and in this way circumvents the no-go theorem of constructing a massive 11-dimensional SUGRA \cite{Bautier:1997yp}. This `massive' Killing vector plays an intrinsic role in the theory. It acts as a gauged isometry not just in the M9-brane worldvolume action, but in all the brane worldvolume actions that are coupled to the theory \cite{Bergshoeff:1997ak,Lozano:1997ee,Ortin:1997jh,Eyras:1998kx}. It is assumed that no fields depend on the co-ordinate parametrising the massive isometry direction, in other words all Lie derivatives with respect to the Killing vector vanish. Denoting this Killing vector by $\hat{\alpha}$, we therefore have the following identity for an arbitrary $p$-form $\hat{Y}^{(p)}$:
\begin{eqnarray}
d(i_{\hat{\alpha}}\hat{Y}^{(p)})=-i_{\hat{\alpha}}d\hat{Y}^{(p)} .
\end{eqnarray}
We will make extensive use of this relation in what follows.

In order to reproduce Romans' massive IIA SUGRA the dimensional reduction must specifically be performed over the massive isometry direction. If the reduction occurs over a different isometry then a different non-covariant massive IIA theory is produced. We will not consider this latter option in the upcoming sections since this non-covariant massive IIA theory is merely a truncation of the theory obtained from reducing the $SL(2,\mathbb{R})$ 11-dimensional SUGRA.

A further example of the special role played by $\hat{\alpha}$ is its presence in the massive terms of the field equations. For example, the field equations for the 3-form potential $\hat{A}$ and its dual 6-form potential $\hat{C}$ were given in \cite{Bergshoeff:1997ak} as\footnote{Our definition of the mass parameter differs from \cite{Bergshoeff:1997ak} by a factor of 2.}
\begin{eqnarray}\label{eq:11-d A}
\hat{F}^{(4)}&=&d\hat{A}+\frac{1}{2}\hat{m}(i_{\hat{\alpha}}\hat{A})^2\\ \label{eq:11-d C}
\hat{F}^{(7)}&=&d\hat{C}-\frac{1}{2}d\hat{A}\wedge \hat{A}+\hat{m}i_{\hat{\alpha}}\hat{A}\wedge i_{\hat{\alpha}}\hat{C}-\frac{1}{3!}\hat{m}\hat{A}\wedge (i_{\hat{\alpha}}\hat{A})^2\nonumber \\ &&+\hat{m}i_{\hat{\alpha}}\hat{N}^{(8)}\end{eqnarray}
where $\hat{N}^{(8)}$ is the 8-form dual potential to $\hat{\alpha}$. We discuss this field in the next subsection. In our conventions closure of the supersymmetry algebra requires the Hodge duality relation $\hat{F}^{(7)}=\hat{\ast}\hat{F}^{(4)}$.

Rather than being gauge invariant, the above field strengths transform covariantly\footnote{The use of the term `covariant' here is completely separate from its use in describing $SL(2,\mathbb{R})$ `covariant' theories, and also `non-covariant` (in the spacetime sense) SUGRAs.} under massive gauge transformations. The covariant transformation is defined as follows for a $p$-form field $\hat{Y}^{(p)}$
\begin{eqnarray}\label{eq:11-d gauge trans}
\delta \hat{Y}^{(p)} = \hat{m}\hat{\lambda}\wedge i_{\hat{\alpha}}\hat{Y}^{(p)}
\end{eqnarray}
where $\hat{\lambda}=i_{\hat{\alpha}}\hat{\chi}$ and $\hat{\chi}$ is the standard 2-form gauge parameter of $\hat{A}$. It is stated in \cite{Bergshoeff:1997ak} that all the $p$-form fields with the exception of $\hat{A}$ and $\hat{C}$ transform according to (\ref{eq:11-d gauge trans}). However in this reference the higher rank potentials were not considered. We find that generally the rule (\ref{eq:11-d gauge trans}) does not apply to gauge potentials, which usually undergo more complicated transformations. Furthermore, we give an example in the next subsection of field strengths that do not transform simply according to (\ref{eq:11-d gauge trans}). The situation is clearer in the $SL(2,\mathbb{R})$ covariant theory, where we propose that the general rules depend on the $SL(2,\mathbb{R})$ representation of the field.
 
The full transformation of $\hat{A}$ is given by
\begin{eqnarray}\label{eq:11-d A gauge}
\delta \hat{A}=d\hat{\chi}+\hat{m}\hat{\lambda}\wedge i_{\hat{\alpha}}\hat{A}
\end{eqnarray} 
and acts as a connection-field for the massive gauge transformations.  The total connection is then given by\footnote{We denote these objects with a tilde to avoid confusion with the bilinear forms we use in later sections.} 
\begin{eqnarray}
\hat{\tilde{\Omega}}=\hat{\tilde{\omega}}(\hat{e})+\hat{\tilde{K}}
\end{eqnarray}
where $\hat{\tilde{K}}$ is the contorsion tensor given by
\begin{eqnarray}
\hat{\tilde{K}}_{\hat{\mu}_1\hat{\mu}_2\hat{\mu}_3}=\frac{1}{2}\hat{m}\biggl[\hat{T}_{\hat{\mu}_1\hat{\mu}_3\hat{\mu}_2}+\hat{T}_{\hat{\mu}_2\hat{\mu}_3\hat{\mu}_1}-\hat{T}_{\hat{\mu}_1\hat{\mu}_2\hat{\mu}_3}\biggr]
\end{eqnarray}
with the torsion tensor given by
\begin{eqnarray}
\hat{\tilde{T}}_{\hat{\mu}_1\hat{\mu}_2}^{\phantom{\hat{\mu}_1\hat{\mu}_2}\hat{\mu}_3}=-\hat{m}(i_{\hat{\alpha}}\hat{A})_{\hat{\mu}_1\hat{\mu}_2}\hat{\alpha}^{\hat{\mu}_3} .
\end{eqnarray}
We define the field equation for the 10-form potential $\hat{A}^{(10)}$ as in \cite{Callister:2007jy}
\begin{eqnarray}\label{eq:11-d F11}
\nonumber i_{\hat{\alpha}}\hat{F}^{(11)}&=&-d(i_{\hat{\alpha}}\hat{A}^{(10)})-i_{\hat{\alpha}}\hat{F}\wedge i_{\hat{\alpha}}\hat{N}^{(8)}+\frac{1}{4!}i_{\hat{\alpha}}(d\hat{A}\wedge \hat{A}\wedge (i_{\hat{\alpha}}\hat{A})^2)\\ && +\frac{1}{5!}\hat{m}(i_{\hat{\alpha}}\hat{A})^5
\end{eqnarray}
where $\hat{F}^{(11)}$ is related to $\hat{m}$ through $\hat{F}^{(11)}=-\hat{\ast}|\hat{\alpha}|^4\hat{m}$. Equation (\ref{eq:11-d F11}) was defined so that it reduces to the field strength equation of the 9-form Ramond-Ramond potential in Romans' IIA after performing a double dimensional reduction. The reduction rules for the potentials are given in Appendix \ref{sec:reducs}. Note that since $\hat{F}^{(11)}$ is of maximum rank it always has a component parallel to $\hat{\alpha}$, which has been made explicit above by the overall contraction with $\hat{\alpha}$. While this is correct, the gauge algebras and field equations seem to have a structure that is independent of the spacetime dimension \cite{Cremmer:1998px,Bergshoeff:2005ac}. This explicit contraction masks the full structure of the field equation and it is possible to write $\hat{F}^{(11)}$ without it. The full structure can be inferred from demanding that $\hat{F}^{(11)}$ be gauge covariant without making use of the dimensionality of the background, which will eventually be required to construct the M9-brane charges in Section \ref{sec:11-d cov charges}. In doing this we find that it is only possible to construct such a field equation if we introduce a 12-form potential, which is included in an analogous fashion to $\hat{N}^{(8)}$ in (\ref{eq:11-d C}). One can then consider the field equation and gauge transformations of this 12-form potential, which in turn leads to the introduction of a 14-form potential. Repeating the process uncovers an infinite tower of gauge potentials. We will delay giving the general structure to these equations until Section \ref{sec:11-d cov field eqns} where they are dealt with more systematically in an $SL(2,\mathbb{R})$ covariant fashion.

In the above discussion we have tried to emphasise the special role $\hat{\alpha}$ plays in the theory. It is therefore important to distinguish between $\hat{\alpha}$ and any other isometry that may be present in a given spacetime solution, for example the Taub-NUT isometry associated with the KK-monopole. In the following we will denote such a non-massive Killing vector by $\hat{\beta}$. 

When considering solutions where both types of Killing vector are in principal present, there are two possibilities: either both Killing vectors are completely distinct and should therefore be treated separately, or they coincide in which case the massive isometry plays both roles. This distinction is important for example when one wants to perform a dimensional reduction of the KK-monopole in massive 11-d SUGRA to find a solution to Romans' IIA as in \cite{Bergshoeff:1998ef,Eyras:1998kx}. If one wants to produce the D6-brane solution then it is sufficient to assume both the Taub-NUT and massive Killing vectors coincide, however if one wants to produce either the IIA KK-monopole or KK6-monopole by performing the reduction over a worldvolume or standard transverse direction then one must distinguish between the Killing vectors in the massive 11-d theory.

It is known that when such isometries are present, the Killing vectors play a similar role to gauge potentials, with the KK-monopole magnetically coupling to the Taub-NUT Killing vector. However, due to the different roles played by $\hat{\alpha}$ and $\hat{\beta}$ in the massive theory, their associated field equations will differ. When constructing the generalised charge for the KK-monopole it is therefore important to check under which of the situations just described it is closed. The case where the Killing vectors are taken to coincide was discussed in \cite{Callister:2007jy} using the partial field strength equation of the 8-form potential dual to $\hat{\alpha}$ given in \cite{Bergshoeff:1997ak,Sato:1999bu,Sato:2000mw}. In the following subsections we will construct the full 8-form potential field equations by comparing the 11-dimensional equations with the equations in Romans' IIA produced after dimensional reduction. We stress that knowing the $\hat{\beta}$-type equations is important for constructing the fields in $SL(2,\mathbb{R})$ covariant 11-d SUGRA.

\subsection{Massive Killing vector field equations}\label{sec:11-d massive field eqns}

The field strength $\hat{G}^{(2)}$ of the massive isometry $\hat{\alpha}$ has been given previously as \cite{Sato:1999bu,Sato:2000mw}
\begin{eqnarray} \label{eq:11-d G2}
\hat{G}^{(2)}&=&d\hat{\alpha}+\hat{m}\hat{R}_{\alpha}^2i_{\hat{\alpha}}\hat{A}
\end{eqnarray}   
where $\hat{R}_{\alpha}^2=|\hat{\alpha}|^2$ is the square of the radius of the massive compact isometry. 

Looking at (\ref{eq:11-d G2}) and using the massive gauge transformation rule of $\hat{\alpha}$ which is $\delta \hat{\alpha}=\hat{m}\hat{R}_{\alpha}^2\hat{\lambda}$, we confirm that $\hat{G}^{(2)}$ transforms according to (\ref{eq:11-d gauge trans}). Furthermore, the definition (\ref{eq:11-d G2}) reproduces the field equation for the 1-form gauge potential in IIA after performing a dimensional reduction along $\hat{\alpha}$. 

We define the 9-form field strength $\hat{G}^{(9)}=\hat{\ast}\hat{G}^{(2)}$ with potential $\hat{N}^{(8)}$. The field equation for $i_{\hat{\alpha}}\hat{N}^{(8)}$ has been given previously \cite{Bergshoeff:1997ak,Sato:1999bu,Sato:2000mw} and is sufficient for constructing the generalised charge for the massive KK-monopole in the instance where $\hat{\alpha}$ and the Taub-NUT isometry coincide \cite{Callister:2007jy}. However it is straightforward to find the full equation by considering its reduction to IIA. 

The reduction rule for $\hat{G}^{(9)}$ can be inferred from the reduction of $\hat{G}^{(2)}$ which is:
\begin{eqnarray} \label{eq:G2 reducs}	
\hat{G}^{(2)} &\rightarrow&e^{\frac{4}{3}\phi}dC^{(1)}+e^{\frac{4}{3}\phi}mB+\frac{4}{3}e^{\frac{4}{3}\phi}d\phi\wedge (C^{(1)}+dz)\nonumber \\ &=& e^{\frac{4}{3}\phi}F^{(2)}+\frac{4}{3}e^{\frac{4}{3}\phi}d\phi\wedge (C^{(1)}+dz)
\end{eqnarray}
where we have used an adapted co-ordinate system where $\hat{\alpha}^{\hat{\mu}}=\hat{\delta}^{\hat{\mu}z}$. Refer to \cite{Callister:2007jy} for our definitions of the standard IIA fields. By working in an orthonormal frame it is then easy enough to deduce the reduction rules for $\hat{G}^{(9)}$ as being
\begin{eqnarray}\label{eq:11-d G9 reducs}
\hat{G}^{(9)}\rightarrow \frac{4}{3}H^{(9)}+F^{(8)}\wedge (C^{(1)} + dz)
\end{eqnarray}
where $H^{(9)}$ is the 9-form field strength dual to the dilaton `field strength', explicitly $H^{(9)}=e^{-2\phi}\ast d\phi$. Varying the massive IIA action with respect to $\phi$ yields the following equation of motion \cite{Cremmer:1998px}
\begin{eqnarray} \label{eq:IIA dH9}
dH^{(9)}&=&\frac{1}{2} H^{(7)}\wedge H+\frac{1}{4}F^{(4)}\wedge F^{(6)}-\frac{3}{4}F^{(2)}\wedge F^{(8)}+\frac{5}{4}mF^{(10)} .
\end{eqnarray}
Using this relation and the Bianchi identity for $F^{(8)}$ we can uplift and determine the Bianchi identity for $\hat{G}^{(9)}$ as being
\begin{eqnarray}\label{eq:11-d G9 bianchi}
d\hat{G}^{(9)}=-\frac{2}{3}\hat{F}^{(7)}\wedge i_{\hat{\alpha}}\hat{F}-\frac{1}{3}\hat{F}\wedge i_{\hat{\alpha}}\hat{F}^{(7)}+\frac{5}{3}\hat{m}i_{\hat{\alpha}}\hat{F}^{(11)}-\hat{m}i_{\hat{\alpha}}\hat{G}^{(9)}\wedge i_{\hat{\alpha}}\hat{A} .
\end{eqnarray}
The last term here is typical for field strengths that transform according to (\ref{eq:11-d gauge trans}) and it is a simple matter to show that the gauge transformations of both sides match if all the field strengths transform according to (\ref{eq:11-d gauge trans}), and using (\ref{eq:11-d A gauge}) for $\hat{A}$. In this paper we do not explicitly determine the full gauge transformation rules for the higher rank potentials since they are not required to construct the generalised charges. Instead we determine the Bianchi identities of the field strengths and check that they transform correctly under the massive gauge transformations. From this we construct the field equations which guarantees that appropriate gauge transformations for the potentials exist. In the current case we use (\ref{eq:11-d G9 bianchi}) to define the 8-form potential $\hat{N}^{(8)}$ as
\begin{eqnarray} \label{eq:11-d G9}
\hat{G}^{(9)}&=&d\hat{N}^{(8)}-\frac{2}{3}i_{\hat{\alpha}}\hat{F}\wedge \hat{C}+\frac{1}{3}\hat{F}\wedge i_{\hat{\alpha}}\hat{C}+\frac{1}{3!}\hat{F}\wedge i_{\hat{\alpha}}\hat{A}\wedge \hat{A}\nonumber \\ &&-\frac{5}{3}\hat{m}i_{\hat{\alpha}}\hat{A}^{(10)}+\hat{m}i_{\hat{\alpha}}\hat{A}\wedge i_{\hat{\alpha}}\hat{N}^{(8)}-\frac{1}{4!}\hat{m}(i_{\hat{\alpha}}\hat{A})^3\wedge \hat{A} .
\end{eqnarray}
Contracting $\hat{\alpha}$ with (\ref{eq:11-d G9}) yields the field equation for $i_{\hat{\alpha}}\hat{G}^{(9)}$ 
\begin{eqnarray} \label{eq:11-d ialphaG9}
i_{\alpha}\hat{G}^{(9)}&=&-d(i_{\hat{\alpha}}\hat{N}^{(8)})+i_{\hat{\alpha}}\hat{F}\wedge i_{\hat{\alpha}}\hat{C}+\frac{1}{3!}i_{\hat{\alpha}}(\hat{F}\wedge \hat{A} \wedge i_{\hat{\alpha}}\hat{A})\nonumber \\ &&-\frac{1}{4!}\hat{m}(i_{\hat{\alpha}}\hat{A})^4
\end{eqnarray}
which agrees with the previous definitions in the literature. Note however that given (\ref{eq:11-d ialphaG9}) then (\ref{eq:11-d G9}) is not the equation one would produce by naively `extracting' the contraction with $\hat{\alpha}$ as proposed in \cite{Bergshoeff:1997ak}. Some terms are only present in the full equation whilst the term involving  $\hat{C}$ actually splits into two terms in the full equation. This is a characteristic that applies to many of the fields considered in this paper.

\subsection{Non-massive Killing vector field equations}\label{sec:11-d beta eqns}

We now repeat the method of the previous subsection for a non-massive Killing vector $\hat{\beta}$ with the assumption that all Lie derivatives with respect to $\hat{\beta}$ vanish. Here we have the added complication that the 11-dimensional field equations now involve both $\hat{\alpha}$ and $\hat{\beta}$. We do know however that the structure of the field equations here should only differ from the analogous ones given above by their massive terms and that these should be the same as those given previously when we take $\hat{\alpha}$ and $\hat{\beta}$ to coincide. In determining these equations we will also find it necessary to introduce a second 10-form potential.

In order to deduce the field equation for $\hat{\beta}$ we must first determine its gauge transformations. These can be inferred from the gauge transformations of the IIA fields that $\hat{\beta}$ reduces to. The reduction rules for $\hat{\beta}$ differ from those of $\hat{\alpha}$ and are given by
\begin{eqnarray}
\hat{\beta}\rightarrow e^{-\frac{2}{3}\phi}\beta+e^{\frac{4}{3}\phi}i_{\beta}C^{(1)}(C^{(1)}+dz) .
\end{eqnarray}
In Romans' massive IIA theory $\beta$ and $\phi$ do not have massive gauge transformations while $C^{(1)}$ transforms according to $\delta C^{(1)}=m\lambda$ \cite{Bergshoeff:2006qw} where $\lambda$ is obtained from the dimensional reduction of $\hat{\lambda}$ (specifically we have $\hat{\lambda}_{\hat{\mu}}\rightarrow -\lambda_{\mu}$) and is the gauge parameter of the NS-NS 2-form potential. We therefore conclude
\begin{eqnarray}\label{eq:11-d beta gauge}
\delta \hat{\beta}=\hat{m}\hat{\lambda} \ i_{\hat{\beta}}\hat{\alpha}  + \hat{m}i_{\hat{\beta}}\hat{\lambda} \ \hat{\alpha} .
\end{eqnarray}
We note that $\hat{\beta}$ does not transform covariantly according to (\ref{eq:11-d gauge trans}) but rather has an extra term present involving $i_{\hat{\beta}}\hat{\lambda}$. Therefore we would not necessarily expect the field strength of $\hat{\beta}$, which we denote $\hat{S}^{(2)}$,  to transform covariantly either. In fact, given (\ref{eq:11-d beta gauge}) we were unable to define $\hat{S}^{(2)}$ such that it did transform covariantly. We therefore propose the following definition\footnote{We use the shorthand notation $i_{\hat{\beta}\hat{\alpha}} \hat{A}_{\hat{\mu}}$ to express the double contraction $i_{\hat{\beta}}(i_{\hat{\alpha}}\hat{A})_{\hat{\mu}}=\hat{\beta}^{\hat{\rho}}\hat{\alpha}^{\hat{\nu}}\hat{A}_{\hat{\nu}\hat{\rho}\hat{\mu}}$.}
\begin{eqnarray}\label{eq:11-d S2}
\hat{S^{(2)}}=d\hat{\beta}+\hat{m}i_{\hat{\beta}} \hat{\alpha} \ i_{\hat{\alpha}}\hat{A} -\hat{m}i_{\hat{\beta}\hat{\alpha}}\hat{A}\wedge \hat{\alpha}
\end{eqnarray}
which transforms as
\begin{eqnarray}
\delta \hat{S}^{(2)}=\hat{m}\hat{\lambda}\wedge i_{\hat{\alpha}}\hat{S}^{(2)}+\hat{m}i_{\hat{\beta}}\hat{\lambda} \ \hat{G}^{(2)}
\end{eqnarray}
which is the field strength generalisation of (\ref{eq:11-d beta gauge}). Note that we have taken $\hat{\alpha}^{\hat{\mu}}$ and $\hat{\beta}^{\hat{\mu}}$ to be invariant which can also be shown to be correct from considering their dimensional reduction to Romans' IIA.

We mention that in \cite{Eyras:1998kx} it was proposed that $\hat{\beta}$ should transform according to (\ref{eq:11-d gauge trans}) with there being a compensating transformation for $\hat{\beta}^{\hat{\mu}}$. It was argued that these transformations were necessary to construct a gauge invariant kinetic term in the worldvolume action of the massive KK-monopole. However, we have found that it is also possible to construct an appropriate term using the rule (\ref{eq:11-d beta gauge}). We give the explicit details of how to do this in Appendix \ref{sec:KK action}. Furthermore, when considering the higher rank potentials below we were only able to construct field equations with consistent gauge properties if transformations analogous to (\ref{eq:11-d beta gauge}) were used. We will return to this point in Section \ref{sec:11-d cov SUGRA} when we consider the $SL(2,\mathbb{R})$ covariant theory. 

The reduction rule for $\hat{S}^{(2)}$ is calculated as
\begin{eqnarray}\label{eq:11-d S2 reducs}
\hat{S}^{(2)}&\rightarrow& d(e^{-\frac{2}{3}\phi}\beta+e^{\frac{4}{3}\phi}i_{\beta}C^{(1)}C^{(1)})+me^{\frac{4}{3}\phi}i_{\beta}C^{(1)}B\nonumber \\ &&-me^{\frac{4}{3}\phi}i_{\beta}B\wedge C^{(1)}+(d(e^{\frac{4}{3}\phi}i_{\beta}C^{(1)})-me^{\frac{4}{3}\phi}i_{\beta}B)\wedge dz\nonumber \\
&=&e^{-\frac{2}{3}\phi}G^{(2)}-\frac{2}{3}e^{-\frac{2}{3}\phi}d\phi \wedge \beta+e^{\frac{4}{3}\phi}i_{\beta}C^{(1)}F^{(2)}\nonumber \\ &&+e^{\frac{4}{3}\phi}(\frac{4}{3}i_{\beta}C^{(1)}d\phi+X^{(1)})\wedge (C^{(1)}+dz)
\end{eqnarray}
where we have defined $G^{(2)}=d\beta$ and $X^{(1)}=d(i_{\beta}C^{(1)})-mi_{\beta}B$. Due to our assumption that no fields depend on $\hat{\beta}$ in 11 dimensions, it follows that ${\cal L}_{\beta}C^{(1)}=0$ and therefore $X^{(1)}=-i_{\beta}F^{(2)}$, which we treat as an independent field. Such a field is not intrinsically present in Romans' theory but is sourced by the KK6-monopole which magnetically couples to the scalar potential $i_{\beta}C^{(1)}$. The situation is essentially the same as the KK-monopole magnetically coupling to the Taub-NUT Killing isometry which also is not an intrinsic field of the theory. Note that $X^{(1)}$ is gauge invariant since $F^{(2)}$ is gauge invariant. 

Now following the previous case we define a 9-form field strength as $\hat{S}^{(9)}=\hat{\ast}\hat{S}^{(2)}$ and use (\ref{eq:11-d S2 reducs}) to determine how it reduces. Once again this is most easily seen by working in an orthonormal frame. However it seems like the second term in the second equation of (\ref{eq:11-d S2 reducs}) could be difficult to dualise. In fact due to the way this term involves $\beta$, it does dualise and we find
\begin{eqnarray}\label{eq:11-d S9 reducs}
\hat{S}^{(9)}&\rightarrow& X^{(9)}+\frac{4}{3}i_{\beta}C^{(1)}H^{(9)}+(G^{(8)}+i_{\beta}C^{(1)}F^{(8)}+\frac{2}{3}i_{\beta}H^{(9)})\wedge C^{(1)}\nonumber \\ && +(G^{(8)}+i_{\beta}C^{(1)}F^{(8)}+\frac{2}{3}i_{\beta}H^{(9)})\wedge dz
\end{eqnarray}
where we have defined the field strengths $X^{(9)}=e^{-2\phi}\ast X^{(1)}$ and $G^{(8)}=e^{-2\phi}\ast G^{(2)}$.

We now set about determining the field equations for the three unknown field strengths in play: $G^{(8)}$ and $X^{(9)}$ in IIA, and $\hat{S}^{(9)}$ in 11-dimensions. Finding these equations is slightly more problematic than for the analogous fields in the previous subsection. This stems from the fact that it is not straightforward to determine the Bianchi identities of the IIA fields by using a variational principal as was done for $H^{(9)}$, since in this case we are not dealing with intrinsic fields of the theory.

One can attempt to construct the equations directly by demanding gauge invariance but this also is insufficient. Firstly, for $G^{(8)}$ gauge invariance alone does not fully determine the field equation since there is effectively another gauge invariant 8-form field, namely $i_{\beta}H^{(9)}$, and so any combination of these two fields is gauge invariant. The field equation for $i_{\beta}G^{(8)}$ has been given in the literature previously, but this does not solve the problem since we cannot uniquely determine the structure of the full equation from this as we discussed at the end of the previous subsection for the case of $\hat{G}^{(9)}$. Secondly, when considering $X^{(9)}$ it appears impossible to construct a gauge invariant field strength using just the IIA fields we have so far considered. A similar situation also occurs for $\hat{S}^{(9)}$ irrespective of whether we assume the gauge properties to follow (\ref{eq:11-d gauge trans}) or (\ref{eq:11-d beta gauge}). To solve this problem it is necessary to introduce a second 10-form potential in the 11-d theory which appears in the definition of $\hat{S}^{(9)}$ and reduces to a 9-form potential in IIA that appears in the field equation for $X^{(9)}$.

We now outline a method which can be used to determine these equations. This method determines the field equations piece by piece and is fairly cumbersome, we therefore do not give the explicit results of the intermediate calculations.

Since $\hat{G}^{(9)}$ is fully known, the massless equation for $\hat{S}^{(9)}$ is also known, since these fields can only differ in their massive terms. Using this fact and the reduction rule (\ref{eq:11-d S9 reducs}), we see that a double dimensional reduction of $\hat S^{(9)}$ will give us the massless terms of $G^{(8)}$ since the equations for $F^{(8)}$ and $i_{\beta}H^{(9)}$ are already known. The massive terms are then uniquely determined by demanding gauge invariance under the massive gauge transformations of IIA. Equivalently we find that the only suitable massive Bianchi identity is given by
\begin{eqnarray}\label{eq:IIA G8 bianchi}
dG^{(8)}&=&\frac{1}{2}i_{\beta}F^{(2)}\wedge F^{(8)}-\frac{1}{2}F^{(2)}\wedge i_{\beta}F^{(8)}-i_{\beta}H\wedge H^{(7)} +\frac{1}{2}F^{(4)}\wedge i_{\beta}F^{(6)}\nonumber \\ && -\frac{1}{2}i_{\beta}F^{(4)}\wedge F^{(6)} +\frac{1}{2}mi_{\beta}F^{(10)} .
\end{eqnarray}
We give the equation for $G^{(8)}$ in the next section.

Given $G^{(8)}$ we can uplift to 11 dimensions and determine the full structure of $i_{\hat{\alpha}}\hat{S}^{(9)}$, i.e. those components parallel to $\hat{\alpha}$. This puts constraints on the massive terms in $\hat{S}^{(9)}$ but it does not determine them fully. We then proceed by considering the components parallel to $\hat{\beta}$, i.e. $i_{\hat{\beta}}\hat{S}^{(9)}$, and their dimensional reduction to IIA which we see from (\ref{eq:11-d S9 reducs}) will produce the terms in the field equation for $i_{\beta}X^{(9)}$. We can then follow the example of $G^{(8)}$ by trying to determine the full structure of the massive terms by using the massless terms and demanding gauge invariance. However in this case, it seems impossible to achieve gauge invariance without the introduction of a new 9-form potential. We illustrate this point by considering the Bianchi identities. We find that the only gauge invariant massive Bianchi identity for $i_{\beta}X^{(9)}$ consistent with the massless terms is given by
\begin{eqnarray}\label{eq:IIA X9 bianchi}
i_{\beta}dX^{(9)}&=&-2i_{\beta}F^{(2)}\wedge i_{\beta}H^{(9)}-i_{\beta}(G^{(8)}\wedge F^{(2)}) -i_{\beta}F^{(4)}\wedge i_{\beta}H^{(7)}\nonumber \\ &&
\end{eqnarray}
which turns out to not explicitly include any massive terms.

Uplifting (\ref{eq:IIA X9 bianchi}) to 11 dimensions using (\ref{eq:11-d S9 reducs}) along with (\ref{eq:IIA G8 bianchi}) and the other appropriate IIA equations will give the form of $i_{\hat{\beta}}d\hat{S}^{(9)}$. Then taking into account the terms we have already calculated for $i_{\hat{\alpha}}\hat{S}^{(9)}$ one determines the following Bianchi identity
\begin{eqnarray}\label{eq:11-d S9 bianchi}
d\hat{S}^{(9)}&=&-\frac{1}{3}\hat{F}\wedge i_{\hat{\beta}}\hat{F}^{(7)}+\frac{2}{3}i_{\hat{\beta}}\hat{F}\wedge \hat{F}^{(7)}-\frac{1}{3}\hat{m}i_{\hat{\beta}}\hat{F}^{(11)}\nonumber \\ &&+2\hat{m}i_{\hat{\alpha}}\hat{H}^{(11)}-\hat{m}i_{\hat{\alpha}}\hat{A}\wedge i_{\hat{\alpha}}\hat{S}^{(9)}-\hat{m}i_{\hat{\alpha}\hat{\beta}}\hat{A}\wedge \hat{G}^{(9)} .
\end{eqnarray}
In order to achieve consistency with (\ref{eq:IIA X9 bianchi}) we have been forced to introduce the new 11-form field strength $\hat{H}^{(11)}$ which is defined so that it obeys the following reduction rule to IIA
\begin{eqnarray}
i_{\hat{\alpha}}\hat{H}^{(11)}\rightarrow i_{\beta}C^{(1)}F^{(10)} .
\end{eqnarray}
From this we can conclude that it must be related to the 11-dimensional mass parameter via $\hat{H}^{(11)}=-(|\hat{\alpha}|^2i_{\hat{\beta}}\hat{\alpha}) \hat{\ast} \hat{m}$, and therefore seems to be some kind of generalisation of the standard 11-form field strength $\hat{F}^{(11)}$. We will show in Section \ref{sec:11-d cov field eqns} that these two fields along with a third field given in (\ref{eq:11-d D10}) form a triplet in $SL(2,\mathbb{R})$ covariant 11-dimensional SUGRA. Note that in the instance where $\hat{\alpha}$ and $\hat{\beta}$ coincide we have $\hat{H}^{(11)}\rightarrow\hat{F}^{(11)}$ and $\hat{S}^{(9)}\rightarrow\hat{G}^{(9)}$ and then (\ref{eq:11-d S9 bianchi}) becomes equivalent to (\ref{eq:11-d G9 bianchi}) which is a non-trivial check on the massive terms.
 
It is straightforward to show that both sides of (\ref{eq:11-d S9 bianchi}) have the same massive gauge transformations if $\hat{S}^{(9)}$ and $\hat{H}^{(11)}$ transform in analogy to (\ref{eq:11-d beta gauge}) as
\begin{eqnarray}
\delta\hat{S}^{(9)}&=&\hat{m}\hat{\lambda}\wedge i_{\hat{\alpha}}\hat{S}^{(9)}+\hat{m}i_{\hat{\beta}}\hat{\lambda} \ \hat{G}^{(9)} \\ \label{eq:11-d H11 gauge}
\delta \hat{H}^{(11)}&=&\hat{m}\hat{\lambda}\wedge i_{\hat{\alpha}}\hat{H}^{(11)}+\hat{m}i_{\hat{\beta}}\hat{\lambda} \ \hat{F}^{(11)} .
\end{eqnarray}
The first term on the RHS of (\ref{eq:11-d H11 gauge}) is identically zero since it is a maximum rank equation and must therefore have a free index tangential to $\hat{\alpha}$, however we have included it to emphasise the algebraic structure of the transformations.

Using (\ref{eq:11-d S9 bianchi}) and the correspondence between $\hat{S}^{(9)}$ and $\hat{G}^{(9)}$ as well as that between $\hat{H}^{(11)}$ and $\hat{F}^{(11)}$ we can define $\hat{S}^{(9)}$ in terms of an 8-form potential $\hat{T}^{(8)}$
\begin{eqnarray}\label{eq:11-d S9}
\hat{S}^{(9)}&=&d\hat{T}^{(8)}+\frac{1}{3}\hat{F}\wedge i_{\hat{\beta}}\hat{C}-\frac{2}{3}i_{\hat{\beta}}\hat{F}\wedge \hat{C}+\frac{1}{6}\hat{F}\wedge i_{\hat{\beta}}\hat{A}\wedge \hat{A}\nonumber \\ 
&& +\frac{1}{3}\hat{m}i_{\hat{\beta}}\hat{A}^{(10)} -2\hat{m}i_{\hat{\alpha}}\hat{B}^{(10)} +\hat{m}i_{\hat{\alpha}\hat{\beta}}\hat{A}\wedge \hat{N}^{(8)}\nonumber \\ 
&&  +\hat{m}i_{\hat{\alpha}}\hat{A}\wedge i_{\hat{\alpha}}\hat{T}^{(8)} -\frac{1}{4!}\hat{m}i_{\hat{\beta}}\hat{A}\wedge (i_{\hat{\alpha}}\hat{A})^2\wedge \hat{A}
\end{eqnarray}
where $\hat{B}^{(10)}$ is the gauge potential of $\hat{H}^{(11)}$ which we define below. Due to the way that this equation was determined it is possible that there could also be massive terms present of the type $\hat{m}i_{\hat{\alpha}\hat{\beta}}(\textrm{11-form})$ which would have been missed by our method. However, adding any such term would appear to spoil the gauge transformation of (\ref{eq:11-d S9}) and we therefore do not think that any such terms should be present. This is further supported when we construct the generalised charges in Section \ref{sec:11-d cov charges} where (\ref{eq:11-d S9}) is precisely the field equation necessary for the charges to be closed.

In order to arrive at the Bianchi identity (\ref{eq:11-d S9 bianchi}) from (\ref{eq:11-d S9}) we must use the field equations of $i_{\hat{\beta}}\hat{A}^{(10)}$ and $i_{\hat{\alpha}}\hat{B}^{(10)}$. Carrying this out however does not determine the individual field strengths, but rather the specific combination that appears in (\ref{eq:11-d S9}). Furthermore, even if they could be individually determined, due to the contractions with the Killing vectors they would not reflect the full structure of the field equations as we discussed when we gave (\ref{eq:11-d F11}). Because of this fact, here we will only give the field equation for $i_{\hat{\beta}\hat{\alpha}}\hat{B}^{(10)}$. This is sufficient for most purposes since in practice the equation will always have two components parallel to the isometry directions. It is given by
\begin{eqnarray}\label{eq:11-d H11}
i_{\hat{\beta}\hat{\alpha}}\hat{H}^{(11)} &=&d(i_{\hat{\beta}\hat{\alpha}}\hat{B}^{(10)}) -\frac{1}{2}i_{\hat{\beta}}(i_{\hat{\alpha}}\hat{F}\wedge i_{\hat{\alpha}}\hat{T}^{(8)}) +\frac{1}{2}i_{\hat{\alpha}}(i_{\hat{\beta}}\hat{F}\wedge i_{\hat{\beta}}\hat{N}^{(8)})\nonumber \\
&& +\frac{1}{4!}i_{\hat{\beta}\hat{\alpha}}(\hat{F}\wedge i_{\hat{\alpha}}\hat{A}\wedge i_{\hat{\beta}}\hat{A}\wedge \hat{A}) -\hat{m}i_{\hat{\beta}\hat{\alpha}}\hat{A}\wedge i_{\hat{\beta}\hat{\alpha}}\hat{A}^{(10)} \nonumber \\
&& -\frac{1}{16}\hat{m}i_{\hat{\beta}}\hat{A}\wedge (i_{\hat{\alpha}} \hat{A})^3\wedge i_{\hat{\beta}\hat{\alpha}}\hat{A}) .
\end{eqnarray}
The full structure of the 11-form field equations is given in an $SL(2,\mathbb{R})$ covariant fashion in Section \ref{sec:11-d cov SUGRA}.

\section{Field equations in 10-dimensional SUGRAs}\label{sec:10-d field eqns}

\subsection{Field equations in IIA from 11-dimensions}

We now explicitly give the IIA equations that are produced from dimensionally reducing the equations of the last section. We will then T-dualise these in the next subsection. It is a simple matter to perform these calculations using the reduction rules for the potentials given in Appendix \ref{sec:reducs}. 

From (\ref{eq:11-d G9 reducs}) and (\ref{eq:11-d G9}) we find the equation for $H^{(9)}$ is given by
\begin{eqnarray}\label{eq:IIA H9}
H^{(9)}&=&d\phi^{(8)}+\frac{1}{2}H\wedge B^{(6)}+\frac{1}{4}(F^{(4)}+H\wedge C^{(1)})\wedge C^{(5)}-\frac{3}{4}F^{(8)}\wedge C^{(1)} \nonumber \\ &&+\frac{5}{4}mC^{(9)}-m\frac{3}{4}C^{(7)}\wedge B+\frac{1}{4!}mC^{(3)}\wedge (B)^3
\end{eqnarray}
where $\phi^{(8)}$ is the 8-form magnetic dual of the dilaton. This equation has been previously given in \cite{Cremmer:1998px,Bergshoeff:2006qw}.

From (\ref{eq:11-d S9 reducs}) and (\ref{eq:11-d S9}) we find the equation for $G^{(8)}$ is given by 
\begin{eqnarray}\label{eq:IIA G8}
G^{(8)}&=&dN^{(7)}-\frac{1}{2}i_{\beta}C^{(1)}\wedge F^{(8)}+\frac{1}{2}i_{\beta}F^{(8)}\wedge C^{(1)}-i_{\beta}H\wedge B^{(6)}\nonumber \\ &&-\frac{1}{2}(F^{(4)}+H\wedge C^{(1)})\wedge i_{\beta}C^{(5)}+\frac{1}{2}i_{\beta}(F^{(4)}+H\wedge C^{(1)})\wedge C^{(5)}\nonumber \\ &&-\frac{1}{2}H\wedge i_{\beta}C^{(3)}\wedge C^{(3)}-\frac{1}{2}mi_{\beta}B\wedge C^{(7)}+\frac{1}{2}mB\wedge i_{\beta}C^{(7)}-\frac{1}{2}mi_{\beta}C^{(9)}\nonumber \\ &&-\frac{1}{4}mi_{\beta}B\wedge (B)^2\wedge C^{(3)}+\frac{1}{12}mi_{\beta}C^{(3)}\wedge (B)^3
\end{eqnarray}
where $N^{(7)}$ is the 7-form magnetic dual of the Killing vector $\beta$. Under contraction with $\beta$ (\ref{eq:IIA G8}) simplifies to the equation for $i_{\beta}G^{(8)}$ given in \cite{Callister:2007jy}.\footnote{In that reference the Killing isometry was denoted by $\alpha$.}

From (\ref{eq:11-d S9 reducs}) and (\ref{eq:11-d S9}) we find the equation for $i_{\beta}X^{(9)}$ is given by
\begin{eqnarray}\label{eq:IIA X9}
i_{\beta}X^{(9)}&=&-d(i_{\beta}N^{(8)})-2i_{\beta}C^{(1)}i_{\beta}H^{(9)}-i_{\beta}(G^{(8)}\wedge C^{(1)})-i_{\beta}C^{(1)}i_{\beta}(F^{(8)}\wedge C^{(1)})\nonumber \\ &&+i_{\beta}(F^{(4)}+H\wedge C^{(1)})\wedge (i_{\beta}B^{(6)}+\frac{1}{6}i_{\beta}C^{(3)}\wedge C^{(3)})\nonumber \\ &&+\frac{1}{6}(F^{(4)}+H\wedge C^{(1)})\wedge (i_{\beta}C^{(3)})^2 +2mi_{\beta}B\wedge i_{\beta}\phi^{(8)}\nonumber \\ &&+2mi_{\beta}B^{(9)}-mi_{\beta}(N^{(7)}\wedge B)+\frac{1}{6}m(i_{\beta}C^{(3)})^2\wedge (B)^2\nonumber \\ &&-\frac{1}{3}mC^{(3)}\wedge i_{\beta}C^{(3)}\wedge B\wedge i_{\beta}B
\end{eqnarray}
where $N^{(8)}$ is the 8-form magnetic dual the scalar potential $i_{\beta}C^{(1)}$. We only require this equation with an overall contraction with $\beta$ since we only consider its double reduction T-duality transformation to IIB in the next subsection. Note the presence of the new potential $B^{(9)}$ originating from $\hat{B}^{(10)}$ in 11-dimensions.

Finally, we give the field equation for $i_{\beta}B^{(9)}$ which is obtained from reducing (\ref{eq:11-d H11})
\begin{eqnarray}\label{eq:IIA B9}
i_{\beta}C^{(1)}i_{\beta}F^{(10)}&=&-d(i_{\beta}B^{(9)})-\frac{1}{2}i_{\beta}(H\wedge N^{(7)})+i_{\beta}\phi^{(8)}\wedge i_{\beta}H\nonumber \\ &&-\frac{1}{2}i_{\beta}(F^{(4)}+H\wedge C^{(1)})\wedge i_{\beta}C^{(7)}\nonumber \\ && +mi_{\beta}C^{(9)}\wedge i_{\beta}B+\frac{1}{12}mi_{\beta}C^{(3)}\wedge i_{\beta}B \wedge (B)^3 .
\end{eqnarray}

\subsection{Field equations from T-duality}\label{sec:10-d T-duality field eqns}

We now T-dualise the IIA fields to the IIB theory. It is to be understood that we always perform the T-duality transformation along the isometry described by $\beta$. This is important since this vector is used in the definition of the fields $N^{(7)}$ and $N^{(8)}$. They would therefore exhibit different T-duality transformations if we were using a different isometry and we do not consider this option in this paper. Performing the T-duality produces the IIB field equations for the two 8-form potentials that transform as a part of a triplet along with the standard Ramond-Ramond 8-form potential. These equations were given in an $SU(1,1)$ covariant form in \cite{Dall'Agata:1998va,Bergshoeff:2005ac}. This triplet maps to a triplet of 10-form potentials in 11-d SUGRA by performing direct reduction T-duality transformations to IIA and then uplifting. The potentials $\hat{A}^{(10)}$ and $\hat{B}^{(10)}$ both belong to this triplet and we can deduce the third via its relation to one of the IIB 8-forms.

The T-duality rules transforming the more common fields between massless IIA and IIB have been given in \cite{Bergshoeff:1995as,Eyras:1998hn}. In the present case however we are considering Romans' massive IIA and therefore need to adapt these rules slightly. This situation was considered in \cite{Bergshoeff:1996ui} (see also \cite{Eyras:1998kx}). Since the IIB theory has no standard massive deformation, the mass parameter is produced by performing a Scherk-Schwarz dimensional reduction using the subgroup of the global $SL(2,\mathbb{R})$ symmetry that involves shifts of the axion. In the IIB theory the mass parameter is therefore expressed by the potentials having a dependence on the compact isometry direction. We can then simply use the massless T-duality rules but account for this dependence by allowing the IIB potentials to have non-vanishing Lie derivatives w.r.t. $\beta$. The axion ($l$) has a linear dependence which is expressed as ${\cal L}_{\beta}l=m$ \cite{Bergshoeff:1996ui}. Consistency with the IIB gauge algebra then requires the following dependencies for the rest of the potentials
\begin{eqnarray}\label{eq:IIB axion lie}
{\cal L}_{\beta} {\cal C}^{(2n)}&=&\frac{1}{n!}m({\cal B})^n\\
{\cal L}_{\beta} {\cal B}&=&0 \\
{\cal L}_{\beta} {\cal B}^{(6)}&=&m(-{\cal C}^{(6)}+{\cal C}^{(4)}\wedge {\cal B})\\ \label{eq:IIB dilaton lie}
{\cal L}_{\beta} \varphi &=& 0
\end{eqnarray}
where ${\cal C}^{(2n)}$ are the Ramond-Ramond potentials\footnote{We use both the notation $l$ and $C^{(0)}$ to refer to the axion.} ($n=0,1,2,3,4,5$), ${\cal B}$ and ${\cal B}^{(6)}$ are the NS-NS 2-form and its magnetic dual respectively and $\varphi$ is the dilaton. We define our potentials as in \cite{Callister:2007jy}.

\subsubsection{T-dualising from IIA to IIB} 

We will now T-dualise the IIA fields $\phi^{(8)}$, $N^{(8)}$ and $N^{(7)}$. We restrict ourselves to the case where all of the IIA potentials are independent of the T-duality isometry. 

We can determine the T-duality rules for the potentials by first finding how their field strengths transform. This can be done by using essentially the same technique used in the previous section, where it was determined how the higher rank field strengths dimensionally reduced by first considering how their lower rank duals reduced. For the purposes of the T-duality we split the co-ordinate system into \{$\mu_i,y$\} where $y$ parametrises the compact direction and the $\mu_i$ represent the other 9 directions. Starting with $N^{(7)}_{\mu_1 \ldots \mu_7}$ we find\footnote{The case of $N^{(7)}_{\mu_1 \ldots \mu_6 y}$ was considered in \cite{Callister:2007jy}.}
\begin{eqnarray}
G^{(8)}_{\mu_1\ldots\mu_8}\rightarrow (-{\cal G}^{(8)}+{\cal R}_{\beta}^{-2}i_{\beta}{\cal G}^{(8)}\wedge \beta+(-{\cal H}^{(7)}+{\cal R}_{\beta}^{-2} i_{\beta}{\cal H}^{(7)}\wedge \beta)\wedge i_{\beta}{\cal B})_{\mu_1\ldots \mu_8}\nonumber \\ 
\end{eqnarray}
where ${\cal R}_{\beta}$ is the radius of the compact isometry defined by $\beta$ and the definitions of the field strengths are given in \cite{Callister:2007jy}.

We then find an equation for ${\cal G}^{(8)}$ by T-dualising (\ref{eq:IIA G8}). In order for this equation to be well-formed we find that $N^{(7)}_{\mu_1 \ldots \mu_7}$ transforms as
\begin{eqnarray}
N^{(7)}_{\mu_1\ldots \mu_7}&\rightarrow& \biggl(-{\cal N}^{(7)}+{\cal R}_{\beta}^{-2}i_{\beta}{\cal N}^{(7)}\wedge \beta -\frac{1}{2}i_{\beta}{\cal C}^{(6)}\wedge {\cal C}^{(2)}\nonumber \\ 
&& -\frac{1}{2}{\cal R}_{\beta}^{-2}i_{\beta}{\cal C}^{(6)}\wedge i_{\beta}{\cal C}^{(2)}\wedge \beta  +(-{\cal B}^{(6)} -{\cal R}_{\beta}^{-2}i_{\beta}{\cal B}^{(6)}\wedge \beta  \nonumber \\
&& +{\cal C}^{(4)}\wedge {\cal C}^{(2)} +{\cal R}_{\beta}^{-2}i_{\beta}{\cal C}^{(4)}\wedge {\cal C}^{(2)}\wedge \beta \nonumber \\ 
&& +{\cal R}_{\beta}^{-2}{\cal C}^{(4)}\wedge i_{\beta}{\cal C}^{(2)}\wedge \beta )\wedge i_{\beta}{\cal B} \biggr)_{\mu_1\ldots \mu_7}
\end{eqnarray}
where ${\cal N}^{(7)}$ is the IIB magnetic dual of $\beta$. We state the corresponding transformation from IIB to IIA for completeness:
\begin{eqnarray}
{\cal N}^{(7)}_{\mu_1\ldots \mu_7}&\rightarrow&(-N^{(7)}+R_{\beta}^{-2}i_{\beta}N^{(7)}\wedge \beta-B^{(6)}\wedge i_{\beta}B + R_{\beta}^{-2}i_{\beta}B^{(6)}\wedge i_{\beta}B \wedge \beta \nonumber \\ &&+\frac{1}{2}C^{(5)}\wedge i_{\beta}C^{(3)} - \frac{1}{2}R_{\beta}^{-2}i_{\beta}C^{(5)}\wedge i_{\beta}C^{(3)}\wedge \beta)_{\mu_1\ldots \mu_7} .
\end{eqnarray}
The IIB field equation is then given by
\begin{eqnarray}\label{eq:IIB G8}
{\cal G}^{(8)}&=&d{\cal N}^{(7)}+\frac{1}{2} {\cal C}^{(0)} i_{\beta}{\cal F}^{(9)} - \frac{1}{2}{\cal F}^{(7)} \wedge i_{\beta}{\cal C}^{(2)} -\frac{1}{2}i_{\beta}{\cal F}^{(7)}\wedge {\cal C}^{(2)}\nonumber \\ &&+\frac{1}{2}i_{\beta}{\cal F}^{(5)} \wedge {\cal C}^{(4)} -\frac{1}{2}{\cal H}\wedge i_{\beta} {\cal C}^{(2)}\wedge {\cal C}^{(4)}+\frac{1}{2}i_{\beta}{\cal H}\wedge {\cal C}^{(4)}\wedge {\cal C}^{(2)}\nonumber \\ &&-i_{\beta}{\cal H}\wedge {\cal B}^{(6)}+\frac{1}{2}i_{\beta}{\cal F}^{(1)}({\cal C}^{(8)}-{\cal C}^{(6)}\wedge {\cal B})
\end{eqnarray}
which can be seen to be valid for all components by using the equation for $i_{\beta}{\cal G}^{(8)}$ given in \cite{Callister:2007jy}.\footnote{Note that in \cite{Callister:2007jy} only the massless T-duality relations were used and as such terms including $i_{\beta}{\cal F}^{(1)}={\cal L}_{\beta}l$ were neglected.} The Bianchi identity reads
\begin{eqnarray}
d{\cal G}^{(8)}&=&\frac{1}{2}i_{\beta}{\cal F}^{(1)}{\cal F}^{(9)}+\frac{1}{2} {\cal F}^{(1)}\wedge i_{\beta}{\cal F}^{(9)}-\frac{1}{2}{\cal F}^{(7)}\wedge i_{\beta}{\cal F}^{(3)}\nonumber \\ &&-\frac{1}{2}i_{\beta}{\cal F}^{(7)}\wedge {\cal F}^{(3)}+\frac{1}{2}i_{\beta}{\cal F}^{(5)}\wedge {\cal F}^{(5)}-i_{\beta}{\cal H}\wedge {\cal H}^{(7)}
\end{eqnarray}
which shows that the above definition of ${\cal G}^{(8)}$ is gauge invariant.

Next we consider the potentials $N^{(8)}$ and $\phi^{(8)}$. Since the dual potentials for these cases are both scalars and we are assuming independence of $y$, the components parallel to and transverse to $y$ of the 8-forms are not independent. Because of this the corresponding 9-form field strengths do not map to independent field strengths in IIB when direct reduction T-duality transformations are performed.\footnote{The situation is different if the global $SO(1,1)$ symmetry is used to perform a Scherk-Schwarz reduction from the IIA side \cite{Bergshoeff:2002mb,Bergshoeff:2002nv} which would allow for a dependence on $y$. This then maps to a non-covariant form of IIB in which such field strengths do exist. Although we do not consider this option in any great detail, we make some qualitative observations in Appendix \ref{sec:mapping 10-forms}.} For this reason we only consider the double reduction form of the T-duality transformations, namely that of $N^{(8)}_{y\mu_1\ldots \mu_7}=i_{\beta}N^{(8)}_{\mu_1\ldots \mu_7}$ etc.  The relevant field strengths transform as 
\begin{eqnarray}
i_{\beta}H^{(9)}_{\mu_1\ldots \mu_8}\rightarrow (i_{\beta}{\cal H}^{(9)}+\frac{1}{2}{\cal G}^{(8)}+\frac{1}{2}{\cal R}_{\beta}^{-2}i_{\beta}{\cal G}^{(8)}\wedge \beta)_{\mu_1\ldots \mu_8}
\end{eqnarray}
and
\begin{eqnarray}\label{eq:X9 T-dual}
i_{\beta}X^{(9)}_{\mu_1\ldots \mu_8}\rightarrow -e^{-2\varphi}i_{\beta}{\cal F}^{(9)}_{\mu_1\ldots \mu_8}
\end{eqnarray}
where ${\cal H}^{(9)}$ is the dual of the dilaton `field strength', explicitly ${\cal H}^{(9)}=e^{-2\varphi}\ast d\varphi$.

We then transform (\ref{eq:IIA H9}) and (\ref{eq:IIA X9}) using the following rules for the 8-form potentials
\begin{eqnarray}
i_{\beta}\phi^{(8)}_{\mu_1\ldots \mu_7}&\rightarrow& (i_{\beta}\varphi^{(8)}-\frac{1}{2}{\cal N}^{(7)}+\frac{1}{2}{\cal R}_{\beta}^{-2}i_{\beta}{\cal N}^{(7)}\wedge \beta+\frac{1}{4}i_{\beta}{\cal C}^{(6)}\wedge {\cal C}^{(2)}\nonumber \\ &&+\frac{1}{4}{\cal R}_{\beta}^{-2}i_{\beta}{\cal C}^{(6)}\wedge i_{\beta}{\cal C}^{(2)}\wedge \beta)_{\mu_1\ldots \mu_7}\\
i_{\beta}N^{(8)}_{\mu_1\ldots \mu_7}&\rightarrow&(-i_{\beta}{\cal N}^{(8)}+{\cal B}^{(6)}\wedge i_{\beta}{\cal C}^{(2)}-{\cal R}_{\beta}^{-2}i_{\beta}{\cal B}^{(6)}\wedge i_{\beta}{\cal C}^{(2)}\wedge \beta\nonumber \\ &&+\frac{1}{3}i_{\beta}{\cal C}^{(4)}\wedge ({\cal C}^{(2)})^2+\frac{2}{3}{\cal R}_{\beta}^{-2}i_{\beta}{\cal C}^{(4)}\wedge i_{\beta}{\cal C}^{(2)}\wedge {\cal C}^{(2)}\wedge \beta)_{\mu_1\ldots \mu_7} \nonumber \\
 & & 
\end{eqnarray}
with inverse rules given by 
\begin{eqnarray}
\i_{\beta}\varphi^{(8)}_{\mu_1\ldots \mu_7}&\rightarrow&(i_{\beta}\phi^{(8)}-\frac{1}{2}N^{(7)}+\frac{1}{2}R_{\beta}^{-2}i_{\beta}N^{(7)}\wedge \beta \nonumber \\ && + \frac{1}{2}C^{(5)}\wedge i_{\beta}C^{(3)}-\frac{1}{2}R_{\beta}^{-2}i_{\beta}C^{(5)}\wedge i_{\beta}C^{(3)}\wedge \beta)_{\mu_1\ldots \mu_7} \nonumber \\
 & & \\
i_{\beta}{\cal N}^{(8)}_{\mu_1\ldots \mu_7}&\rightarrow&(-i_{\beta}N^{(8)}+i_{\beta}N^{(7)}\wedge (C^{(1)}-R_{\beta}^{-2}i_{\beta}C^{(1)}\beta) \nonumber \\ && -i_{\beta}C^{(3)}\wedge i_{\beta}C^{(5)}\wedge (C^{(1)}-R_{\beta}^{-2}i_{\beta}C^{(1)}\beta) \nonumber \\ && - \frac{1}{3}(i_{\beta}C^{(3)})^2\wedge (C^{(3)}-R_{\beta}^{-2}i_{\beta}C^{(3)} \wedge \beta))_{\mu_1\ldots \mu_7} .
\end{eqnarray}
The rules for the field $N^{(8)}$ have been given in \cite{Eyras:1999at} with different field definitions.

The following IIB field equations are produced
\begin{eqnarray}\label{eq:IIB phi8}
{\cal H}^{(9)}&=&d\varphi^{(8)}-l{\cal F}^{(9)}+\frac{1}{2}{\cal H}\wedge {\cal B}^{(6)}+\frac{1}{2}{\cal C}^{(2)}\wedge {\cal F}^{(7)}
\end{eqnarray}
and
\begin{eqnarray}\label{eq:IIB N8}
e^{-2\varphi}{\cal F}^{(9)}&=&d{\cal N}^{(8)}+l^2{\cal F}^{(9)}+2l{\cal H}^{(9)}-\frac{1}{2}{\cal F}^{(5)}\wedge ({\cal C}^{(2)})^2\nonumber \\ &&-({\cal F}^{(3)}+l{\cal H})\wedge {\cal B}^{(6)}-\frac{1}{3}{\cal H}\wedge ({\cal C}^{(2)})^3
\end{eqnarray}
where $\varphi^{(8)}$ is the 8-form magnetic dual of the dilaton and ${\cal N}^{(8)}$ is dual to a particular combination of the dilaton and axion \cite{Meessen:1998qm,Dall'Agata:1998va}.

The Bianchi identities can be easily calculated. The results are
\begin{eqnarray}
d{\cal H}^{(9)}&=&-\frac{1}{2}{\cal H}\wedge {\cal H}^{(7)}-{\cal F}^{(1)}\wedge {\cal F}^{(9)}+\frac{1}{2}{\cal F}^{(3)}\wedge {\cal F}^{(7)}\\
d(e^{-2\varphi}{\cal F}^{(9)})&=&2{\cal F}^{(1)}\wedge {\cal H}^{(9)}+{\cal F}^{(3)}\wedge {\cal H}^{(7)}
\end{eqnarray}
which shows that the above field definitions are gauge invariant.

Note that from the T-duality transformation we actually only produce the double dimensional reduction of the above field equations. However since neither field depends on $\beta$ in its definition, the full equations can be deduced trivially. This is not the case if we are dealing with fields that do depend on $\beta$ in their definition for example ${\cal N}^{(7)}$; here the field equation for ${\cal N}^{(7)}_{\mu_1 \ldots \mu_7}$ cannot be naively deduced from the field equation of $i_{\beta}{\cal N}^{(7)}_{\mu_1 \ldots \mu_6}$. This is the same situation as what was found in 11-dimensional SUGRA for $\hat{N}^{(8)}$. 

In carrying out the T-duality, we find that the IIB potentials are required to have the following Lie derivatives
\begin{eqnarray}
{\cal L}_{\beta}\varphi^{(8)}&=&m({\cal C}^{(8)}-\frac{1}{2}{\cal C}^{(6)}\wedge {\cal B})\\
{\cal L}_{\beta}{\cal N}^{(8)}&=&m(-2\varphi^{(8)}-{\cal C}^{(6)}\wedge {\cal C}^{(2)}+{\cal C}^{(4)}\wedge {\cal C}^{(2)}\wedge {\cal B})\\
{\cal L}_{\beta}{\cal N}^{(7)}&=&m(\frac{1}{2}i_{\beta}{\cal C}^{(8)} -\frac{1}{2}i_{\beta}{\cal C}^{(6)}\wedge {\cal B} +\frac{1}{2}{\cal C}^{(6)}\wedge i_{\beta}{\cal B} \nonumber \\
&& +\frac{1}{4}i_{\beta}{\cal C}^{(4)}\wedge ({\cal B})^{2})
\end{eqnarray}
which agrees with what is found by demanding consistency between their field equations and (\ref{eq:IIB axion lie})-(\ref{eq:IIB dilaton lie}).

As previously mentioned the fields $\varphi^{(8)}$ and ${\cal N}^{(8)}$ form a triplet along with the Ramond-Ramond 8-form potential ${\cal C}^{(8)}$. We will re-express them in an $SL(2,\mathbb{R})$ covariant form in Section \ref{sec:cov IIB charges}. Since they are magnetically dual to only two scalars, the axion and dilaton, they only have two independent degrees of freedom. This is usually expressed by a constraint on the three 9-form field strengths but is manifest in the field equations given above since they are only given in terms of the two independent 9-form field strengths, ${\cal F}^{(9)}$ and ${\cal H}^{(9)}$.

\subsubsection{T-dualising from IIB to IIA}

We now T-dualise the IIB 8-form potentials via a direct dimensional reduction to obtain 9-form potentials in Romans' IIA.

Repeating the process above we use the well known result for the transformation of the Ramond-Ramond 9-form field strength
\begin{eqnarray}\label{eq:RR T-dual}
{\cal F}^{(2n+1)}_{\mu_1\ldots \mu_9}\rightarrow (-i_{\beta}F^{(2n+2)}+F^{(2n)}\wedge i_{\beta}B -R_{\beta}^{-2}i_{\beta}F^{(2n)}\wedge i_{\beta}B\wedge \beta)_{\mu_1\ldots \mu_9} .
\end{eqnarray}

Since in the Scherk-Schwarz reduction we are only using the subgroup of $SL(2,\mathbb{R})$ that describes shifts of the axion, we have $i_{\beta}d\varphi=0$.\footnote{This restriction will be lifted in Section \ref{sec:cov T-duality} when we consider the full $SL(2,\mathbb{R})$ group.} The situation here is similar to that of the IIA 8-forms described above where the components ${\cal H}^{(9)}_{\mu_1\ldots \mu_9}$ are not independent of ${\cal H}^{(9)}_{y\mu_1\ldots \mu_8}$ and therefore do not have an independent mapping to a 10-form field strength in Romans' IIA. The actual transformation is given by
\begin{eqnarray}
{\cal H}^{(9)}_{\mu_1\ldots \mu_9}&=&{\cal R}_{\beta}^{-2}(i_{\beta}{\cal H}^{(9)}\wedge\beta)_{\mu_1\ldots \mu_9}\nonumber \\ &\rightarrow&\left[ (i_{\beta}H^{(9)}+\frac{1}{2}G^{(8)}+\frac{1}{2}R_{\beta}^{-2}i_{\beta}G^{(8)}\wedge \beta)\wedge i_{\beta} B \right]_{\mu_1\ldots \mu_9} .
\end{eqnarray}
We then transform (\ref{eq:IIB phi8}) and (\ref{eq:IIB N8}) using the following rules for the 8-form potentials
\begin{eqnarray}
\varphi^{(8)}_{\mu_1\ldots \mu_8}&\rightarrow& \biggl(i_{\beta}B^{(9)} - \frac{1}{2}i_{\beta}C^{(3)}\wedge i_{\beta}C^{(7)} +(-i_{\beta}\phi^{(8)}  + \frac{1}{2}N^{(7)} \nonumber \\ 
&& - \frac{1}{2}R_{\beta}^{-2}i_{\beta}N^{(7)}\wedge \beta-\frac{1}{2}i_{\beta}C^{(3)}\wedge C^{(5)} \nonumber \\ && +\frac{1}{2}R_{\beta}^{-2}i_{\beta}C^{(3)}\wedge i_{\beta}C^{(5)} \wedge \beta )\wedge i_{\beta}B \biggr)_{\mu_1\ldots \mu_8} 
\\
{\cal N}^{(8)}_{\mu_1\ldots \mu_8}&\rightarrow& \biggl(-i_{\beta}D^{(9)} +\frac{1}{2}i_{\beta}C^{(5)}\wedge (i_{\beta}C^{(3)})^2 + (i_{\beta}N^{(8)}\nonumber \\ 
&& -i_{\beta}N^{(7)} \wedge (C^{(1)}-R_{\beta}^{-2}i_{\beta}C^{(1)}\beta)\nonumber \\ &&+i_{\beta}C^{(5)}\wedge i_{\beta}C^{(3)}\wedge (C^{(1)}-R_{\beta}^{-2}i_{\beta}C^{(1)}\beta)\nonumber \\ 
&& +\frac{1}{3}(i_{\beta}C^{(3)})^2\wedge (C^{(3)}-R_{\beta}^{-2}i_{\beta}C^{(3)}\wedge \beta))\wedge i_{\beta}B \biggr)_{\mu_1\ldots \mu_8}
\end{eqnarray}
with inverse rules given by
\begin{eqnarray}
i_{\beta}B^{(9)}_{\mu_1\ldots \mu_8}&\rightarrow & \biggl(\varphi^{(8)} 
+ {\cal R}_{\beta}^{-2}i_{\beta}\varphi^{(8)}\wedge \beta 
+ \frac{1}{2}{\cal C}^{(6)}\wedge {\cal C}^{(2)} \nonumber \\ 
&& + \frac{1}{2}{\cal R}_{\beta}^{-2}i_{\beta}{\cal C}^{(6)}\wedge {\cal C}^{(2)}\wedge \beta 
+ \frac{1}{2}{\cal R}_{\beta}^{-2}{\cal C}^{(6)}\wedge i_{\beta}{\cal C}^{(2)}\wedge \beta   \biggr)_{\mu_1\ldots \mu_8} \\
i_{\beta}D^{(9)}_{\mu_1\ldots \mu_8}&\rightarrow & \biggl( -{\cal N}^{(8)} 
- {\cal R}_{\beta}^{-2}i_{\beta}{\cal N}^{(8)}\wedge \beta
+ \frac{1}{2}{\cal C}^{(4)}\wedge ({\cal C}^{(2)})^2 \nonumber \\
&& + \frac{1}{2}{\cal R}_{\beta}^{-2}i_{\beta}{\cal C}^{(4)}\wedge ({\cal C}^{(2)})^2
+ {\cal R}_{\beta}^{-2}{\cal C}^{(4)}\wedge i_{\beta}{\cal C}^{(2)} \wedge {\cal C}^{(2)}
\biggr)_{\mu_1\ldots \mu_8} .
\end{eqnarray}
This produces the following IIA field equations
\begin{eqnarray}\label{eq:IIA B9 II}
i_{\beta}C^{(1)}i_{\beta}F^{(10)}&=&-d(i_{\beta}B^{(9)})-\frac{1}{2}i_{\beta}(H\wedge N^{(7)})+i_{\beta}\phi^{(8)}\wedge i_{\beta}H\nonumber \\ &&-\frac{1}{2}i_{\beta}(F^{(4)}+H\wedge C^{(1)})\wedge i_{\beta}C^{(7)}\nonumber \\ && +mi_{\beta}C^{(9)}\wedge i_{\beta}B+\frac{1}{12}mi_{\beta}C^{(3)}\wedge i_{\beta}B \wedge (B)^3\\ \label{eq:IIA D9}
e^{-2\phi}R_{\beta}^2i_{\beta}F^{(10)}&=&d(i_{\beta}D^{(9)})+(i_{\beta}C^{(1)})^2i_{\beta}F^{(10)}-i_{\beta}N^{(8)}\wedge i_{\beta}H\nonumber \\ &&+i_{\beta}N^{(7)}\wedge i_{\beta}(F^{(4)}+H\wedge C^{(1)})\nonumber \\ &&+\frac{1}{3!}i_{\beta}(H\wedge C^{(3)})\wedge (i_{\beta}C^{(3)})^2 -2mi_{\beta}B^{(9)}\wedge i_{\beta}B\nonumber \\ &&-\frac{1}{4}m(i_{\beta}C^{(3)})^2\wedge (B)^2\wedge i_{\beta}B .
\end{eqnarray}
The first equation was also calculated from the dimensional reduction of $\hat{H}^{(11)}$ and given by (\ref{eq:IIA B9}). The agreement between both these methods is a non-trivial consistency check on (\ref{eq:11-d S9}) and (\ref{eq:11-d H11}) in 11-dimensions. On the other hand, the potential $D^{(10)}$ did not arise in our treatment of the 11-dimensional fields. However uplifting (\ref{eq:IIA D9}) using the rules in Appendix \ref{sec:reducs} gives the following 11-d equation
\begin{eqnarray}\label{eq:11-d D10}
(\hat{R}_{\hat{\beta}}^2\hat{R}_{\hat{\alpha}}^{-2}&-&2\hat{R}_{\hat{\alpha}}^{-4}(i_{\hat{\alpha}}\hat{\beta})^2)i_{\hat{\beta}\hat{\alpha}}\hat{F}^{(11)}=-d(i_{\hat{\beta}\hat{\alpha}}\hat{D}^{(10)})+i_{\hat{\alpha}}(i_{\hat{\beta}}\hat{T}^{(8)}\wedge i_{\hat{\beta}}\hat{F})\nonumber \\ 
&& -\frac{1}{4!}i_{\hat{\beta}\hat{\alpha}}(\hat{F}\wedge \hat{A} \wedge (i_{\hat{\beta}}\hat{A})^2)
 +2\hat{m}i_{\hat{\beta}\hat{\alpha}}\hat{A}\wedge i_{\hat{\beta}\hat{\alpha}}\hat{B}^{(10)}\nonumber \\ 
&& +\frac{1}{16} \hat{m}(i_{\hat{\beta}}\hat{A})^2\wedge (i_{\hat{\alpha}}\hat{A})^2\wedge i_{\hat{\beta}\hat{\alpha}}\hat{A}   
\end{eqnarray}
where $\hat{R}_{\hat{\alpha}}$ is the radius of the compact isometry direction described by $\hat{\alpha}$. In the next section we will show that $\hat{D}^{(10)}$ forms a triplet along with $\hat{A}^{(10)}$ and $\hat{B}^{(10)}$.

We note that $B^{(9)}$ and $D^{(9)}$ are the T-duals of the potentials that couple to the non-Dirichlet IIB 7-branes. The IIA theory that contains the 8-branes T-dual to these is the non-covariant IIA theory discussed in the introduction \cite{Meessen:1998qm}. However since we are just considering Romans' IIA here we have effectively ignored a 10-form field strength that is sourced by these 8-branes. As such the equations (\ref{eq:IIA B9 II}) and (\ref{eq:IIA D9}) are not complete since this field strength is missing. These equations are then actually truncations of the full equations which is why the `field strength' $i_{\beta}C^{(1)}i_{\beta}F^{(10)}$ in (\ref{eq:IIA B9 II}) is not gauge invariant. They are however still useful in deducing the field equations in $SL(2,\mathbb{R})$ covariant 11-dimensional SUGRA, which in turn can be used to determine the missing terms.

\section{$SL(2,\mathbb{R})$ covariant massive 11-dimensional supergravity}\label{sec:11-d cov SUGRA}

We will now consider the extension of the massive 11-dimensional theory considered so far in this paper to the $SL(2,\mathbb{R})$ covariant massive 11-dimensional SUGRA discussed in the introduction. It was first considered in \cite{Meessen:1998qm} where it was interpreted as containing two M9-branes, and later generalised to the case of $n$ M9-branes in \cite{AlonsoAlberca:2002tb}. Due to the M9-branes there are two compact isometry directions in the theory which we assume all the fields to be independent of. There are therefore two mutually commuting Killing vectors that define a $T^2$ manifold. Dimensional reduction over this manifold produces the triplet of $SL(2,\mathbb{R})$ 9-dimensional massive SUGRAs obtained by performing a Scherk-Schwarz reduction of the IIB theory containing 7-branes using its full $SL(2,\mathbb{R})$ symmetry group. In this way the M9-branes are mapped to the IIB 7-branes. The $SL(2,\mathbb{R})\subset GL(2,\mathbb{R})$ symmetry of performing this reduction corresponds to the symmetry group of the IIB theory \cite{Bergshoeff:1995as,Bergshoeff:1995cg,Bergshoeff:1995cw,Aspinwall:1995fw}. To determine how the multiplets of IIB states map to the 11-dimensional theory it is therefore important to express the theory in an $SL(2,\mathbb{R})$ covariant manner.

In this paper our analysis of this theory simply amounts to the $SL(2,\mathbb{R})$ covariantisation of the massive 11-dimensional SUGRA. We now adopt the more systematic notation $\hat{k}_a^{\phantom{a}\hat{\mu}}$ to denote the Killing vectors of the theory. The index $a=1,2$ is an $SL(2,\mathbb{R})$ index. Each field will contain a number of symmetrised $SL(2,\mathbb{R})$ indices depending on their representation. The Killing vectors therefore lie in the doublet representation. Instead of a single mass parameter we now have a $2\times 2$ symmetric mass matrix $\hat{Q}^{ab}$ which we will parametrise as
\begin{eqnarray}\label{eq:11-d Q}
\hat{Q}^{ab}=\left(
\begin{array}{cc}
\hat{m}_2+\hat{m}_3 & \hat{m}_1\\
\hat{m}_1 & -(\hat{m}_2-\hat{m}_3)
\end{array} \right)
=\left(
\begin{array}{cc}
\hat{m}_+ & \hat{m}_1\\
\hat{m}_1 & -\hat{m}_-
\end{array} \right) .
\end{eqnarray}

The massive 11-dimensional SUGRA considered in the previous sections is obtained by the following truncation
\begin{eqnarray}\label{eq:11-d Q trunc}
\hat{Q}^{ab}=\left(
\begin{array}{cc}
\hat{m} & 0\\
0 & 0
\end{array} \right)
\end{eqnarray}
and setting $\hat{k}_1=\hat{\alpha}$ and $\hat{k}_2=\hat{\beta}$. In this instance $\hat{\beta}$ is only considered as a standard Killing vector as opposed to a massive Killing vector due to the truncation of $\hat{Q}^{ab}$.

\subsection{$SL(2,\mathbb{R})$ covariant Field Equations}\label{sec:11-d cov field eqns}

We will now present the field equations of this theory. The $SL(2,\mathbb{R})$ covariant field equation of the 3-form potential was given in \cite{Meessen:1998qm} as
\begin{eqnarray}\label{eq:11-d cov F4}
\hat{F}&=&d\hat{A}+\frac{1}{2}\hat{Q}^{ab}i_{\hat{k}_{a}}\hat{A}\wedge i_{\hat{k}_{b}}\hat{A}
\end{eqnarray}
and simply amounts to the straightforward covariantisation of (\ref{eq:11-d A}). It is also a trivial task to deduce the covariant field equation for the 6-form potential from (\ref{eq:11-d C}) as
\begin{eqnarray}\label{eq:11-d cov F7} 
\hat{F}^{(7)}&=&d\hat{C}^{(6)}-\frac{1}{2}F\wedge \hat{A}+\hat{Q}^{ab}\biggl[ i_{\hat{k}_{a}}\hat{A}\wedge i_{\hat{k}_{b}}\hat{C}\nonumber \\ && +\frac{1}{12}\hat{A}\wedge i_{\hat{k}_{a}}\hat{A}\wedge i_{\hat{k}_{b}}\hat{A}+i_{\hat{k}_{a}}\hat{N}_{b}^{(8)} \biggr] .
\end{eqnarray}
Where $\hat{N}_a$ are the doublet of 8-form potentials which we consider below. Both these equations are easily deduced since the potentials $\hat{A}$ and $\hat{C}$ are $SL(2,\mathbb{R})$ scalars. The massive gauge transformations (\ref{eq:11-d gauge trans}) generalises to 
\begin{eqnarray}\label{eq:11-d massive gauge scalar} 
\delta \hat{Y}^{(p)}=\hat{Q}^{ab}\hat{\lambda}_a\wedge i_{\hat{k}_b}\hat{Y}^{(p)}
\end{eqnarray}
for $SL(2,\mathbb{R})$ scalars where $\hat{\lambda}_a=i_{\hat{k}_a}\hat{\chi}$. We re-emphasise the point that the gauge potentials generally exhibit more complicated massive gauge transformations than those specified by the above rule and the generalisations thereof given below. In fact the gauge transformation of $\hat{A}$ is now given by
\begin{eqnarray}\label{eq:11-d cov 3-form gauge}
\delta\hat{A}=d\hat{\chi}+\hat{Q}^{ab}\hat{\lambda}_{a}\wedge i_{\hat{k}_b}\hat{A}
\end{eqnarray}
and therefore the torsion tensor becomes
\begin{eqnarray}\label{eq:11-d cov connection}
\hat{\tilde{T}}_{\hat{\mu}\hat{\nu}}^{\phantom{\hat{\mu}\hat{\nu}}\hat{\rho}}=-\hat{Q}^{ab}(i_{\hat{k}_a}\hat{A})_{\hat{\mu}\hat{\nu}}\hat{k}_b^{\phantom{n}\hat{\rho}}
\end{eqnarray}
from which the contorsion tensor and total connection can be calculated.

Next we give the covariant field equations for the Killing vectors $\hat{k}_a$ which form $SL(2,\mathbb{R})$ doublets. There is not enough information in (\ref{eq:11-d G2}) to deduce the covariant form of the equation since loosely speaking it is not possible to distinguish the $SL(2,\mathbb{R})$ free indices from the dummy indices. This was why it was important for us to construct (\ref{eq:11-d S2}), from which we can deduce the following covariant field equation
\begin{eqnarray}\label{eq:11-d G2 cov}
\hat{G}^{(2)}_{a}=d\hat{k}_{a}+\hat{Q}^{bc}\biggl[(\hat{k}_{b}\cdot \hat{k}_{a})i_{\hat{k}_{c}}\hat{A}+i_{\hat{k}_{b}\hat{k}_{a}}\hat{A}\wedge \hat{k}_{c}\biggr]
\end{eqnarray}
which gives both (\ref{eq:11-d G2}) and (\ref{eq:11-d S2}) when the truncation (\ref{eq:11-d Q trunc}) is applied. Note that the last term on the LHS vanishes in (\ref{eq:11-d G2}). Due to $\hat{Q}^{ab}$ being symmetric there can be no other terms present in (\ref{eq:11-d G2 cov}) since they would have had to show up in (\ref{eq:11-d S2}), hence (\ref{eq:11-d G2 cov}) is uniquely determined.

The massive gauge transformations of $\hat{k}_a$ are given by the covariantisation of (\ref{eq:11-d beta gauge})
\begin{eqnarray}\label{eq:11-d massive isom gauge}
\delta \hat{k}_a=\hat{Q}^{bc}\biggl[(\hat{k}_{b}\cdot \hat{k}_{a})\hat{\lambda}_c +({\hat{k}_a}\cdot \hat{\lambda}_b) \hat{k}_c\biggr] .
\end{eqnarray}
It is possible to deduce the general massive gauge transformation rule for $SL(2,\mathbb{R})$ doublets from (\ref{eq:11-d massive gauge scalar}) by considering objects such as $i_{\hat{k}_a}\hat{Y}^{(p+1)}=\hat{Y}^{(p)}_a$, where $\hat{Y}^{(p+1)}$ is a $(p+1)$-form $SL(2,\mathbb{R})$ scalar. One then finds the gauge covariant rule for $SL(2,\mathbb{R})$ doublet $p$-forms $\hat{Y}^{(p)}_a$ to be
\begin{eqnarray}\label{eq:11-d massive gauge doublet}
\delta \hat{Y}^{(p)}_a=\hat{Q}^{bc}\biggl[\hat{\lambda}_b\wedge i_{\hat{k}_c}\hat{Y}^{(p)}_a +({\hat{k}_a}\cdot \hat{\lambda}_b) \hat{Y}^{(p)}_c\biggr] .
\end{eqnarray}
This is consistent with (\ref{eq:11-d beta gauge}) and also makes both sides of (\ref{eq:11-d G2 cov}) transform consistently. The first term on the RHS of (\ref{eq:11-d massive gauge doublet}) is analogous to the first term in (\ref{eq:11-d massive gauge scalar}) whereas the second term is specific to $SL(2,\mathbb{R})$ doublets. This explains why $\hat{\beta}$ did not transform according to (\ref{eq:11-d gauge trans}) in Section \ref{sec:11-d beta eqns}. From now on we will use the term `gauge covariant' to apply to fields transforming by this type of $SL(2,\mathbb{R})$ representation dependent rule.

Next we consider the 8-form potentials $\hat{N}^{(8)}_a$ which also form $SL(2,\mathbb{R})$ doublets. In terms of the notation used in Section \ref{sec:11-d field eqns} we have $\hat{N}^{(8)}_1=\hat{N}^{(8)}$ and $\hat{N}^{(8)}_2=\hat{T}^{(8)}$, and also $\hat{G}^{(9)}_1=\hat{G}^{(9)}$ and $\hat{G}^{(9)}_2=\hat{S}^{(9)}$. In analogy to the case of the Killing vectors we find that (\ref{eq:11-d G9}) does not contain enough information to deduce the covariant equation whereas (\ref{eq:11-d S9}) does. The covariant equation is then uniquely determined to be
\begin{eqnarray}\label{eq:11-d cov G9}
\hat{G}_{a}^{(9)}&=&d\hat{N}^{(8)}_{a} +\frac{1}{3}\hat{F}\wedge i_{\hat{k}_{a}}\hat{C}-\frac{2}{3}i_{\hat{k}_{a}}\hat{F}\wedge \hat{C} +\frac{1}{3!}\hat{F}\wedge i_{\hat{k}_{a}}\hat{A}\wedge \hat{A} \nonumber \\ &&+\hat{Q}^{bc}\biggl[ i_{\hat{k}_{b}\hat{k}_{a}}\hat{A}\wedge \hat{N}_{c}^{(8)} +i_{\hat{k}_{b}}\hat{A}\wedge i_{\hat{k}_{c}}\hat{N}_{a}^{(8)} +\frac{1}{3}i_{\hat{k}_{a}}\hat{A}_{bc}^{(10)}\nonumber \\ &&-2i_{\hat{k}_{b}}\hat{A}^{(10)}_{ac} -\frac{1}{4!}i_{\hat{k}_{a}}\hat{A}\wedge i_{\hat{k}_{b}}\hat{A}\wedge i_{\hat{k}_{c}}\hat{A}\wedge \hat{A}\biggr]
\end{eqnarray}
where $\hat{A}^{(10)}_{ab}$ are the triplet of 10-form potentials which we consider below. Applying the truncation (\ref{eq:11-d Q trunc}) it is easy to see that (\ref{eq:11-d G9}) and (\ref{eq:11-d S9}) are recovered. The Bianchi identity is found by covariantising (\ref{eq:11-d S9 bianchi}) and is given by
\begin{eqnarray}\label{eq:11-d cov G9 bianchi}
d\hat{G}_a^{(9)}&=&-\frac{1}{3}\hat{F}^{(4)}\wedge i_{\hat{k}_a}\hat{F}^{(7)}
+\frac{2}{3}i_{\hat{k}_a}\hat{F}^{(4)}\wedge \hat{F}^{(7)}\nonumber \\ &&
+\hat{Q}^{bc}\biggl[ -\frac{1}{3}i_{\hat{k}_a}\hat{F}^{(11)}_{bc}
+2i_{\hat{k}_b}\hat{F}^{(11)}_{ac}
-i_{\hat{k}_b}\hat{A}\wedge i_{\hat{k}_c}\hat{G}^{(9)}_a \nonumber \\ &&
-i_{\hat{k}_b\hat{k}_a}\hat{A}\wedge \hat{G}^{(9)}_c \biggr] .
\end{eqnarray}
We can extend the rules (\ref{eq:11-d massive gauge scalar}) and (\ref{eq:11-d massive gauge doublet}) to find the covariant gauge transformation rule for $SL(2,\mathbb{R})$ triplets by considering objects such as $i_{\hat{k}_{(a}}\hat{Y}^{(p+1)}_{b)}=\hat{Y}^{(p)}_{ab}$, where $\hat{Y}^{(p+1)}$ is a $(p+1)$-form $SL(2,\mathbb{R})$ doublet. One then finds the gauge covariant rule for $SL(2,\mathbb{R})$ triplet $p$-forms $\hat{Y}^{(p)}_a$ as
\begin{eqnarray}\label{eq:triplet gauge}
\delta \hat{Y}^{(p)}_{ab}=\hat{Q}^{cd}\biggl[ \hat{\lambda}_c\wedge i_{\hat{k}_d}\hat{Y}^{(p)}_{ab} 
+i_{\hat{k}_a}\hat{\lambda}_c \hat{Y}^{(p)}_{bd} +i_{\hat{k}_b}\hat{\lambda}_c \hat{Y}^{(p)}_{ad} \biggr] .
\end{eqnarray}
Using this rule show together with (\ref{eq:11-d massive gauge scalar}) and (\ref{eq:11-d massive gauge doublet}) we see that both sides (\ref{eq:11-d cov G9 bianchi}) transform consistently which guarantees that (\ref{eq:11-d cov G9}) has the correct structure. 

From (\ref{eq:11-d G2 cov}) it is easy to show that we have the identity $\epsilon^{ab}i_{\hat{k}_{a}}\hat{G}^{(2)}_{b}=0$ where $\epsilon^{ab}$ is the $SL(2,\mathbb{R})$ antisymmetric symbol and $\epsilon^{12}=+1$. This then leads to the following constraint on the 9-form field strengths
\begin{eqnarray}\label{eq:11-d 9-form constraint}
\epsilon^{ab}\epsilon^{cd}i_{\hat{k}_a\hat{k}_b}(\hat{G}^{(9)}_c\wedge \hat{k}_d)=0
\end{eqnarray}
This constraint is related to the constraint on the 9-form field strengths in IIB.

Finally we consider the three 10-forms $\hat{A}^{(10)}_{ab}$ which transform as a triplet under $SL(2,\mathbb{R})$. In terms of the notation used in the earlier sections we have $\hat{A}^{(10)}_{11}=\hat{A}^{(10)}$, $\hat{A}^{(10)}_{22}=\hat{D}^{(10)}$ and $\hat{A}^{(10)}_{12}=\hat{A}^{(10)}_{21}=\hat{B}^{(10)}$. In this case, we cannot fully determine the covariant field equation merely from the non-covariant equations (\ref{eq:11-d H11}) and (\ref{eq:11-d D10}) we have so far given since they do not reflect the full structure due to the overall contractions with the Killing vectors. However, they do provide constraints on the full equations. Further constraints are found by relating (\ref{eq:11-d cov G9}) and (\ref{eq:11-d cov G9 bianchi}), although we stress that this alone does not fully determine (\ref{eq:11-d cov F11}) largely due to the fact that in (\ref{eq:11-d cov G9}) the $SL(2,\mathbb{R})$ indices on the 10-form potentials involve contractions with the mass matrix. Then using these various constraints, demanding gauge covariance and neglecting the dimensionality of the spacetime the following field equation can be determined 
\begin{eqnarray}\label{eq:11-d cov F11}
\hat{F}^{(11)}_{ab}&=&d\hat{A}^{(10)}_{ab} +\frac{3}{4}i_{\hat{k}_{(a}}\hat{F}\wedge \hat{N}^{(8)}_{b)}  -\frac{1}{4}\hat{F}\wedge i_{\hat{k}_{(a}}\hat{N}^{(8)}_{b)}+\frac{1}{4!}\hat{F}\wedge i_{\hat{k}_{a}}\hat{A}\wedge i_{\hat{k}_{b}}\hat{A}\wedge \hat{A} \nonumber \\ 
&& +\hat{Q}^{cd}\biggl[-3i_{\hat{k}_c}\hat{A}^{(12)}_{abd}+\frac{3}{4}i_{\hat{k}_{(a}}\hat{A}^{(12)}_{b)cd}+2i_{\hat{k}_{c}\hat{k}_{(a}}\hat{A}\wedge \hat{A}^{(10)}_{b)d} \nonumber \\ 
&&+i_{\hat{k}_c}\hat{A}\wedge i_{\hat{k}_{d}}\hat{A}^{(10)}_{ab}  -\frac{1}{80}i_{\hat{k}_{a}}\hat{A}\wedge i_{\hat{k}_{b}}\hat{A}\wedge i_{\hat{k}_{c}}\hat{A}\wedge i_{\hat{k}_{d}}\hat{A}\wedge \hat{A}\biggr] .
\end{eqnarray}
Note the presence of a quadruplet of 12-form potentials $\hat{A}^{(12)}_{abc}$. Obviously these identically vanish in $D=11$, however they appear as part of the full gauge algebra and in principal they would appear explicitly if the spacetime dimension were greater than eleven. This observation is in agreement with \cite{Bergshoeff:2006qw} where Romans' IIA SUGRA was considered and a 10-form potential was found to exist which contained a 10-form gauge parameter in its gauge transformations. This demonstrates that an 11-form potential must appear in its full field strength equation, the same conclusion is reached by dimensionally reducing the (1,1) component of (\ref{eq:11-d cov F11}). 

Given the presence of the potentials $\hat{A}^{(12)}_{abc}$ in (\ref{eq:11-d cov F11}), one should determine whether the required gauge transformations in that case (which are only partly determined) are consistent overall by constructing a gauge covariant 13-form field equation. The situation here is essentially the same situation as was just encountered in determining the field equation of the $\hat{A}^{(10)}_{ab}$ fields, since they appear in (\ref{eq:11-d cov G9}) which only partly determined their gauge properties. It is found that in order to construct the 13-form field equation it is necessary to introduce an $SL(2,\mathbb{R})$ quintuplet of 14-form potentials in an analogous fashion as done previously, and so the situation repeats itself. This process then reveals the existence of an infinite tower of potentials whose field strengths can be written in the general form
\begin{eqnarray}
\hat{F}^{(2j+7)}_{a_1\ldots a_j}&=&d\hat{A}^{(2j+6)}_{a_1\ldots a_j} +\frac{j+1}{j+2}i_{\hat{k}_{(a_1}}\hat{F}^{(4)}\wedge \hat{A}^{(2j+4)}_{a_2\ldots a_j)} -\frac{1}{j+2}\hat{F}^{(4)}\wedge i_{\hat{k}_{(a_1}}\hat{A}^{(2j+4)}_{a_2\ldots a_j)} \nonumber \\
&& +\frac{1}{(j+2)!}\hat{F}^{(4)}\wedge i_{\hat{k}_{(a_1}}\hat{A}^{(3)}\wedge \ldots \wedge i_{\hat{k}_{a_j)}}\hat{A}^{(3)}\wedge \hat{A}^{(3)}\nonumber \\
&& +\hat{Q}^{bc}\biggl[ -(j+1)i_{\hat{k}_b}\hat{A}^{(2j+8)}_{a_1\ldots a_j c} +\frac{j(j+1)}{2(j+2)}i_{\hat{k}_{(a_1}}\hat{A}^{(2j+8)}_{a_2\ldots a_j) bc} \nonumber \\
&& +ji_{\hat{k}_b\hat{k}_{(a_1}}\hat{A}^{(3)}\wedge \hat{A}^{(2j+6)}_{a_2\ldots a_j)c} +i_{\hat{k}_{b}}\hat{A}^{(3)}\wedge i_{\hat{k}_c
}\hat{A}^{(2j+6)}_{a_1\ldots a_j}\nonumber \\
&& -\frac{(j+1)}{2(j+3)(j+2)!}i_{\hat{k}_{b}}\hat{A}^{(3)}\wedge i_{\hat{k}_c
}\hat{A}^{(3)}\wedge i_{\hat{k}_{(a_1}}\hat{A}^{(3)}\wedge \ldots \wedge i_{\hat{k}_{a_j)}}\hat{A}^{(3)} \biggr]\nonumber \\ &&
\end{eqnarray}
where we have indicated the rank of the 4-form field strength and 3-form potential for convenience, and the $a_j$ are $SL(2,\mathbb{R})$ indices. By equating $\hat{C}=-\hat{A}^{(6)}$ and $\hat{N}^{(8)}_a=\hat{A}^{(8)}_a$ we see that the field equations (\ref{eq:11-d cov F7}), (\ref{eq:11-d cov G9}) and (\ref{eq:11-d cov F11}) have this structure for the cases of $j=0,1,2$ respectively. Furthermore we determine the general Bianchi identity to be 
\begin{eqnarray}
d\hat{F}^{(2j+7)}_{a_1\ldots a_j}&=& -\frac{j+1}{j+2}i_{\hat{k}_{(a_1}}\hat{F}^{(4)}\wedge \hat{F}^{(2j+5)}_{a_2\ldots a_j)} +\frac{1}{j+2}\hat{F}^{(4)}\wedge i_{\hat{k}_{(a_1}}\hat{F}^{(2j+5)}_{a_2\ldots a_j)}\nonumber \\
&& +\hat{Q}^{bc}\biggl[ (j+1)i_{\hat{k}_b}\hat{F}^{(2j+9)}_{a_1\ldots a_j c} -\frac{j(j+1)}{2(j+2)}i_{\hat{k}_{(a_1}}\hat{F}^{(2j+9)}_{a_2\ldots a_j) bc} \nonumber \\
&& -ji_{\hat{k}_b\hat{k}_{(a_1}}\hat{A}^{(3)}\wedge \hat{F}^{(2j+7)}_{a_2\ldots a_j)c} -i_{\hat{k}_{b}}\hat{A}^{(3)}\wedge i_{\hat{k}_c
}\hat{F}^{(2j+7)}_{a_1\ldots a_j}\biggr] .
\end{eqnarray}
This can be seen to be gauge covariant by generalising the result (\ref{eq:triplet gauge}) to a $p$-form in a general $SL(2,\mathbb{R})$ representation
\begin{eqnarray}
\delta \hat{Y}^{(p)}_{a_1\ldots a_j}=\hat{Q}^{bc}\biggl[ \hat{\lambda}_b\wedge i_{\hat{k}_c}\hat{Y}^{(p)}_{a_1\ldots a_j} 
+ji_{\hat{k}_{(a_1}}\hat{\lambda}_{|b|} \hat{Y}^{(p)}_{a_2\ldots a_j)c} \biggr] .
\end{eqnarray}

Returning to the 11-form field strengths, we find that they are related to the mass parameters by
\begin{eqnarray}
\hat{F}^{(11)}_{ab}=\hat{\ast} \hat{Q}^{cd}\biggl[ (\hat{k}_a\cdot \hat{k}_b)(\hat{k}_c\cdot \hat{k}_d) -2(\hat{k}_a\cdot \hat{k}_c)(\hat{k}_b\cdot \hat{k}_d) \biggr]
\end{eqnarray}
which follows on from the Hodge duality definitions of $\hat{F}^{(11)}$ and $\hat{H}^{(11)}$ as well as the field equation for $\hat{D}^{(10)}$, (\ref{eq:11-d D10}). This relation could also be derived from the action which contains a cosmological-type term of the form \cite{AlonsoAlberca:2002tb}
\begin{eqnarray}
\hat{Q}^{ab}\hat{Q}^{cd}\biggl[ \frac{1}{2}(\hat{k}_a \cdot \hat{k}_b)(\hat{k}_c \cdot \hat{k}_d) -(\hat{k}_a\cdot \hat{k}_c)(\hat{k}_b\cdot \hat{k}_d) \biggr]
\end{eqnarray}
by introducing the 10-forms as auxiliary fields along the lines of \cite{Bergshoeff:1996ui,Sato:1999bu}.

Explicit calculation reveals that the three 11-form field strengths satisfy the following constraint
\begin{eqnarray}\label{eq:11-d 11-form constraint}
\epsilon^{ab}\epsilon^{cd}(\hat{k}_a\cdot \hat{k}_c)\hat{F}^{(11)}_{bd}=0 .
\end{eqnarray}
This constraint shows that there are in fact only two independent 11-forms which is related to the constraint found in IIB on the 9-form field strengths.

Although the equations given in this section were calculated for the $SL(2,\mathbb{R})$ case, the results should generalise to $n$ Killing vectors. In this case the theory would be $SL(n,\mathbb{R})$ covariant, where this group is a subgroup of the full U-duality group, and there would be a symmetric $n\times n$ mass matrix \cite{AlonsoAlberca:2002tb}. The exception to this are the constraints (\ref{eq:11-d 9-form constraint}) and (\ref{eq:11-d 11-form constraint}) which use the $SL(2,\mathbb{R})$ antisymmetric symbol and are therefore specific to the $n=2$ theory.

\section{$SL(2,\mathbb{R})$ covariant charges}\label{sec:11-d cov charges}

We now present the $SL(2,\mathbb{R})$ covariant charges. These generalise the charges constructed in \cite{Callister:2007jy,HackettJones:2003vz}. The charges consist of bilinear forms and gauge potentials. They do not explicitly depend on the mass parameters and therefore the basic structure of the charges is unaltered going to the $SL(2,\mathbb{R})$ covariant theory. The bilinears are constructed from Killing spinors and antisymmetric combinations of $p$ Dirac $\Gamma$-matrices, $\overline{\hat{\epsilon}}\hat{\Gamma}_{(p)}\hat{\epsilon}$. Specifically there are six non-zero bilinears which, following the conventions of \cite{Callister:2007jy}, we denote by $\hat{K}_{(1)}$, $\hat{\omega}_{(2)}$, $\hat{\Sigma}_{(5)}$, $\hat{\Lambda}_{(6)}$, $\hat{\Pi}_{(9)}$ and $\hat{\Upsilon}_{(10)}$. Here we have labelled the bilinears by their rank but we will drop this label when we write the actual charges. Note that all these bilinears are $SL(2,\mathbb{R})$ scalars.

The essential characteristic of the generalised charges is that they are closed and we must make sure the expressions given in \cite{Callister:2007jy,HackettJones:2003vz} still retain this property in the $SL(2,\mathbb{R})$ covariant theory. In order to check this it is essential that we know the exterior derivatives of the bilinears which, for the purely bosonic states we consider, are determined from the Killing spinor equation. This was given in \cite{Gheerardyn:2001jj} as
\begin{eqnarray}\label{eq:11-d cov Kill spin}
\delta_{\hat{\epsilon}} \hat{\psi}_{\hat{\mu}}=\hat{\tilde{D}}_{\hat{\mu}}\hat{\epsilon}=0
\end{eqnarray}
where
\begin{eqnarray}
\hat{\tilde{D}}_{\hat{\mu}}&=&\hat{\nabla}_{\hat{\mu}}(\hat{\tilde{\Omega}})+\frac{1}{288}\biggl[\hat{\Gamma}_{\hat{\mu}}^{\phantom{\hat{\mu}}\hat{\nu}_1\ldots \hat{\nu}_4} -8\hat{\delta}_{\hat{\mu}}^{\hat{\nu}_1}\hat{\Gamma}^{\hat{\nu}_2\hat{\nu}_3\hat{\nu}_4}\biggr]\hat{F}^{(4)}_{\hat{\nu}_1\ldots \hat{\nu}_4}\nonumber \\
&& -\frac{1}{12}\hat{k}_{a\hat{\nu}}\hat{Q}^{ab}\hat{k}_{b}^{\phantom{a}\hat{\nu}}\hat{\Gamma}_{\hat{\mu}}+\frac{1}{2}\hat{k}_{a\hat{\mu}}\hat{Q}^{ab} \hat{k}_{b\hat{\nu}}\hat{\Gamma}^{\hat{\nu}} .
\end{eqnarray} 
To obtain the exterior derivatives of the bilinears we hit (\ref{eq:11-d cov Kill spin}) from the left with $\overline{\hat{\epsilon}}\hat{\Gamma}_{(p)}$ for various $p$, and then antisymmetrise the free indices. See \cite{Callister:2007jy,HackettJones:2003vz} for more details. 

We briefly discuss the M2- and M5-brane cases. The generalisation of these charges to the covariant theory is trivial since both charges are $SL(2,\mathbb{R})$ scalars. The relevant differential relations for the bilinears are found to be
\begin{eqnarray}\label{eq:11-d cov domega}
d\hat{\omega}&=&i_{\hat{K}}\hat{F}^{(4)}-\hat{Q}^{ab}i_{\hat{k}_{a}}\hat{A}\wedge i_{\hat{k}_{b}}\hat{\omega}\\ \label{eq:11-d cov dsigma}
d\hat{\Sigma}&=&i_{\hat{K}}\hat{F}^{(7)}-\hat{\omega}\wedge \hat{F}^{(4)}-\hat{Q}^{ab}\biggl[(\hat{k}_{a}\cdot \hat{k}_{b})\hat{\Lambda} +i_{\hat{k}_{a}}\hat{\Lambda}\wedge \hat{k}_{b} \nonumber \\ &&+i_{\hat{k}_{a}}\hat{A}\wedge i_{\hat{k}_{b}}\hat{\Sigma}\biggr] .
\end{eqnarray}
The M2- and M5-brane charges are given by \cite{HackettJones:2003vz,Callister:2007jy}
\begin{eqnarray}\label{eq:11-d 2-charge}
\hat{L}^{(2)}&=&\hat{\omega}+i_{\hat{K}}\hat{A}\\  \label{eq:11-d 5-charge}
\hat{L}^{(5)}&=&\hat{\Sigma}+i_{\hat{K}}\hat{C}+(\hat{\omega}+\frac{1}{2}i_{\hat{K}}\hat{A})\wedge \hat{A}
\end{eqnarray}
respectively, where we have chosen gauges for the potentials that satisfy the following Lie derivatives
\begin{eqnarray}\label{eq:11-d cov A gauge}
{\cal L}_{\hat{K}}\hat{A}&=&\hat{Q}^{ab} i_{\hat{k}_a}\hat{L}^{(2)}\wedge i_{\hat{k}_b}\hat{A}\\ \label{eq:11-d cov C gauge}
{\cal L}_{\hat{K}}\hat{C}&=&\hat{Q}^{ab}\biggl[ i_{\hat{k}_a}\hat{L}^{(2)}\wedge i_{\hat{k}_b}\hat{C}  +\hat{L}^{(KK)}_{ab} \biggr] .
\end{eqnarray}
Using these conditions together with (\ref{eq:11-d cov F4}), (\ref{eq:11-d cov F7}), (\ref{eq:11-d cov domega}) and (\ref{eq:11-d cov dsigma}) it is a simple matter to show that the above charges are closed.

Showing that these gauge conditions are possible relies on using the charges themselves, and the general arguments were given in \cite{Callister:2007jy}. Note the presence of the $\hat{L}^{(KK)}_{ab}$ terms in (\ref{eq:11-d cov C gauge}) which are the generalised charges of the KK-monopoles and are given in the next subsection. The presence of these terms is related to the fact that $\hat{C}$ is a Stueckelberg field with $i_{\hat{k}_{(a}} \hat{N}^{(8)}_{b)}$ being the corresponding massive fields which can be seen from (\ref{eq:11-d cov F7}). The consistency of (\ref{eq:11-d cov A gauge}) and (\ref{eq:11-d cov C gauge}) can be checked by considering the Lie derivative of each term in (\ref{eq:11-d cov F7}) w.r.t. $\hat{K}$. This process is essentially the same as what one would do to calculate the gauge transformations of the potentials. As a result we find that the algebraic structure of the massive gauge transformations is the same as that of the Lie derivatives of the potentials w.r.t. $\hat{K}$. In the above examples we have the equivalences 
\begin{eqnarray}
\hat{L}^{(2)} \sim \hat{\chi}^{(2)} \qquad \hat{L}^{(KK)}_{ab}\sim i_{\hat{k}_{(a}} \hat{\chi}^{(7)}_{b)}
\end{eqnarray} 
where $\hat{\chi}^{(7)}_a$ are the 7-form gauge parameters of $\hat{N}^{(8)}_a$.

\subsection{KK-monopole charges}

The KK-monopole charge was presented in \cite{Callister:2007jy} for the case of a massless and single M9-brane background. It is trivial to covariantise to the $SL(2,\mathbb{R})$ covariant theory which gives
\begin{eqnarray}\label{eq:11-d KK cov charge}
\hat{L}^{(KK)}_{ab}&=&(\hat{k}_a\cdot \hat{k}_b)\hat{\Lambda} +i_{\hat{k}_{(a}}\hat{\Lambda}\wedge \hat{k}_{b)} -i_{\hat{K}}(i_{\hat{k}_{(a}}\hat{N}^{(8)}_{b)}-\frac{1}{3!}\hat{A}\wedge i_{\hat{k}_{a}}\hat{A}\wedge i_{\hat{k}_{b}}\hat{A}) \nonumber \\ &&
-i_{\hat{k}_{(a}}\hat{\omega}\wedge (i_{\hat{k}_{b)}}\hat{C} +\frac{1}{2}\hat{A}\wedge i_{\hat{k}_{b)}}\hat{A})
+i_{\hat{k}_{(a}}\hat{L}^{(5)}\wedge i_{\hat{k}_{b)}}\hat{A} \nonumber \\ &&
-\frac{1}{2}\hat{L}^{(2)}\wedge i_{\hat{k}_{a}}\hat{A}\wedge i_{\hat{k}_{b}}\hat{A} .
\end{eqnarray}
To show that this expression is closed the following differential relation for $\hat{\Lambda}$ is required
\begin{eqnarray}\label{eq:11-d dlambda7 cov}
d[(\hat{k}_{a}\cdot \hat{k}_{b})\hat{\Lambda}]&=&i_{\hat{K}\hat{k}_{(a}}\hat{G}^{(9)}_{b)}  +d(\hat{k}_{(a}\wedge i_{\hat{k}_{b)}}\hat{\Lambda})\nonumber +i_{\hat{k}_{(a}}\hat{\omega}\wedge i_{\hat{k}_{b)}}\hat{F}^{(7)}  \nonumber \\ && +i_{\hat{k}_{(a}}\hat{\Sigma}\wedge i_{\hat{k}_{b)}}\hat{F}^{(4)} 
+\hat{Q}^{cd}\biggr[ 2(\hat{k}_{c}\cdot \hat{k}_{(a})i_{\hat{k}_{b)}\hat{k}_d}\hat{A}\wedge \hat{\Lambda} \nonumber \\ && 
-i_{\hat{k}_{c}\hat{k}_{(a}}\hat{A}\wedge i_{\hat{k}_{b)}}\hat{\Lambda}\wedge \hat{k}_d 
+(\hat{k}_{c}\cdot \hat{k}_{(a})i_{\hat{k}_{b)}}\hat{\Lambda}\wedge i_{\hat{k}_d}\hat{A} \nonumber \\ &&
+i_{\hat{k}_{c}\hat{k}_{(a}}\hat{A}\wedge \hat{k}_{b)}\wedge i_{\hat{k}_{d}}\hat{\Lambda} 
-i_{\hat{k}_{c}}\hat{A}\wedge i_{\hat{k}_{d}\hat{k}_{(a}}\hat{\Lambda}\wedge \hat{k}_{b)} \nonumber \\ &&
-(\hat{k}_{a}\cdot \hat{k}_{b})i_{\hat{k}_{c}}\hat{A}\wedge i_{\hat{k}_d}\hat{\Lambda}
+2(\hat{k}_{c}\cdot \hat{k}_{(a})i_{\hat{k}_{b)}\hat{k}_d}\hat{\Pi}\biggr] .
\end{eqnarray}
In order to calculate this relation one must make use of the fact that the background spacetime contains isometries. One then must consider the components of (\ref{eq:11-d cov Kill spin}) along the isometry co-ordinates in isolation which, in this case, produces two algebraic Killing spinor equations after one decomposes the spin connection and uses the fact that the partial derivatives w.r.t. the isometry co-ordinates vanish. Then in the same way that it is possible to produce a differential relation for $\Lambda$ from (\ref{eq:11-d cov Kill spin}), it is also possible to produce algebraic relations from these additional Killing spinor equations. Combining these relations yields (\ref{eq:11-d dlambda7 cov}). This process is fairly long winded since, due to the singling out of the isometry directions, the calculation must be performed in a non-tensorial manner and so essentially must be repeated for several components in order to ultimately produce a fully tensorial expression. We therefore do not explicitly show the details here, however details of the simpler case where just one isometry is present were given in \cite{Callister:2007jy}.

It is then a straightforward task to check that (\ref{eq:11-d KK cov charge}) is closed using (\ref{eq:11-d dlambda7 cov}) and (\ref{eq:11-d cov G9}) together with the Lie derivative of $i_{\hat{k}_{(a}}\hat{N}^{(8)}_{b)}$ w.r.t. $\hat{K}$ which is given below in equation (\ref{eq:11-d cov N8 gauge}).

From the $SL(2,\mathbb{R})$ index structure of (\ref{eq:11-d KK cov charge}) we see that the KK-monopole charges form a triplet and we will show in the next sections that they are mapped to the triplet of 7-branes charges in IIB. In the current set up we have a Taub-NUT Killing vector from the KK-monopole as well as the two massive Killing vectors from the M9-branes. The charges $\hat{L}^{(KK)}_{11}$ and $\hat{L}^{(KK)}_{22}$ are each associated with the cases where the Taub-NUT Killing vector coincides with one of the massive Killing vectors and so are in this sense equivalent. However if we make the truncation (\ref{eq:11-d Q trunc}) then this equivalence disappears. In this scenario the charge $\hat{L}^{(KK)}_{11}$ represents the charge presented in \cite{Callister:2007jy} where the Taub-NUT Killing vector and the single massive Killing vector coincide. On the other hand $\hat{L}^{(KK)}_{22}$ is associated to the case where the two Killing vectors are distinct, such a configuration was considered in \cite{Eyras:1998kx} where the KK-monopole was considered in Romans' IIA theory. This charge was not considered in \cite{Callister:2007jy} and so it is a new result to show that it is closed.

The third charge in the triplet is $\hat{L}^{(KK)}_{12}$, and we will refer to the state that this corresponds to as the 11-dimensional $r7$-brane. From looking at the charge we see that it explicitly involves both Killing vectors, which suggests that the brane itself should have a 7-dimensional worldvolume with two isometries in its transverse space. As discussed in \cite{Callister:2007jy}, the tension of a brane is related to the factor multiplying the leading bilinear(s) in the charge, which is related to the fact that the charge can be interpreted as a generalised calibration \cite{Gutowski:1999tu}. We therefore determine that the tension of the $r7$-brane must scale as $\hat{k}_{1}\cdot \hat{k}_{2}$. 

We are not aware of this state having been directly investigated in the literature. However the triplet of 7-branes in IIB has been discussed \cite{Bergshoeff:2007aa,Bergshoeff:2006jj,Bergshoeff:2002mb,Meessen:1998qm,Eyras:1999at,Bergshoeff:2006ic} and many of the results should map to the triplet of states here. The charge matrix plays a key role in describing these states,
see e.g./ the IIB 7-branes given in \cite{Bergshoeff:2006ic}.
A general 7-brane is characterised by a set of integers which form a symmetric
matrix $q^{ab}$. Such a brane has a charge given by
\begin{equation}
\hat{L} = q^{ab}\hat{L}^{(KK)}_{ab}
\end{equation}
Due to the constraint on the field strengths (\ref{eq:11-d 9-form constraint}) there will be restrictions on the combinations of charges that can occur.

There exist three conjugacy classes of states defined by whether $\textrm{det}(q)$ is positive, negative or zero. Each conjugacy class has a family of solutions that are related by $SL(2,\mathbb{R})$ transformations, with the single KK-monopole belonging to the $\textrm{det}(q)=0$ class. However, since these transformations preserve $\textrm{det}(q)$, they do not map between conjugacy classes. It was shown in \cite{Bergshoeff:2006jj,Bergshoeff:2007aa} that globally well defined IIB solutions can be constructed for the cases of $\textrm{det}(q)$ zero or positive, but not negative. Such a restriction does not necessarily apply here due to the different co-dimension of the branes, but it would be interesting to explore this. Furthermore the worldvolume action for the $\textrm{det}(q)=0$ class was constructed in \cite{Bergshoeff:2006gs} with the more general case being given in \cite{Bergshoeff:2007aa} to second order in the Born-Infeld field strength. In Appendix \ref{sec:KK action} we construct the $SL(2,\mathbb{R})$ covariant worldvolume kinetic term for the $\textrm{det}(q)=0$ branes and discuss a first step to generalising this for arbitrary $\textrm{det}(q)$.  

We now offer another interpretation of this triplet of charges. By following a similar method to that used to construct (\ref{eq:11-d dlambda7 cov}) it is also possible to derive the following differential relation
\begin{eqnarray}\label{eq:11-d dlambda8 cov}
d(\hat{\Lambda}\wedge \hat{k}_{a})&=&i_{\hat{K}}\hat{G}^{(9)}_{a}-\frac{1}{3}i_{\hat{k}_{a}}\hat{\Sigma}\wedge \hat{F}-\frac{2}{3}\hat{\Sigma}\wedge i_{\hat{k}_{a}}\hat{F}-\frac{1}{3}\hat{\omega}\wedge i_{\hat{k}_{a}}\hat{F}^{(7)}\nonumber \\ &&+\frac{2}{3}i_{\hat{k}_{a}}\hat{\omega}\wedge \hat{F}^{(7)} 
+\hat{Q}^{bc}\biggl[ -(\hat{k}_{a}\cdot \hat{k}_{b})i_{\hat{k}_c}\hat{A}\wedge \hat{\Lambda}  
+i_{\hat{k}_a\hat{k}_b}\hat{A}\wedge \hat{\Lambda}\wedge \hat{k}_c \nonumber \\ &&
-i_{\hat{k}_b}\hat{A}\wedge i_{\hat{k}_c}\hat{\Lambda}\wedge \hat{k}_a
+\frac{1}{3}(\hat{k}_{b}\cdot \hat{k}_{c})i_{\hat{k}_a}\hat{\Pi}
-2(\hat{k}_{a}\cdot \hat{k}_{b})i_{\hat{k}_c}\hat{\Pi} \biggr] .
\end{eqnarray}
Note that (\ref{eq:11-d dlambda7 cov}) is merely the contraction of the above expression with a Killing vector followed by the symmetrisation of the $SL(2,\mathbb{R})$ indices. Using the above expression and (\ref{eq:11-d cov G9}) one then finds that the following expression is closed
\begin{eqnarray}\label{eq:11-d L7}
\hat{L}^{(7)}_{a}&=&\hat{\Lambda}\wedge \hat{k}_a+i_{\hat{K}}\hat{N}^{(8)}_a+\frac{1}{3}i_{\hat{k}_a}\hat{L}^{(5)}\wedge \hat{A}+\frac{2}{3}\hat{L}^{(5)}\wedge i_{\hat{k}_a}\hat{A}\nonumber \\ 
&& -\frac{1}{2}\hat{L}^{(2)}\wedge i_{\hat{k}_a}\hat{A}\wedge \hat{A} -\frac{1}{3}\hat{\omega} \wedge i_{\hat{k}_a}\hat{C} +\frac{2}{3}i_{\hat{k}_a}\hat{\omega} \wedge \hat{C} \nonumber \\ 
&& +\frac{1}{6}i_{\hat{K}}\hat{A}\wedge i_{\hat{k}_a}\hat{A}\wedge \hat{A}
\end{eqnarray}
where the gauge condition for $\hat{N}^{(8)}_a$ is
\begin{eqnarray}\label{eq:11-d cov N8 gauge}
{\cal L}_{\hat{K}}\hat{N}^{(8)}_a&=&\hat{Q}^{bc}\biggl[ i_{\hat{k}_b} \hat{L}_{(2)}\wedge i_{\hat{k}_c}\hat{N}^{(8)}_a +i_{\hat{k}_a\hat{k}_b}\hat{L}_{(2)}\wedge \hat{N}^{(8)}_c  \nonumber \\ 
&& -\frac{2}{3} \hat{L}^{(KK)}_{bc}\wedge i_{\hat{k}_a}\hat{A}
+\frac{1}{3} i_{\hat{k}_a}\hat{L}^{(KK)}_{bc}\wedge \hat{A} \nonumber \\ 
&& -\frac{1}{3}i_{\hat{k}_a}\hat{L}^{(9)}_{bc}+2i_{\hat{k}_b}\hat{L}^{(9)}_{ca} \biggr] .
\end{eqnarray}
Here $\hat{L}^{(9)}_{ab}$ is related to the M9-brane charges which are discussed in the following subsection. Its presence here is a consequence of the fact that $\hat{N}^{(8)}_a$ are Stueckelberg fields with the corresponding massive fields being certain components of $\hat{A}^{(10)}_{ab}$, which is seen from (\ref{eq:11-d cov G9}). This can be equated with the massive gauge transformations by making the equivalence
\begin{eqnarray}
\hat{L}^{(9)}_{ab}\sim -\hat{\chi}^{(9)}_{ab}
\end{eqnarray}
where $\hat{\chi}^{(9)}_{ab}$ are the principal gauge parameters of the $\hat{A}^{(10)}_{ab}$ fields.

Note that the leading term of $\hat{L}^{(7)}_a$ has a different structure from that usually found in the generalised charges. In all the other examples this term had the simple structure of a bilinear and a multiplying factor which reflected the brane's tension and allowed the charge to be interpreted as a generalised calibration. In this instance though such an interpretation is not obvious. However, we observe the following relation with the KK-monopole charges
\begin{eqnarray}\label{eq:11-d L7 LKK rel}
\hat{L}^{(KK)}_{ab}=i_{\hat{k}_{(a}}\hat{L}^{(7)}_{b)}
\end{eqnarray}
which corresponds precisely to the (spacetime) components of $\hat{L}^{(7)}_a$ which contain a leading bilinear term of the standard form. This does not offer a spacetime interpretation of the other components, which naively would seem to correspond to some sort of generalised KK-monopole state. We suspect however that the existence of $\hat{L}^{(7)}$ is a necessary requirement in order to construct the KK-monopole charges due to the fact that they contain Killing vectors, and does not therefore have a more general spacetime interpretation. We find a similar example of this for the M9-branes in the next subsection. It would be interesting to understand precisely why this expression does occur. Expressing the KK-monopole charges in terms of $\hat{L}^{(7)}$ is however useful since it makes identities such as $i_{\hat{k}_1}\hat{L}^{(KK)}_{11}=0$ explicit which is important in determining the types of multiplets that exist which we now briefly discuss.

The index structure of the charges is obviously related to the type of $SL(2,\mathbb{R})$ multiplet which they form. In addition to this, the index structure also determines how the charges should be mapped to IIB up to a discrete S-duality transformation (or alternatively to the 9-dimensional theory), i.e. whether a direct or double dimensional reduction should be performed etc. Therefore, as for the branes themselves, a given charge can belong to different multiplets depending on how it is related to IIB. An example of this was given above (\ref{eq:11-d L7 LKK rel}) where the doublet $\hat{L}^{(7)}_a$ was used to form the triplet of KK-monopole charges. The form of this triplet then contains the information about how the KK-monopoles are be mapped to IIB to form the triplet of charges of the 7-branes. The KK-monopoles can also be mapped to IIB to form the doublet of states that consist of the D5 and NS5-branes. In terms of the charges this is expressed by
\begin{eqnarray}\label{eq:11-d KK-charge doublet}
i_{\hat{k}_b\hat{k}_c}\hat{L}^{(7)}_a= \pm i_{\hat{k}_1\hat{k}_2}\hat{L}^{(7)}_a .
\end{eqnarray}
This charge is interpreted as a doublet since the indices on the Killing vectors are fixed by their antisymmetry. Note that this is only true for the $SL(2,\mathbb{R})$ covariant theory. No other charge multiplets can be formed from $\hat{L}^{(7)}_a$. The quadruplet $i_{\hat{k}_{(a}\hat{k}_{b}}\hat{L}^{(7)}_{c)}$ is identically zero, the doublet $\epsilon^{bc}i_{\hat{k}_a\hat{k}_b}\hat{L}^{(7)}_c$ merely reduces to (\ref{eq:11-d KK-charge doublet}), and the scalar $\epsilon^{ab}i_{\hat{k}_a}\hat{L}^{(7)}_b$ has the same problem as $\hat{L}^{(7)}_a$ in that it does not have the correct structure of a charge. We therefore find that the only multiplets we can construct correspond precisely to those found in the IIB theory as we would expect.

\subsection{M9-brane charges}\label{sec:11-d M9 charges}

We now consider the M9-brane charge. The charge for a single M9-brane was constructed in a non-covariant fashion in \cite{Callister:2007jy} and is given by
\begin{eqnarray}\label{eq:11-d M9-charge}
\nonumber i_{\hat{\alpha}}\hat{L}^{(9)}_{11}&=&\hat{R}_{\hat{\alpha}}^2i_{\hat{\alpha}}\hat{\Pi}+i_{\hat{K}}(i_{\hat{\alpha}}\hat{A}^{(10)}-\frac{1}{4!}\hat{A}\wedge (i_{\hat{\alpha}}\hat{A})^3)\\ \nonumber  &&-i_{\hat{\alpha}}\hat{\omega}\wedge (i_{\hat{\alpha}}\hat{N}^{(8)}-\frac{1}{3!}\hat{A}\wedge (i_{\hat{\alpha}}\hat{A})^2)+\hat{L}^{(KK)}_{11}\wedge i_{\hat{\alpha}}\hat{A}\\ &&-\frac{1}{2}i_{\hat{\alpha}}\hat{L}^{(5)}\wedge (i_{\hat{\alpha}}\hat{A})^2+\frac{1}{3!}\hat{L}^{(2)}\wedge (i_{\hat{\alpha}}\hat{A})^3
\end{eqnarray}
where $\hat{\alpha}$ is interpreted as both the massive Killing vector of the theory and as the Killing vector that appears in the spacetime solution of the brane. The contraction of $\hat{\alpha}$ with the leading bilinear reflects the fact that it lies parallel to the brane worldvolume. 

Each term here will contain three $SL(2,\mathbb{R})$ free indices when written in a covariant fashion and therefore this form of the charge forms part of a quadruplet which maps to the charges of the quadruplet of 9-branes in IIB. However, we expect that there should also be a doublet and triplet of charges that map to the charges of the doublet of 9-branes and triplet of 7-branes in IIB. These multiplets are not directly attainable from the above form of the charge. To determine these multiplets we need to find the full $\hat{L}^{(9)}_{ab}$ expression, which we will refer to as the principal M9-brane charge. The situation here is similar to that in the last section where the KK-monopole charges were found to be more naturally expressed in terms of the expression $\hat{L}^{(7)}_a$. 

This expression can be determined by the usual method whereby all the possible terms that could be present are collected and it is explicitly found which combination is closed. However an alternative method is to use (\ref{eq:11-d cov N8 gauge}) from which the full expression can also be deduced. Either way we find it to be given by
\begin{eqnarray}\label{eq:11-d cov L9}
\hat{L}^{(9)}_{ab}&=&(\hat{k}_a\cdot \hat{k}_b)\hat{\Pi}-i_{\hat{K}}\hat{A}^{(10)}_{ab}+ \frac{3}{4}i_{\hat{k}_{(a}}\hat{\omega} \wedge \hat{N}^{(8)}_{b)} - \frac{1}{4}\hat{\omega} \wedge i_{\hat{k}_{(a}}\hat{N}^{(8)}_{b)}\nonumber \\ 
&& +\frac{3}{4}\hat{L}^{(7)}_{(a}\wedge i_{\hat{k}_{b)}}\hat{A}+\frac{1}{4}\hat{L}^{(KK)}_{ab}\wedge \hat{A} -\frac{1}{4}\hat{L}^{(5)}\wedge i_{\hat{k}_a}\hat{A}\wedge i_{\hat{k}_b}\hat{A}\nonumber \\ 
&& -\frac{1}{4}i_{\hat{k}_{(a}}\hat{L}^{(5)}\wedge i_{\hat{k}_{b)}}\hat{A}\wedge \hat{A} +\frac{1}{3!}\hat{L}^{(2)}\wedge \hat{A}\wedge i_{\hat{k}_a}\hat{A}\wedge i_{\hat{k}_b}\hat{A}\nonumber \\ 
&& -\frac{1}{4!}i_{\hat{K}}\hat{A}\wedge \hat{A}\wedge i_{\hat{k}_a}\hat{A}\wedge i_{\hat{k}_b}\hat{A} .
\end{eqnarray}
This expression can be shown to be closed by determining the differential relation for $(\hat{k}_a\cdot \hat{k}_b) \hat{\Pi}$. This is calculated using the same technique which produced (\ref{eq:11-d dlambda7 cov}) and (\ref{eq:11-d dlambda8 cov}). The result is 
\begin{eqnarray}
d[(\hat{k}_{a}\cdot \hat{k}_{b})\hat{\Pi}]&=&-\frac{1}{4}\hat{\omega}\wedge i_{\hat{k}_{(a}}\hat{G}_{b)}^{(9)} +\frac{3}{4}i_{\hat{k}_{(a}}\hat{\omega}\wedge \hat{G}_{b)}^{(9)} -\frac{1}{4}(\hat{k}_{a}\cdot \hat{k}_{b})\hat{\Lambda}\wedge \hat{F}\nonumber \\
&& -\frac{1}{4}i_{\hat{k}_{(a}}\hat{\Lambda}\wedge \hat{F}\wedge \hat{k}_{b)} +\frac{3}{4}\hat{\Lambda}\wedge i_{\hat{k}_{(a}}\hat{F}\wedge \hat{k}_{b)} -i_{\hat{K}}\hat{F}^{(11)}_{ab}\nonumber \\ 
&& +\hat{Q}^{cd}\biggl[ -\frac{3}{4}(\hat{k}_{a}\cdot \hat{k}_{b})(\hat{k}_{c}\cdot \hat{k}_{d})\hat{\Upsilon} -\frac{3}{2}(\hat{k}_{a}\cdot \hat{k}_{c})(\hat{k}_{b}\cdot \hat{k}_{d})\hat{\Upsilon}\nonumber \\
&& -(\hat{k}_{a}\cdot \hat{k}_{b})i_{\hat{k}_{c}}\hat{\Upsilon}\wedge \hat{k}_{d} +\frac{1}{4}(\hat{k}_{c}\cdot \hat{k}_{d})i_{\hat{k}_{(a}}\hat{\Upsilon}\wedge \hat{k}_{b)}\nonumber \\
&& +\frac{1}{2}(\hat{k}_{c}\cdot \hat{k}_{(a})i_{\hat{k}_{b)}}\hat{\Upsilon}\wedge \hat{k}_{d} -2(\hat{k}_{c}\cdot \hat{k}_{(a})i_{\hat{k}_{|d|}}\hat{\Upsilon}\wedge \hat{k}_{b)}
\nonumber  \\ && 
+2(\hat{k}_{c}\cdot \hat{k}_{(a})i_{\hat{k}_{b)}\hat{k}_{d}}\hat{ A}\wedge \hat{ \Pi} -(\hat{k}_{a}\cdot \hat{k}_{b})i_{\hat{k}_{c}}\hat{A}\wedge i_{\hat{k}_{d}}\hat{ \Pi}\biggr] .
\end{eqnarray}
Note that there are several equivalent variations of this expression, however this form is the most natural for showing that $\hat{L}^{(9)}_{ab}$ is closed. The gauge condition for the $\hat{A}^{(10)}_{ab}$ fields is given by 
\begin{eqnarray}\label{eq:11-d cov A10 gauge}
{\cal L}_{\hat{K}}\hat{A}^{(10)}_{ab}&=&\hat{Q}^{cd}\biggl[ i_{\hat{k}_c}\hat{L}^{(2)}\wedge i_{\hat{k}_d}\hat{A}^{(10)}_{ab} -2i_{\hat{k}_c\hat{k}_{(a}}\hat{L}^{(2)}\wedge \hat{A}^{(10)}_{b)d} \nonumber \\
&&  -\frac{1}{4}\hat{L}^{(KK)}_{cd}\wedge i_{\hat{k}_{(a}}\hat{A}\wedge i_{\hat{k}_{b)}}\hat{A}  +\frac{1}{4}i_{\hat{k}_{(a}}\hat{L}^{(KK)}_{|cd|}\wedge i_{\hat{k}_{b)}}\hat{A}\wedge \hat{A}  \nonumber \\ 
&& -\frac{1}{4} i_{\hat{k}_{(a}}\hat{L}^{(9)}_{|cd|}\wedge i_{\hat{k}_{b)}}\hat{A}+\frac{3}{2} i_{\hat{k}_{c}}\hat{L}^{(9)}_{d(a}\wedge i_{\hat{k}_{b)}}\hat{A} +\frac{1}{2} i_{\hat{k}_{c}\hat{k}_{(a}}\hat{L}^{(9)}_{b)d}\wedge \hat{A}\nonumber \\
&& -3i_{\hat{k}_c}\hat{L}^{(11)}_{abd} +\frac{3}{4}i_{\hat{k}_{(a}}\hat{L}^{(11)}_{b)cd} \biggr] .
\end{eqnarray}

Note the presence of the quadruplet of terms $\hat{L}^{(11)}_{abc}$ in the above expression. This is analogous to the gauge conditions for $\hat{C}$ and $\hat{N}^{(8)}_a$. In this case $\hat{A}^{(10)}_{ab}$ are Stueckelberg fields with the corresponding massive fields being certain components of $\hat{A}^{(12)}_{abc}$ which can be seen from (\ref{eq:11-d cov F11}). Equation (\ref{eq:11-d cov A10 gauge}) can be related to the massive gauge transformations of $\hat{A}^{(10)}_{ab}$ by making the identification
\begin{eqnarray}
\hat{L}^{(11)}_{abc}\sim \hat{\chi}^{(11)}_{abc}
\end{eqnarray}
where $\hat{\chi}^{(11)}_{abc}$ are the principal gauge parameters of the $\hat{A}^{(12)}_{abc}$ fields. 

The structure of the $\hat{L}^{(11)}_{abc}$ terms were inferred from demanding that $\hat{L}^{(9)}_{ab}$ should be closed. They are given by 
\begin{eqnarray}
\hat{L}^{(11)}_{abc}&=&(\hat{k}_{(a}\cdot \hat{k}_b) \hat{\Upsilon}\wedge \hat{k}_{c)} +i_{\hat{K}}\hat{A}^{(12)}_{abc}+\frac{1}{5}\hat{\omega}\wedge i_{\hat{k}_{(a}}\hat{A}^{(10)}_{bc)} -\frac{4}{5}i_{\hat{k}_{(a}}\hat{\omega}\wedge \hat{A}^{(10)}_{bc)}\nonumber \\
&& +\frac{4}{5}\hat{L}^{(9)}_{(ab}\wedge i_{\hat{k}_{c)}}\hat{A} +\frac{1}{5}i_{\hat{k}_{(a}}\hat{L}^{(9)}_{bc)}\wedge \hat{A} -\frac{3}{10}\hat{L}^{(7)}_{(a}\wedge i_{\hat{k}_b}\hat{A} \wedge i_{\hat{k}_{c)}}\hat{A}\nonumber \\
&& -\frac{1}{5}i_{\hat{k}_{(a}}\hat{L}^{(7)}_b\wedge i_{\hat{k}_{c)}}\hat{A}\wedge \hat{A} +\frac{1}{15}\hat{L}^{(5)}\wedge i_{\hat{k}_a}\hat{A}\wedge i_{\hat{k}_b}\hat{A} \wedge i_{\hat{k}_c}\hat{A}\nonumber \\ 
&& +\frac{1}{10}i_{\hat{k}_{(a}}\hat{L}^{(5)}\wedge i_{\hat{k}_b}\hat{A}\wedge i_{\hat{k}_{c)}}\hat{A}\wedge \hat{A} -\frac{1}{4!}\hat{L}^{(2)}\wedge i_{\hat{k}_a}\hat{A} \wedge i_{\hat{k}_b}\hat{A} \wedge i_{\hat{k}_c}\hat{A}\wedge \hat{A}\nonumber \\ 
&&+\frac{1}{5!}i_{\hat{K}}\hat{A}\wedge i_{\hat{k}_a}\hat{A} \wedge i_{\hat{k}_b}\hat{A} \wedge i_{\hat{k}_c}\hat{A}\wedge \hat{A} .
\end{eqnarray}
They have the form of a quadruplet of 11-form charges and in principal would correspond to some type of 11-branes. Obviously spacetime solutions of these branes do not exist but the fact that such charge structures do exist seems to be related to the fact that gauge algebra can be treated independently of the dimensionality of the background spacetime. Note that double dimensional reduction of the $(1,1,1)$ component, using 
\begin{eqnarray}
i_{\hat{\alpha}}\hat{A}^{(12)}&\rightarrow&-C^{(11)} + \frac{1}{5!}C^{(3)}\wedge (B)^4
\end{eqnarray}
where $C^{(11)}$ is an 11-form Ramond-Ramond potential, gives an expression in IIA with the structure of a D10-brane charge in IIA, i.e. a 10-form charge that has the same general structure as the D-brane charges presented in \cite{Callister:2007jy}.

We now discuss the expression $\hat{L}^{(9)}_{ab}$. The situation here is analogous to $\hat{L}^{(7)}_a$ although in this instance we observe that the leading bilinear term of $\hat{L}^{(9)}_{ab}$ is of the form found in the other charges. However, like for $\hat{L}^{(7)}_a$ we suspect that not all the spacetime components correspond to supersymmetric spacetime solutions. Certain combinations do however seem to map to the charges of known states in IIB and so we propose that these correspond to states in the 11-dimensional $SL(2,\mathbb{R})$ theory as well. These consist of the following triplet and quadruplet:
\begin{eqnarray*}
\text{Triplet}&=&i_{\hat{k}_1\hat{k}_2}\hat{L}^{(9)}_{ab}\\ \label{eq:11-d M9 quad}
\text{Quadruplet}&=&i_{\hat{k}_{(a}}\hat{L}^{(9)}_{bc)} .
\end{eqnarray*}
The (1,1,1) component of the quadruplet was given by (\ref{eq:11-d M9-charge}). In the following sections we will show that these multiplets map to the charges of the triplet of 7-branes and quadruplet of 9-branes in IIB respectively.

Considering the triplet first, the charges $i_{\hat{k}_1\hat{k}_2}\hat{L}^{(9)}_{11}$ and $i_{\hat{k}_1\hat{k}_2}\hat{L}^{(9)}_{22}$ merely correspond to the usual M9-branes with an extra isometry parallel to the worldvolume. The extra isometry in each case is not however intrinsic to the charge and therefore neither to the brane itself. They are only present in order for the dimensional reduction or T-duality to be performed in the mapping to IIB. On the other hand the charge $i_{\hat{k}_1\hat{k}_2}\hat{L}^{(9)}_{12}$ does contain two Killing vectors that are intrinsic to the charge and suggest that the corresponding state, which we will refer to as the 11-dimensional r9-brane, contains two isometries parallel to its worldvolume. This therefore corresponds to a different brane which maps to the IIB r7-brane and which we are not aware of having been discussed in the literature. As for the triplet of monopoles, much of the discussion for the IIB 7-branes should be applicable to the states here.

Similarly the charges $i_{\hat{k}_1}\hat{L}^{(9)}_{11}$ and $i_{\hat{k}_2}\hat{L}^{(9)}_{22}$ of the quadruplet simply correspond to the usual M9-brane charges, this time however with no extra isometry. The other two charges $i_{\hat{k}_{(1}}\hat{L}^{(9)}_{12)}$ and $i_{\hat{k}_{(1}}\hat{L}^{(9)}_{22)}$ involve combinations of the M9-brane and r9-brane charges, but with different contractions with the Killing vectors. In each case, when compared to the triplet, the difference involves the contraction of a Killing vector which is intrinsic to the main charge expression. This might therefore mean that the corresponding states are inherently different to those associated with the triplet, despite the charges all being based on the principal M9-brane charge. One would have to investigate the spacetime solutions to resolve these issues. In \cite{Bergshoeff:2006ic} it was shown that, unlike for the triplet of 7-branes, there is only a single conjugacy class for the quadruplet of 9-branes, and also that there are two constraints on the combinations of states that correspond to supersymmetric spacetime solutions, reducing the independent degrees of freedom to two. One would expect these observations to apply here as well, but it might have an obvious geometrical interpretation in terms of the spacetime solutions.

The remaining task is to determine the doublet of states and charges that map to the doublet of 9-branes in IIB. A natural candidate would be the following doublet
\begin{eqnarray}\label{eq:M9 charge doublet}
\epsilon^{ab}i_{\hat{k}_{a}}\hat{L}^{(9)}_{bc} .
\end{eqnarray}
However, on mapping this expression to IIB it is found that this produces a doublet of charges that each intrinsically depend on a Killing vector. This suggests that the corresponding IIB states are some sort of KK9-monopole. Although we do not consider these states in any detail in this paper, in Appendix \ref{sec:mapping 10-forms} we map the 11-dimensional 10-form potentials to IIB and in doing so argue that the KK9-monopoles might source a doublet of mass parameters in a non-covariant (in the spacetime sense) IIB SUGRA. Furthermore, it seems that this doublet might be the IIB origin of the doublet of mass parameters $(m_4,\tilde{m}_4)$ discussed in \cite{Bergshoeff:2002nv}.

\subsection{11-dimensional charge doublet}

Since no more charge multiplets can be constructed from $\hat{L}^{(9)}_{ab}$ we must look elsewhere. In order to construct an appropriate charge doublet it seems necessary to introduce a doublet of 11-form gauge potentials $\hat{A}^{(11)}_a$. Neglecting mass terms, the field equation of this doublet is given simply by
\begin{eqnarray}\label{eq:11-d F12}
d\hat{A}^{(11)}_a&=&\hat{F}^{(12)}_a .
\end{eqnarray}
Taking the $a=1$ component and performing a double dimensional reduction one obtains the field equation given in \cite{Bergshoeff:2006qw} for a 10-form potential in IIA. Furthermore, it is a simple task to show that this doublet maps to the field equation for the 10-form doublet in IIB given in \cite{Bergshoeff:2005ac}, which we show in the following sections. The gauge algebra of this doublet seems to be independent of the other potentials in the theory already considered. A related observation was made in \cite{Bergshoeff:2006gs} where the worldvolume action of the doublet of 9-branes in IIB also seems to follow a different structure than the other branes in the theory.  We do however suspect that (\ref{eq:11-d F12}) is incomplete and should in fact include massive terms. The full structure could in principal be determined by investigating the closure of the SUSY transformations as done in \cite{Bergshoeff:2005ac,Bergshoeff:2006gs} but we do not investigate this in this paper.
 
We then propose the following structure for a doublet of 10-form `charges'
\begin{eqnarray}\label{eq:11-d L10}
\hat{L}^{(10)}_a=\hat{\Pi}\wedge \hat{k}_a +i_{\hat{K}}\hat{A}^{(11)}_a .
\end{eqnarray}
It is problematic to show that (\ref{eq:11-d L10}) is closed generally without making use of the dimensionality of the background spacetime. This is because the differential relation for the bilinear term is calculated using the Killing spinor equation (\ref{eq:11-d cov Kill spin}) which is derived from supersymmetry transformations and therefore must take into account the spacetime dimension. We therefore restrict our attention to the following doublet
\begin{eqnarray}\label{eq:11-d ikL10}
i_{\hat{k}_1\hat{k}_2}\hat{L}^{(10)}_a
\end{eqnarray}
for which the differential relation of the bilinear term is calculable. We find that this term is independently closed if we neglect the massive terms. We then conjecture that if it were possible to calculate the differential relation while neglecting the dimensionality of the background spacetime then one would find it to be given in the massless case by
\begin{eqnarray}\label{eq:11-d dPi doublet}
d(\hat{\Pi}\wedge \hat{k}_a)&=&i_{\hat{K}}\hat{F}^{(12)}_a .
\end{eqnarray}

The expression (\ref{eq:11-d ikL10}) is then easily shown to be closed in the massless case, and we will show how this maps to the charges of the 9-brane doublet in IIB in the following sections. The massive case requires us to take account of the the massive terms that would appear in (\ref{eq:11-d dPi doublet}) and (\ref{eq:11-d F12}) and also the Lie derivative of the $\hat{A}^{(11)}_a$ w.r.t $\hat{K}$ which would appear to include higher rank charge structures just as for the other potentials already considered. Due to the difficulties in obtaining the full algebraic structures of the differential relations of the high rank bilinears, we do not explore the massive case in this paper.

From the structure of (\ref{eq:11-d L10}) we would expect the charge doublet (\ref{eq:11-d ikL10}) to represent a doublet of 9-branes each with a single isometry direction transverse to the worldvolume (since each charge contains only one intrinsic Killing vector) and tensions which scale as $|\hat{k}_1|^2$ and $|\hat{k}_2|^2$. The situation here is essentially just a higher dimensional version of $i_{\hat{k}_1\hat{k}_2}\hat{L}^{(7)}_a$.

\section{Charges in non-covariant IIA}\label{sec:IIA charges}

We now give the IIA charges produced from dimensionally reducing the 11-dimensional charges in the previous section. These charges are guaranteed to be closed due to their 11-dimensional origin. Therefore, for the sake of brevity we do not give the full field strength equations, differential relations for the bilinears or Lie derivatives of the potentials w.r.t. $K$, but these are easily calculated. The reduction is performed over $\hat{k}_1=\hat{\alpha}$ and the reduction rules for the fields and bilinears are given in Appendix \ref{sec:reducs} where the non-covariant notation of Section \ref{sec:11-d field eqns} is used. The definitions of the IIA fields and lower rank charges which are used were given in \cite{Callister:2007jy}.

\subsection{Dimensionally reducing KK-monopole charges}

We begin by considering the KK-monopole charges. Reducing $\hat{L}^{(KK)}_{11}$ gives the D6-brane charge which was presented in \cite{Callister:2007jy} so we do not restate it here. Reducing $\hat{L}^{(KK)}_{22}$ on the other hand produces the charge for the KK6-monopole which is given by
\begin{eqnarray}\label{eq:IIA KK6 charge}
M^{(KK6)}&=&\biggl(e^{-3\phi}R_{\beta}^2 \ +e^{-\phi}(i_{\beta}C^{(1)})^2\biggr) \ \Lambda + e^{-3\phi}i_{\beta}\Lambda\wedge \beta - e^{-2\phi}i_{\beta}C^{(1)} \ \tilde{\Sigma}\wedge \beta \nonumber \\ 
&& + (e^{-2\phi}R_{\beta}^2 \ \tilde{\Sigma} - e^{-2\phi}i_{\beta}\tilde{\Sigma} \wedge \beta +e^{-\phi}i_{\beta}C^{(1)} \ i_{\beta}\Lambda)\wedge C^{(1)} \nonumber \\ 
&&  + i_{\beta}M^{(NS5)}\wedge i_{\beta}C^{(3)} - i_{\beta}(e^{-\phi}\Omega +\tilde{K}\wedge C^{(1)})\wedge i_{\beta}B^{(6)} \nonumber \\ 
&& -\frac{1}{2}i_{\beta}\biggl((e^{-\phi}\Omega+\tilde{K}\wedge C^{(1)})\wedge C^{(3)}\biggr)\wedge i_{\beta}C^{(3)} - M^{(0)}i_{\beta}N^{(7)} \nonumber \\
&& + i_{\beta K}N^{(8)} -\frac{1}{3}i_{\beta}(i_{K}C^{(3)}\wedge C^{(3)})\wedge i_{\beta}C^{(3)} .
\end{eqnarray}
Here $M^{(p)}$ is the charge of a D$p$-brane and $M^{(NS5)}$ is the charge of the NS5-brane. We see that the KK6-monopole tension scales as $e^{-3\phi}R_{\beta}^2 \ +e^{-\phi}(i_{\beta}C^{(1)})^2$ which is deduced from the leading bilinear term, and also that it minimally couples to $i_{\beta}N^{(8)}$. This is in agreement with \cite{Eyras:1999at}.

Next we consider $\hat{L}^{(KK)}_{12}$ which reduces to the charge of what we will refer to as the r6-brane. We find it to be given by
\begin{eqnarray}\label{eq:IIA r6 charge}
M^{(r6)}&=&e^{-\phi}i_{\beta}C^{(1)}\Lambda + i_{\beta K}\phi^{(8)} + \frac{1}{2}\biggl( i_{K}N^{(7)} - e^{-2\phi}\Sigma\wedge \beta + i_{\beta}M^{(NS5)}\wedge B \nonumber \\ && - M^{(0)}i_{\beta}C^{(7)} + (e^{-\phi}Z-\frac{1}{2}\tilde{K}\wedge C^{(3)} - i_{K}C^{(5)})\wedge i_{\beta}C^{(3)} \\ && - i_{\beta}(e^{-\phi}\Omega + \tilde{K}\wedge C^{(1)})\wedge C^{(5)} + e^{-\phi}i_{\beta}\Lambda\wedge C^{(1)} + \tilde{K}\wedge i_{\beta}B^{(6)}  \biggl) . \nonumber
\end{eqnarray}
We see that the tension here scales as $e^{-\phi}i_{\beta}C^{(1)}$ and the brane minimally couples to the combination of potentials $\phi^{(8)}-\frac{1}{2}N^{(7)}$. In \cite{Bergshoeff:2006qw} the potential $\phi^{(8)}$ was considered and it was found that it does not minimally couple to any supersymmetric state on its own. This does not contradict the result here since in that reference states containing isometries were not considered, nor was the potential $N^{(7)}$.

\subsection{Dimensionally reducing the M9-brane triplet}

We now consider the triplet of M9-brane charges. The charge $i_{\hat{\alpha}\hat{\beta}}\hat{L}^{(9)}_{11}$ reduces to the D8-brane charge given in \cite{Callister:2007jy} with an overall contraction with $\beta$ and so we will not give it here. The $i_{\hat{\alpha}\hat{\beta}}\hat{L}^{(9)}_{22}$ charge reduces to the KK8-monopole charge which is found to be given by
\begin{eqnarray}\label{eq:IIA KK8 charge}
i_{\beta}M^{(KK8)}&=&(e^{-3\phi}R_{\beta}^2+e^{-\phi}(i_{\beta}C^{(1)})^2)i_{\beta}\Psi+i_{K\beta}D^{(9)}+i_{\beta}\tilde{K}i_{\beta}N^{(8)}\nonumber \\ &&-i_{\beta}(e^{-\phi}\Omega+\tilde{K}\wedge C^{(1)})\wedge i_{\beta}N^{(7)}-\frac{1}{2}i_{\beta}(e^{-\phi}Z-i_K C^{(5)})\wedge (i_{\beta}C^{(3)})^2\nonumber \\ &&+\frac{1}{3}i_{\beta}(\tilde{K}\wedge C^{(3)})\wedge (i_{\beta}C^{(3)})^2+M^{(KK6)}\wedge i_{\beta}B + M^{(KK)}\wedge i_{\beta}C^{(3)} \nonumber \\ &&- i_{\beta}M^{(NS5)}\wedge i_{\beta}C^{(3)}\wedge i_{\beta}B
\end{eqnarray}
where $M^{(KK)}$ is the charge of the IIA KK-monopole. The KK8-monopole was considered for example in \cite{Eyras:1999at}. It has a tension that scales as $e^{-3\phi}R_{\beta}^3+e^{-\phi}R_{\beta}(i_{\beta}C^{(1)})^2$. This can be read of from the leading bilinear term in its charge once the contraction of $\beta$ with $\Psi$ is taken into account which causes the factor in this term to differ from the brane tension by a factor of $R_{\beta}$. Furthermore the KK8-monopole minimally couples to the potential $i_{\beta}D^{(9)}$.

Next we reduce the charge $i_{\hat{\alpha}\hat{\beta}}\hat{L}^{(9)}_{12}$. We will denote the corresponding IIA brane as the r8-brane and find its charge to be given by
\begin{eqnarray}\label{eq:IIA r8 charge}
i_{\beta}M^{(r8)}&=&e^{-\phi}i_{\beta}C^{(1)}i_{\beta}\Psi+i_{K\beta}B^{(9)}+i_{\beta}\tilde{K}i_{\beta}\phi^{(8)} + M^{(r6)}\wedge i_{\beta}B\nonumber \\ && +\frac{1}{2}\biggl( i_{\beta}(e^{-\phi}\Lambda + \tilde{K}\wedge C^{(5)} + i_{K}C^{(7)})\wedge i_{\beta}C^{(3)} \nonumber \\ &&- i_{\beta}(e^{-\phi}\Omega+\tilde{K}\wedge C^{(1)})\wedge i_{\beta}C^{(7)} - i_{\beta}(\tilde{K}\wedge N^{(7)}) \nonumber \\ && +M^{(KK)}\wedge B-i_{\beta}M^{(NS5)}\wedge B \wedge i_{\beta}B \biggr) .
\end{eqnarray}
The r8-brane is related by T-duality to the r7-brane in IIB. Its tension can be seen to scale as $e^{-\phi}R_{\beta}i_{\beta}C^{(1)}$ and it minimally couples to $i_{\beta}B^{(9)}$.

\subsection{Dimensionally reducing M9-brane quadruplet}

Next we consider the dimensional reduction of the quadruplet of M9-brane charges. The charge $i_{\hat{\alpha}}\hat{L}_{11}$ merely reduces to the D8-brane charge and so we do not restate it here. The next charge is then $\frac{2}{3}i_{\hat{\alpha}}\hat{L}^{(9)}_{12}+\frac{1}{3}i_{\hat{\beta}}\hat{L}^{(9)}_{11}$, which reduces to the following
\begin{eqnarray}\label{eq:IIA r9 charge}
M^{(r9)}&=&\frac{1}{3}e^{-2\phi}i_{\beta}\Pi +e^{-\phi}i_{\beta}C^{(1)}\Psi +\frac{1}{3}e^{-\phi}i_{\beta}\Psi\wedge C^{(1)} +\frac{1}{3}i_{K\beta}A^{(10)}\nonumber \\
&& -\frac{2}{3}i_K B^{(9)} +\tilde{K}\wedge (\frac{2}{3}i_{\beta}\phi^{(8)}-\frac{1}{3}N^{(7)}) +\frac{1}{3}i_{\beta}(e^{-\phi}\Omega +\tilde{K}\wedge C^{(1)})\wedge C^{(7)}\nonumber \\
&&  +\frac{1}{3}(e^{-\phi}\Lambda +\tilde{K}\wedge C^{(5)} +i_K C^{(7)})\wedge i_{\beta}C^{(3)} +\frac{2}{3}M^{(r6)}\wedge B \nonumber \\
&& -\frac{1}{6}i_{\beta}M^{(NS5)}\wedge (B)^2 +\frac{1}{3}M^{(0)}i_{\beta}C^{(9)} .
\end{eqnarray}
We will refer to the state to which this charge corresponds as the r9-brane. It has a tension which scales as $e^{-\phi}\sqrt{\frac{1}{9}e^{-2\phi}R^2_{\beta}+(i_{\beta}C^{(1)})^2}$ (arising from the coefficients of the two independent leading bilinears \cite{Callister:2007jy}) and minimally couples to the combination of potentials $\frac{1}{3}i_{\beta}A^{(10)}-\frac{2}{3}B^{(9)}$. In \cite{Bergshoeff:2006qw} the potential $A^{(10)}$ was considered and it was found that it does not minimally couple to any supersymmetric state on its own. This does not contradict the result here since in that reference states containing isometries were not considered, nor was the potential $B^{(9)}$.

Next we reduce the charge $\frac{2}{3}i_{\hat{\beta}}\hat{L}^{(9)}_{12}+\frac{1}{3}i_{\hat{\alpha}}\hat{L}^{(9)}_{22}$. In IIA this gives
\begin{eqnarray}\label{eq:IIA s9 charge}
M^{(s9)}&=&\frac{2}{3}e^{-2\phi}i_{\beta}C^{(1)}i_{\beta}\Pi +(\frac{1}{3}e^{-3\phi}R_{\beta}^2+e^{-\phi}(i_{\beta}C^{(1)})^2)\Psi \nonumber \\
&& +\frac{2}{3}e^{-\phi}i_{\beta}C^{(1)}i_{\beta}\Psi\wedge C^{(1)} +\frac{2}{3}i_{K\beta}B^{(10)} -\frac{1}{3}i_K D^{(9)} +\frac{1}{3}\tilde{K}\wedge i_{\beta}N^{(8)} \nonumber \\
&& -\frac{1}{6}(e^{-\phi}Z-i_K C^{(5)}-\frac{2}{3}\tilde{K}\wedge C^{(3)})\wedge (i_{\beta}C^{(3)})^2\nonumber \\
&& +i_{\beta}(e^{-\phi}\Omega +\tilde{K}\wedge C^{(1)})\wedge (-\frac{2}{3}\phi^{(8)} +\frac{1}{3}N^{(7)}) +\frac{1}{3}M^{(KK6)}\wedge B \nonumber \\
&& +\frac{2}{3}M^{(r6)}\wedge i_{\beta}C^{(3)}  -\frac{1}{3}i_{\beta}M^{(NS5)}\wedge i_{\beta}C^{(3)}\wedge B +\frac{2}{3}M^{(0)}i_{\beta}B^{(9)} . \nonumber \\
&&
\end{eqnarray}
We will refer to the state to which this charge corresponds as the s9-brane. It has a tension which scales as $e^{-\phi}\sqrt{\frac{4}{9}e^{-2\phi}R^2_{\beta}(i_{\beta}C^{(1)})^2+(\frac{1}{3}e^{-2\phi}R^2_{\beta} + (i_{\beta}C^{(1)})^2)^2}$ and minimally couples to the combination of potentials $\frac{2}{3}i_{\beta}B^{(10)}-\frac{1}{3}D^{(9)}$.

The last charge in this multiplet is $i_{\hat{\beta}}\hat{L}^{(9)}_{22}$. It reduces to
\begin{eqnarray}\label{eq:IIA q9 charge}
M^{(q9)}&=&(e^{-4\phi}R_{\beta}^2+e^{-2\phi}(i_{\beta}C^{(1)})^2)i_{\beta}\Pi +(e^{-3\phi}R_{\beta}^2+e^{-\phi}(i_{\beta}C^{(1)})^2)i_{\beta}(\Psi\wedge C^{(1)}) \nonumber \\
&& +i_{K\beta}D^{(10)} +i_{\beta}(e^{-\phi}\Omega +\tilde{K}\wedge C^{(1)})\wedge (-i_{\beta}N^{(8)}+\frac{1}{6}(i_{\beta}C^{(3)})^2)\nonumber \\
&&  +\frac{1}{8}i_K C^{(3)}\wedge C^{(3)} \wedge (i_{\beta}C^{(3)})^2 +M^{(KK6)}\wedge i_{\beta}C^{(3)} \nonumber \\
&& -\frac{1}{2}i_{\beta}M^{(NS5)}\wedge (i_{\beta}C^{(3)})^2 +M^{(0)}i_{\beta}D^{(9)} .
\end{eqnarray}
We will refer to the state to which this charge corresponds as the q9-brane. It has a tension which scales as $e^{-\phi}(e^{-2\phi}R^2_{\beta}+(i_{\beta}C^{(1)})^2)^{3/2}$ and minimally couples to $i_{\beta}D^{(10)}$.

\subsection{Dimensionally reducing 9-brane doublet}

We now dimensionally reduce the 9-brane doublet (\ref{eq:11-d ikL10}). We first consider the charge $i_{\hat{\alpha}\hat{\beta}}\hat{L}^{(10)}_1$ which produces
\begin{eqnarray}\label{eq:IIA t9 charge}
M^{(t9)}=e^{-2\phi}\Pi  + i_K\overline{A}^{(10)}
\end{eqnarray}
where we have removed an overall contraction of $\beta$ since this Killing vector is not intrinsically present in the charge. We will refer to the brane that this charge corresponds to as the IIA t9-brane. This brane was discussed in \cite{Bergshoeff:2005ac} where it was shown to have a tension which scales with $e^{-2\phi}$ which can also be read off from the above charge. 

Next we dimensionally reduce the charge $i_{\hat{\alpha}\hat{\beta}}\hat{L}^{(10)}_2$ which produces
\begin{eqnarray}\label{eq:IIA u9 charge}
M^{(u9)}=e^{-2\phi}i_{\beta}C^{(1)}i_{\beta}\Pi - e^{-3\phi}(R^2_{\beta}\Psi +i_{\beta}\Psi \wedge \beta) + i_{\beta K}\overline{B}^{(10)} .
\end{eqnarray}
We will refer to the brane that this charge corresponds to as the IIA u9-brane. We see from the charge structure that it contains an isometry direction that obviously lies parallel to the brane worldvolume, and has a tension that scales as $e^{-2\phi}R_{\beta}\sqrt{(i_{\beta}C^{(1)})^2+e^{-2\phi}R_{\beta}^2}$.

\section{Charges in IIB}\label{sec:IIB charges}

\subsection{T-dualising from IIA}\label{sec:non-cov IIB charges}

We now T-dualise the charges given in the previous section to IIB. Since we were considering the non-covariant IIA SUGRA we must take into account the role of the mass parameters in the T-duality rules. In actual fact, it was shown in \cite{Meessen:1998qm} that the usual massless rules for the potentials can still be used, with the IIB potentials still taken to be independent of the isometry direction. The information of the massive terms is then expressed on the IIB side by making pseudo field strength redefinitions. We use this scheme here to T-dualise the charges and discuss in Section \ref{sec:cov T-duality} in what sense the interpretation of the resulting IIB charges depends on the type of IIA theory they have been T-dualised from. The massless T-duality rules for the bilinears and potentials that are not explicitly given in this paper can be found in \cite{Callister:2007jy}, along with the lower rank IIB charges and definitions of the IIB fields and bilinears we use.

\subsubsection{Triplet of 7-branes}

We start by considering the IIB 7-branes. For recent discussions on the spacetime solutions of these branes see \cite{Bergshoeff:2007aa,Bergshoeff:2006jj,Bergshoeff:2002mb}. The D7-brane charge can be obtained by performing a direct T-duality transformation on the D6-brane charge or alternatively a double T-duality transformation on the D8-brane charge. This case was considered in \cite{Callister:2007jy} and so we do not repeat it here. 

The next charge is that of the NS7-brane. This can be obtained from IIA by either performing a direct T-duality transformation on the KK6-monopole charge (\ref{eq:IIA KK6 charge}) or a double T-duality transformation on the KK8-monopole charge (\ref{eq:IIA KK8 charge}). The result is found to be  
\begin{eqnarray}\label{eq:IIB NS7 charge}
N^{(NS7)}&=&(e^{-3\varphi}+l^2e^{-\varphi})\Pi^{12}+i_{K^+}{\cal N}^{(8)} +(e^{-\varphi}K^{12}-lK^-)\wedge {\cal B}^{(6)}\nonumber \\ && + \biggl( e^{-\varphi}l \Sigma^{12}+e^{-2\varphi}\Sigma^--i_{K+}{\cal B}^{(6)}+\frac{1}{2}e^{-\varphi}\Phi^{12}\wedge {\cal C}^{(2)}\\ \nonumber && + \frac{1}{3!}K^-\wedge ({\cal C}^{(2)})^2\biggr) \wedge {\cal C}^{(2)} .
\end{eqnarray}
The tension of the NS7-brane was shown to scale as $e^{-3\varphi}+l^2e^{-\varphi}$ in \cite{Bergshoeff:2006ic}, which can also be trivially read off from the above charge. Furthermore this brane minimally couples to ${\cal N}^{(8)}$.

We refer to the third brane as the r7-brane. Its charge can be found by performing a direct T-dualisation of the r6-brane charge (\ref{eq:IIA r6 charge}) or double T-dualising the r8-brane charge (\ref{eq:IIA r8 charge}). It is found to be  
\begin{eqnarray}\label{eq:IIB r7 charge}
N^{(r7)}&=&e^{-\varphi}l\Pi^{12}-i_{K^+}\varphi ^{(8)} + \frac{1}{2}\biggl( e^{-\varphi}\Sigma^{12}\wedge {\cal C}^{(2)} - K^-\wedge {\cal B}^{(6)} \nonumber \\ && + N^{(NS5)}\wedge {\cal B} - N^{(1)}\wedge {\cal C}^{(6)} \biggr) 
\end{eqnarray} 
where $N^{(p)}$ is the charge of a D$p$-brane, and $N^{(NS5)}$ is the charge of the NS5-brane. The tension for this brane was also given in \cite{Bergshoeff:2006ic} and was shown to scale as $le^{-\varphi}$, which is in agreement with the above charge. Furthermore, it minimally couples to $\varphi^{(8)}$.

\subsubsection{Quadruplet of 9-branes}\label{sec:IIB quad 9-charges}

We now consider the charges for the quadruplet of 9-branes. These are obtained by T-dualising the IIA charges which are produced from dimensional reduction of (\ref{eq:11-d M9 quad}). In order to do this we must first determine the T-duality rules for the potentials which minimally couple to these branes. We present the details of this task in Appendix \ref{sec:mapping 10-forms} and simply use the results in this section.

The first 9-brane in this quadruplet is the D9-brane, whose charge can be calculated by performing a direct T-duality transformation of the D8-brane charge. It has already been presented in \cite{Callister:2007jy} and so we do not restate it here. 

We will refer to the next brane as the s9-brane. The (double dimensional reduction of the) charge of this brane can be calculated by T-dualising (\ref{eq:IIA s9 charge}) along $\beta$. Doing this yields
\begin{eqnarray}
N^{(s9)}&=&(\frac{1}{3}e^{-3\varphi} +l^2e^{-\varphi})\Omega^{12} +\frac{2}{3}le^{-2\varphi}\Omega^- +\frac{1}{3}(e^{-3\varphi}+l^2e^{-\varphi})\Pi^{12}\wedge {\cal B} \nonumber \\
&& + \frac{2}{3}le^{-\varphi}\Pi^{12}\wedge {\cal C}^{(2)} +\frac{1}{6}e^{-\varphi}\Sigma^{12}\wedge ({\cal C}^{(2)})^2 +\frac{1}{3}(e^{-2\varphi}\Sigma^- +le^{-\varphi}\Sigma^{12})\wedge {\cal C}^{(2)}\wedge {\cal B} \nonumber \\
&& +\frac{1}{6}e^{-\varphi}\Phi^{12}\wedge ({\cal C}^{(2)})^2\wedge {\cal B} +(e^{-\varphi}K^{12}-lK^-)\wedge (\frac{2}{3}\varphi^{(8)} +\frac{1}{3}{\cal B}^{(6)}\wedge {\cal B}) \nonumber \\
&& +K^-\wedge (\frac{1}{3}{\cal N}^{(8)}-\frac{1}{3}{\cal B}^{(6)}\wedge {\cal C}^{(2)} +\frac{1}{18}({\cal C}^{(2)})^3\wedge {\cal B})  +i_{K^+}{\cal A}^{(10)}\nonumber \\
&& +\frac{1}{3}i_{K^+}{\cal N}^{(8)}\wedge {\cal B} +\frac{2}{3}\varphi^{(8)}\wedge i_{K^+}{\cal C}^{(2)} -\frac{1}{3}i_{K^+}{\cal B}^{(6)}\wedge {\cal C}^{(2)}\wedge {\cal B} .
\end{eqnarray}
From this charge we see that the tension of the s9-brane should scale as $e^{-\varphi}\sqrt{\frac{4}{9}l^2e^{-2\varphi}+(\frac{1}{3}e^{-2\varphi} + l^2)^2}$, which is in agreement with \cite{Bergshoeff:2006ic}, and minimally couple to the potential ${\cal A}^{(10)}$. 

The next brane we refer to as the r9-brane. The (double dimensional reduction of the) charge of this brane can be calculated by T-dualising (\ref{eq:IIA r9 charge}) along $\beta$. Doing this yields 
\begin{eqnarray}
N^{(r9)}&=&le^{-\varphi}\Omega^{12}+\frac{1}{3}e^{-2\varphi}\Omega^- +\frac{1}{3}e^{-\varphi}\Pi^{12}\wedge {\cal C}^{(2)} + \frac{2}{3}le^{-\varphi}\Pi^{12}\wedge {\cal B} \nonumber \\
&& +\frac{1}{6}(e^{-2\varphi}\Sigma^- +le^{-\varphi}\Sigma^{12})\wedge ({\cal B})^2 +\frac{1}{3}e^{-\varphi}\Sigma^{12}\wedge {\cal C}^{(2)}\wedge {\cal B}\nonumber \\
&& +\frac{1}{6}e^{-\varphi}\Phi^{12}\wedge {\cal C}^{(2)}\wedge ({\cal B})^2 \nonumber \\
&& +(e^{-\varphi}K^{12}-lK^-)\wedge (\frac{1}{3}{\cal C}^{(8)} -\frac{1}{3}{\cal C}^{(6)}\wedge {\cal B} +\frac{1}{6}{\cal C}^{(4)}\wedge ({\cal B})^2) \nonumber \\
&& +K^-\wedge (-\frac{2}{3}\varphi^{(8)} -\frac{1}{3}{\cal B}^{(6)}\wedge {\cal B} +\frac{1}{12}({\cal C}^{(2)})^2\wedge ({\cal B})^2) +i_{K^+}{\cal B}^{(10)} \nonumber \\
&& -\frac{2}{3}i_{K^+}\varphi^{(8)}\wedge {\cal B} +\frac{1}{3}{\cal C}^{(8)}\wedge i_{K^+}{\cal C}^{(2)} -\frac{1}{6}i_{K^+}{\cal B}^{(6)}\wedge ({\cal B})^2 \nonumber \\
&&-\frac{1}{3}{\cal C}^{(6)}\wedge i_{K^+}{\cal C}^{(2)}\wedge {\cal B} +\frac{1}{6}{\cal C}^{(4)}\wedge i_{K^+}{\cal C}^{(2)}\wedge ({\cal B})^2 .
\end{eqnarray}
From this charge we see that the tension of the s9-brane should scale as $e^{-\varphi}\sqrt{\frac{1}{9}e^{-2\varphi}+l^2}$, which is in agreement with \cite{Bergshoeff:2006ic}, and minimally couple to the potential ${\cal B}^{(10)}$.

We refer to the final brane in the quadruplet as the q9-brane. The (double dimensional reduction of the) charge of this brane can be calculated by T-dualising (\ref{eq:IIA q9 charge}) along $\beta$. Doing this yields 
\begin{eqnarray}
N^{(q9)}&=&(e^{-4\varphi}+l^2e^{-2\varphi})\Omega^- +(le^{-3\varphi}+l^3e^{-\varphi})\Omega^{12} +(e^{-3\varphi}+l^2e^{-\varphi})\Pi^{12}\wedge {\cal C}^{(2)}\nonumber \\
&& +\frac{1}{2}(e^{-2\varphi}\Sigma^- +le^{-\varphi}\Sigma^{12})\wedge ({\cal C}^{(2)})^2 +\frac{1}{6}e^{-\varphi}\Phi^{12}\wedge ({\cal C}^{(2)})^3 \nonumber \\
&& +(e^{-\varphi}K^{12}-lK^-)\wedge (-{\cal N}^{(8)} +{\cal B}^{(6)}\wedge {\cal C}^{(2)}) +\frac{1}{24}K^-\wedge ({\cal C}^{(2)})^4\nonumber \\
&& -i_{K^+}{\cal D}^{(10)} +i_{K^+}{\cal N}^{(8)}\wedge {\cal C}^{(2)} -\frac{1}{2}i_{K^+}{\cal B}^{(6)}\wedge ({\cal C}^{(2)})^2 .
\end{eqnarray}
From this charge we see that the tension of the q9-brane should scale as $e^{-\varphi}(e^{-2\varphi}+l^2)^{3/2}$, which is in agreement with \cite{Bergshoeff:2006ic}, and minimally couple to the potential ${\cal D}^{(10)}$.

As already mentioned, in \cite{Bergshoeff:2006ic} it was shown that there were two constraints which  restrict which combinations of these branes can be supersymmetrically coupled to the IIB SUGRA action. Such constraints are not deducible from the structural form of the charges. Furthermore, when dealing with spacetime filling branes the total charge must vanish for consistency which involves some $N=1$ truncation. It is interesting to note that no such truncation is required in order for the 9-brane charges to be closed. This is presumably related to the fact that the gauge algebra and therefore the charge structures, are independent of the dimensionality of the background spacetime, as already noticed due to the existence of $\hat{L}^{(11)}_{abc}$ in 11-dimensions in Section \ref{sec:11-d M9 charges}.

\subsubsection{Doublet of 9-branes}

We now consider the charges for the doublet of 9-branes. These are obtained by T-dualising the IIA charges which are produced from the dimensional reduction of (\ref{eq:11-d ikL10}). We first perform a double dimensional T-duality on (\ref{eq:IIA t9 charge}) using the following rule for the IIA potential
\begin{eqnarray}
i_{\beta}\overline{A}^{(10)}_{\mu_1\ldots \mu_9}\rightarrow i_{\beta}\overline{\cal A}^{(10)}_{\mu_1\ldots \mu_9}
\end{eqnarray}
where we have split the co-ordinates in the same fashion as done in Section \ref{sec:10-d T-duality field eqns}. The result is then (the double dimensional reduction of)
\begin{eqnarray}\label{eq:IIB t9 charge}
N^{(t9)}=e^{-2\varphi}\Omega^- - i_{K^+}\overline{\cal A}^{(10)} .
\end{eqnarray}
We see that the brane that this charge corresponds to, which we will refer to as the IIB t9-brane, has a tension that scales as $e^{-2\varphi}$. This is in agreement with \cite{Bergshoeff:2006ic}.

Next we T-dualise (\ref{eq:IIA u9 charge}) along $\beta$ using the following rule for the IIA potential
\begin{eqnarray}
i_{\beta}\overline{B}^{(10)}_{\mu_1\ldots \mu_9}\rightarrow i_{\beta}\overline{\cal B}^{(10)}_{\mu_1\ldots \mu_9} .
\end{eqnarray}
The result is 
\begin{eqnarray}\label{eq:IIB u9 charge}
N^{(u9)}=e^{-3\varphi}\Omega^{12} - le^{-2\varphi}\Omega^{-} + i_{K^+}\overline{\cal B}^{(10)} .
\end{eqnarray}
We see that the brane that this charge corresponds to, which we will refer to as the IIB u9-brane, has a tension that scales as $e^{-2\varphi}\sqrt{e^{-2\varphi}+l^2}$. This is in agreement with \cite{Bergshoeff:2006ic}.

\subsection{$SL(2,\mathbb{R})$ covariant IIB charges}\label{sec:cov IIB charges}

We now consider the $SL(2,\mathbb{R})$ covariant version of IIB and show that the charges transform in the same multiplets as the states they correspond to. This involves re-expressing the fields, bilinears and charges given in \cite{Callister:2007jy} and in this paper in an $SL(2,\mathbb{R})$ covariant fashion. The 1, 3 and 5-form field equations have been given previously in \cite{Meessen:1998qm} and the 7, 9 and 11-form ones have been given in an $SU(1,1)$ covariant form in \cite{Dall'Agata:1998va,Bergshoeff:2005ac}.

So far in this paper we have been working with the string frame metric, however the $SL(2,\mathbb{R})$ covariant theory is more natural expressed in the Einstein frame. For the generalised charges the only effect this has is a scaling of the bilinears due to their construction from Dirac matrices. Since this offers little simplification we will remain working in the string frame to avoid excessive changes of notation, and simply write the bilinear multiplets in terms of the string frame bilinears.

The bilinear multiplets are determined by calculating their transformations under the discrete S-duality transformation. This is done by mapping a given bilinear to 11-dimensions, performing the transformation $\hat{k}_1\rightarrow -\hat{k}_2$, $\hat{k}_2\rightarrow \hat{k}_1$, then mapping back to IIB. We find that the bilinears fall into groups that transform according to essentially the same rule. Denoting a general p-form bilinear by $Y_{(p)}$ we find that for $p=1,5,9$ we have
\begin{eqnarray}
Y^{12}_{(p)}&\rightarrow & |\lambda|^{\frac{p-1}{4}}(-e^{-\varphi}Y^-_{(p)}-lY^{12}_{(p)})\nonumber \\ \label{eq:IIB cov bilins 1}
Y^{-}_{(p)}&\rightarrow & |\lambda|^{\frac{p-1}{4}}(e^{-\varphi}Y^{12}_{(p)}-lY^{-}_{(p)})
\end{eqnarray}     
whereas for $p=3,7$ we get
\begin{eqnarray}\label{eq:IIB cov bilins 2}
Y^{12}_{(p)}\rightarrow |\lambda|^{\frac{p+1}{4}}Y^{12}_{(p)}
\end{eqnarray}
where $\lambda=l+ie^{-\varphi}$ is the axion-dilaton. The only other bilinear transformation we need is for $K^+$ with its spacetime index up, which appears in the charges contracted with various gauge potentials. This can be shown to transform as a scalar. Note that also this object is invariant when going between the Einstein and string frame since the scaling of both $K^+_{\mu}$ and the inverse metric cancel.

The scalar fields $\varphi$ and $l$ parametrise the coset $SL(2,\mathbb{R})/SO(2)$ and transform as the following matrix \cite{Meessen:1998qm}
\begin{eqnarray}
{\cal M}_{ab}=e^{\varphi}\left(
\begin{array}{cc}
|\lambda|^2 & l \\
l & 1
\end{array}
\right) .
\end{eqnarray}

From this it can be determined that under the discrete S-duality transformations the scalars obey
\begin{eqnarray}
e^{-\varphi}\rightarrow \frac{e^{-\varphi}}{|\lambda|^2}\qquad l\rightarrow \frac{-l}{|\lambda|^2} .
\end{eqnarray}

\subsubsection{Doublet of 1-branes}   

This multiplet consists of the D1-brane and the F-string, for which the non-covariant generalised charges were given in \cite{Callister:2007jy}. The leading terms were found to be $e^{-\varphi}K^{12}-lK^-$ and ${K^-}$ respectively. From (\ref{eq:IIB cov bilins 1}) it can be shown that these transform as a doublet, specifically we define  
\begin{eqnarray}\label{eq:IIB 1-form bilin doublet}
K_a=\left(
\begin{array}{c}
e^{-\varphi}K^{12}-lK^-\\
-K^-
\end{array}
\right) .
\end{eqnarray}
The two 3-form field strengths and 2-form potentials form the doublets 
\begin{eqnarray}
{\cal H}_a=\left(
\begin{array}{c}
{\cal F}^{(3)}+l{\cal H}\\
{\cal H}
\end{array}
\right)
\qquad 
{\cal B}_a=\left(
\begin{array}{c}
{\cal C}^{(2)}\\
{\cal B}
\end{array}
\right)
\end{eqnarray}
and the field equations become simply
\begin{eqnarray}
{\cal H}_a=d{\cal B}_a .
\end{eqnarray}
We then find that the 1-brane charges can be written as the following doublet
\begin{eqnarray}
N^{(1)}_{a}=K_a+i_{K^+}{\cal B}_a=\left(
\begin{array}{c}
N^{(D1)}\\
-N^{(F1)}
\end{array}
\right)
\end{eqnarray}
where $N^{(F1)}$ is the charge of the F-string. It is trivial to check that this expression is generally closed, however we delay giving the required covariant form of the bilinear differential relations until Section \ref{sec:cov T-duality} when we discuss how the presence of the mass parameters on the IIA side are dealt with in IIB.

\subsubsection{3-brane singlet}
The D3-brane transforms as a singlet under $SL(2,\mathbb{R})$, and the non-covariant charge was given in \cite{Callister:2007jy}. The leading term here is $e^{-\varphi}\Phi^{12}$ which from (\ref{eq:IIB cov bilins 2}) can be seen to transform as a scalar. We therefore define $\Phi=e^{-\varphi}\Phi^{12}$ which is actually equivalent to the Einstein frame definition of the bilinear.

The 5-form field equation is also an $SL(2,\mathbb{R})$ scalar which can be seen by making the following redefinition of the Ramond-Ramond 4-form:
\begin{eqnarray}
{\cal C}^{(4)}\rightarrow \tilde{\cal C}^{(4)}+\frac{1}{2}{\cal C}^{(2)}\wedge {\cal B} .
\end{eqnarray}
We then get the following field equation
\begin{eqnarray}
{\cal F}^{(5)}=d\tilde{\cal C}^{(4)}-\frac{1}{2}\epsilon^{ab}{\cal B}_a\wedge d{\cal B}_b
\end{eqnarray}
where $\epsilon^{12}=+1$. Given these definitions we can write the D3-brane charge as
\begin{eqnarray}
N^{(3)}=\Phi-i_{K^+}\tilde{\cal C}^{(4)}+\epsilon^{ab}K_a\wedge {\cal B}_b+\frac{1}{2}\epsilon^{ab}i_{K^+}{\cal B}_a\wedge {\cal B}_b
\end{eqnarray}
which is an $SL(2,\mathbb{R})$ scalar as expected.

\subsubsection{Doublet of 5-branes}
Next we consider the D5-brane and NS5-brane which form an $SL(2,\mathbb{R})$ doublet. Again the non-covariant forms of the charges were given in \cite{Callister:2007jy}. The respective leading terms here are $e^{-\varphi}\Sigma^{12}$ and $e^{-2\varphi}\Sigma^-+e^{-\varphi}l\Sigma^{12}$. We then define the doublet
\begin{eqnarray}
\Sigma_a=\left(
\begin{array}{c}
e^{-2\varphi}\Sigma^-+e^{-\varphi}l\Sigma^{12}\\
e^{-\varphi}\Sigma^{12}
\end{array}
\right)
\end{eqnarray}
which can be checked to transform correctly using (\ref{eq:IIB cov bilins 1}). Note that this doublet has a qualitatively different structure to (\ref{eq:IIB 1-form bilin doublet}). The two types of doublet can however be mapped to one another using ${\cal M}_a^{\phantom{a}b}$.

The two 7-form field strengths and 6-form potentials can also be written as a doublet. To show this we need to make the following field redefinitions
\begin{eqnarray}
{\cal C}^{(6)}&\rightarrow& -\tilde{\cal C}^{(6)}+\frac{1}{6}{\cal C}^{(2)}\wedge ({\cal B})^2\\
{\cal B}^{(6)}&\rightarrow& \tilde{\cal B}^{(6)}+\tilde{\cal C}^{(4)}\wedge {\cal C}^{(2)}+\frac{1}{6}{\cal B}\wedge ({\cal C}^{(2)})^2 .
\end{eqnarray}
The doublets are then given by
\begin{eqnarray}
{\cal H}^{(7)}_a=\left(
\begin{array}{c}
{\cal H}^{(7)}-l{\cal F}^{(7)}\\
-{\cal F}^{(7)}
\end{array}
\right)
\qquad 
{\cal B}^{(6)}_a=\left(
\begin{array}{c}
\tilde{\cal B}^{(6)}\\
\tilde{\cal C}^{(6)}
\end{array}
\right)
\end{eqnarray}
from which we can write the field equations as
\begin{eqnarray}
{\cal H}^{(7)}_a=d{\cal B}^{(6)}_a+\tilde{\cal C}^{(4)}\wedge d{\cal B}_a+\frac{1}{6}\epsilon^{bc}{\cal B}_a\wedge {\cal B}_b\wedge d{\cal B}_c .
\end{eqnarray}

Using these definitions the D5-brane charge and the NS5-brane charge can be written as the following doublet
\begin{eqnarray}
N^{(5)}_{a}&=&\Sigma_a+\Phi\wedge {\cal B}_a+K_a\wedge \tilde{\cal C}^{(4)}+\frac{1}{2}\epsilon^{bc}K_b\wedge {\cal B}_c\wedge {\cal B}_a-i_{K^+}{\cal B}^{(6)}_a\nonumber \\ 
&& -i_{K^+}\tilde{\cal C}^{(4)}\wedge {\cal B}_a +\frac{1}{6}\epsilon^{bc}i_{K^+}{\cal B}_b\wedge {\cal B}_c\wedge {\cal B}_a\nonumber \\ 
& =& \left(
\begin{array}{c}
N^{(NS5)}\\
N^{(D5)}
\end{array}
\right) .
\end{eqnarray}

\subsubsection{Triplet of 7-branes}

We next consider the triplet of 7-branes which is comprised of the D7-brane, NS7-brane and r7-brane. As discussed in the introduction it is often stated that these branes are an example of a discrepancy between the charges appearing in the SUSY algebra and the spectrum of BPS states that couple to the theory. While this is true for the flatspace SUSY algebra where the 7-form charge transforms as a singlet, we now show that the generalised charges transform as a triplet.

The D7-brane generalised charge was given in \cite{Callister:2007jy} and NS7-brane and r7-brane charges are given by (\ref{eq:IIB NS7 charge}) and (\ref{eq:IIB r7 charge}) respectively. The leading terms in all cases involve the 7-form bilinear $\Pi^{12}$ and some scalar factor consisting of the axion and dilaton. From (\ref{eq:IIB cov bilins 2}) we see that the combination $\Pi=e^{-2\varphi}\Pi^{12}$ is an $SL(2,\mathbb{R})$ scalar. To form a term that transforms as a triplet we must use ${\cal M}_{ab}$. It is the presence of this term, which is neglected for flat spacetimes, which causes the discrepancy between the flatspace SUSY algebra and the brane multiplet.
 
The three 8-form potentials ${\cal N}^{(8)}$, $\varphi^{(8)}$ and the Ramond-Ramond potential ${\cal C}^{(8)}$ form a triplet. Their field strength equations are given by (\ref{eq:IIB N8}), (\ref{eq:IIB phi8}) along with the standard Ramond-Ramond field strength. In order to express these as a triplet we need to make the following redefinitions
\begin{eqnarray}
{\varphi}^{(8)}&\rightarrow&\tilde{\varphi}^{(8)}+\frac{1}{2}{\cal C}^{(2)}\wedge \tilde{\cal C}^{(6)}-\frac{1}{48}({\cal C}^{(2)})^2\wedge ({\cal B})^2\\
{\cal C}^{(8)}&\rightarrow&\tilde{\cal C}^{(8)}+\frac{1}{24}{\cal C}^{(2)}\wedge ({\cal B})^3 \\
{\cal N}^{(8)}&\rightarrow&-\tilde{\cal N}^{(8)}+\frac{1}{2}\tilde{\cal C}^{(4)}\wedge ({\cal C}^{(2)})^2+\frac{1}{8}({\cal C}^{(2)})^3\wedge {\cal B} .
\end{eqnarray}
We then find the following triplet structures
\begin{eqnarray}\label{eq:IIB 9-form field triplet}
{\cal F}^{(9)}_{ab}&=&\left(
\begin{array}{cc}
2l{\cal H}^{(9)}+(l^2-e^{-2\varphi}){\cal F}^{(9)} & {\cal H}^{(9)}+l{\cal F}^{(9)} \\
{\cal H}^{(9)}+l{\cal F}^{(9)} & {\cal F}^{(9)}
\end{array}
\right)
\\ 
{\cal C}^{(8)}_{ab}&=&\left(
\begin{array}{cc}
\tilde{\cal N}^{(8)} & \tilde{\varphi}^{(8)} \\
\tilde{\varphi}^{(8)} & \tilde{\cal C}^{(8)}
\end{array}
\right) .
\end{eqnarray}
The 9-form field equations can then be written as
\begin{eqnarray}
{\cal F}^{(9)}_{ab}=d{\cal C}^{(8)}_{ab}+ {\cal B}^{(6)}_{(a}\wedge d{\cal B}_{b)}-\frac{1}{4!}\epsilon^{cd}{\cal B}_{a}\wedge {\cal B}_{b}\wedge {\cal B}_{c}\wedge d{\cal B}_{d} 
\end{eqnarray}
with the constraint
\begin{eqnarray}
{\cal M}^{ab}{\cal F}^{(9)}_{ab}=0
\end{eqnarray}
which follows from (\ref{eq:IIB 9-form field triplet}) and demonstrates that there are only two independent 9-form field strengths. We use the following convention to raise $SL(2,\mathbb{R})$ indices
\begin{eqnarray}\label{eq:IIB 9-form field constraint}
\epsilon^{ac}{\cal M}_{cb}={\cal M}^a_{\phantom{a}b} .
\end{eqnarray} 

The 7-form charges (\ref{eq:IIB NS7 charge}) and (\ref{eq:IIB r7 charge}) together with the D7-brane charge can then be seen to form the following triplet
\begin{eqnarray}
N^{(7)}_{ab}&=&{\cal M}_{ab}{\Pi}  
+\Sigma_{(a}\wedge {\cal B}_{b)} +\frac{1}{2}{\Phi}\wedge {\cal B}_a\wedge {\cal B}_b +\frac{1}{6}\epsilon^{cd}K_{c}\wedge {\cal B}_{d}\wedge {\cal B}_{a}\wedge {\cal B}_{b}\nonumber \\ 
&&+ K_{(a}\wedge {\cal B}^{(6)}_{b)} + K_{(a}\wedge \tilde{C}^{(4)}\wedge {\cal B}_{b)}-i_{K^+}{\cal B}^{(6)}_{(a}\wedge {\cal B}_{b)} -i_{K^+}{\cal C}^{(8)}_{ab} \nonumber \\ 
&& -\frac{1}{2}i_{K^+}\tilde{\cal C}^{(4)}\wedge {\cal B}_{a}\wedge {\cal B}_{b} +\frac{1}{4!}\epsilon^{cd}i_{K^+}{\cal B}_{c}\wedge {\cal B}_{d}\wedge {\cal B}_{a}\wedge {\cal B}_{b}\nonumber \\
&=&\left(
\begin{array}{cc}
N^{(NS7)}&N^{(r7)}\\
N^{(r7)}&N^{(D7)}
\end{array}
\right) .
\end{eqnarray}

\subsubsection{Quadruplet of 9-branes}\label{sec:IIB cov 9-charge quad}

We now discuss the quadruplet of 9-branes which was discussed in \cite{Bergshoeff:2006ic}. It consists of the D9-brane along with other branes that we refer to as the r9, s9 and q9-branes. The non-covariant form of the D9-brane charge was given in \cite{Callister:2007jy}, and the other three in Section \ref{sec:IIB quad 9-charges}. In each case the leading bilinear terms consist of combinations of $\Omega^{12}$ and $\Omega^{-}$ with factors made up from the dilaton and axion. Using (\ref{eq:IIB cov bilins 2}) we define the following bilinear doublet
\begin{eqnarray}\label{eq:IIB omega doublet}
\Omega_a=\left(
\begin{array}{c}
e^{-3\varphi}\Omega^-+e^{-2\varphi}l\Omega^{12}\\
e^{-2\varphi}\Omega^{12}
\end{array}
\right) .
\end{eqnarray}
We then redefine the 10-form potentials according to
\begin{eqnarray}
{\cal A}^{(10)}&\rightarrow &\tilde{\cal A}^{(10)} -\frac{2}{3}\tilde{\varphi}^{(8)}\wedge {\cal C}^{(2)} -\frac{1}{6}\tilde{\cal C}^{(6)}\wedge ({\cal C}^{(2)})^2 +\frac{1}{360}({\cal C}^{(2)})^3\wedge ({\cal B})^2\nonumber \\ && \\
{\cal B}^{(10)} &\rightarrow &\tilde{\cal B}^{(10)} -\frac{1}{3}\tilde{\cal C}^{(8)}\wedge {\cal C}^{(2)} -\frac{1}{360}({\cal C}^{(2)})^2 \wedge ({\cal B})^3\\
{\cal C}^{(10)}&\rightarrow &\tilde{\cal C}^{(10)} +\frac{1}{5!}{\cal C}\wedge ({\cal B})^4\\
{\cal D}^{(10)}&\rightarrow &-\tilde{\cal D}^{(10)} +\frac{1}{6}\tilde{\cal C}^{(4)}\wedge ({\cal C}^{(2)})^3 +\frac{1}{20}{\cal B}\wedge ({\cal C}^{(2)})^4 .
\end{eqnarray}
Then using (\ref{eq:IIB 11-form 1})-(\ref{eq:IIB 11-form 2}) we can write the quadruplet 11-form field strength equations as
\begin{eqnarray}
{\cal F}^{(11)}_{abc}=d{\cal C}^{(10)}_{abc} - {\cal C}^{(8)}_{(ab}\wedge d{\cal B}_{c)} -\frac{1}{5!}\epsilon^{de}{\cal B}_a \wedge {\cal B}_b \wedge {\cal B}_c \wedge {\cal B}_d\wedge d{\cal B}_e  
\end{eqnarray}
where we have defined the components of the quadruplet of field strengths ${\cal F}^{(11)}_{abc}$ as
\begin{eqnarray}
{\cal F}^{(11)}_{111}&=&l^3{\cal F}^{(11)} +2l^2{\cal H}^{(11)} +2l{\cal X}^{(11)} -{\cal G}^{(11)}\nonumber \\
{\cal F}^{(11)}_{112}&=&l^2{\cal F}^{(11)} +\frac{4}{3}l{\cal H}^{(11)} +\frac{2}{3}{\cal X}^{(11)}\nonumber \\
{\cal F}^{(11)}_{122}&=& l{\cal F}^{(11)} +\frac{2}{3}{\cal H}^{(11)}\nonumber \\
{\cal F}^{(11)}_{222}&=&{\cal F}^{(11)}
\end{eqnarray}
and potentials ${\cal C}^{(10)}_{abc}$ as
\begin{eqnarray}
{\cal C}^{(10)}_{111}=\tilde{\cal D}^{(10)} &\qquad & {\cal C}^{(10)}_{112}=\tilde{\cal A}^{(10)} \nonumber \\
{\cal C}^{(10)}_{122}=\tilde{\cal B}^{(10)} &\qquad & {\cal C}^{(10)}_{222}=\tilde{\cal C}^{(10)} .
\end{eqnarray}

The quadruplet of 9-form charges is then given by
\begin{eqnarray}
N_{abc}^{(9)}&=&{\cal M}_{(ab}\Omega_{c)} +{\cal M}_{(ab}\Pi\wedge {\cal B}_{c)} +\frac{1}{2}\Sigma_{(a}\wedge {\cal B}_b\wedge {\cal B}_{c)}\nonumber \\
&& +\frac{1}{3!}\Phi\wedge {\cal B}_{(a} \wedge {\cal B}_b \wedge {\cal B}_{c)} +\frac{1}{4!}\epsilon^{de}K_d \wedge {\cal B}_e \wedge {\cal B}_{a} \wedge {\cal B}_b \wedge {\cal B}_{c} \nonumber \\
&& +K_{(a}\wedge {\cal C}^{(8)}_{bc)} +K_{(a}\wedge {\cal B}^{(6)}_b \wedge {\cal B}_{c)} +\frac{1}{2}K_{(a}\wedge \tilde{\cal C}^{(4)}\wedge {\cal B}_{b}\wedge {\cal B}_{c)}\nonumber \\
&& +i_{K^+}{\cal C}^{(10)}_{abc} -i_{K^+}{\cal C}^{(8)}_{(ab}\wedge {\cal B}_{c)} -\frac{1}{2}i_{K^+}{\cal B}^{(6)}_{(a}\wedge {\cal B}_b\wedge {\cal B}_{c)} \nonumber \\
&&-\frac{1}{3!}i_{K^+}\tilde{\cal C}^{(4)}\wedge {\cal B}_{(a} \wedge {\cal B}_b \wedge {\cal B}_{c)} \nonumber \\
&& +\frac{1}{5!}\epsilon^{ab}i_{K^+}{\cal B}_d\wedge{\cal B}_e \wedge {\cal B}_{(a} \wedge {\cal B}_b \wedge {\cal B}_{c)} 
\end{eqnarray} 
where we have
\begin{eqnarray}
N_{111}^{(9)}=N^{(q9)} &\qquad & N_{112}^{(9)}=N^{(s9)} \nonumber \\
N_{122}^{(9)}=N^{(r9)} &\qquad & N_{222}^{(9)}=N^{(9)} .
\end{eqnarray}

\subsubsection{Doublet of 9-branes}

It is a simple task to show that the doublet of 9-branes (\ref{eq:IIB t9 charge}) and (\ref{eq:IIB u9 charge}) can be written as a doublet. The 10-form potentials directly transform as a doublet. We therefore define
\begin{eqnarray}
{\cal C}^{(10)}_a =\left(
\begin{array}{c}
\overline{\cal B}^{(10)}\\
\overline{\cal A}^{(10)}
\end{array}
\right)
\end{eqnarray}
which gives rise to the field equation
\begin{eqnarray}
{\cal F}^{(11)}_a=d{\cal C}^{(10)}
\end{eqnarray}
which agrees with the field equations implicitly given in \cite{Bergshoeff:2005ac}. Then using (\ref{eq:IIB omega doublet}) we can write the charges as
\begin{eqnarray}
N^{(9)}_a={\cal M}_a^{\phantom{a}b}\Omega_b - i_{K^+}{\cal C}^{(10)}_a=\left(
\begin{array}{c}
-N^{(u9)}\\
N^{(t9)}
\end{array}
\right) .
\end{eqnarray}

\subsection{$SL(2,\mathbb{R})$ T-duality and `massive' IIB}\label{sec:cov T-duality}

We now discuss how the IIB theory is related to the non-covariant IIA theory, and hence also to the $SL(2,\mathbb{R})$ covariant 11-dimensional theory, through T-duality. Our first motivation for doing this is to justify the use of the massless T-duality rules to map the IIA charges to IIB as done in the previous sections. A second reason is to show how 11-dimensional field equations are related to those given in IIB which provides a check on the massive terms.

This topic was discussed in \cite{Meessen:1998qm} where, as already mentioned, it was shown that the theories are related via the massive $SL(2,\mathbb{R})$ covariant 9-dimensional SUGRAs which are obtained on the IIB side by performing a Scherk-Schwarz dimensional reduction using the full global $SL(2,\mathbb{R})$ symmetry in IIB. In this scheme the IIB potentials are given specific dependencies on the compact isometry direction depending on their $SL(2,\mathbb{R})$ representation. However, when performing the T-duality the direct interpretation of these dependencies are not clear on the IIA side. Such an issue arises when mapping the generalised charges since they explicitly depend on the potentials.

In \cite{Meessen:1998qm} an alternative method to the Scherk-Schwarz reduction for producing the 9-dimensional masses was discussed in which all the fields remain independent of the isometry direction. The general idea here is to gauge the global symmetry in IIB using a gauge field that takes values of the mass parameters in one direction only. Specifically, for an arbitrary $p$-form $Y_{a_1\ldots a_n}$ the following substitution is made
\begin{eqnarray}\label{eq:IIB massive cov deriv}
dY_{a_1\ldots a_n}\rightarrow DY_{a_1\ldots a_n}=dY_{a_1\ldots a_n}+nm_{(a_1}^{\phantom{(a_1}b}dy \wedge Y_{a_2\ldots a_n)b}^{}
\end{eqnarray}
where
\begin{eqnarray}
m_a^{\phantom{a}b}=\left(
\begin{array}{cc}
m_1&m_+\\
m_-&m_1
\end{array}
\right) 
\end{eqnarray}
is the mass matrix, which here acts as the generator of an $SL(2,\mathbb{R})$ transformation. Here $y$ is the co-ordinate that parametrises the isometry direction over which the T-duality is being performed.

Making this substitution effectively modifies the field equations, the new ones being given by
\begin{eqnarray}
\check{\cal H}_a&=&{\cal H}_a+m_a^{\phantom{a}b}{\cal B}_b\wedge dy\\
\check{\cal F}^{(5)}&=&{\cal F}^{(5)}-\frac{1}{2}m^{ab}{\cal B}_a\wedge {\cal B}_b \wedge dy\\
\check{{\cal H}}^{(7)}_a     &=&{\cal H}^{(7)}_a + m_a^{\phantom{a}b}({\cal B}^{(6)}_b+\tilde{\cal C}^{(4)}\wedge {\cal B}_b)\wedge dy \nonumber \\ && 
+\frac{1}{3!}m^{bc}{\cal B}_b\wedge {\cal B}_c\wedge {\cal B}_a\wedge dy\\
\check{\cal F}^{(9)}_{ab}&=&{\cal F}^{(9)}_{ab} + m_{(a}^{\phantom{(a}c}(2{\cal C}^{(8)}_{b)c} + {\cal B}^{(6)}_{b)}\wedge {\cal B}_c)\wedge dy \nonumber \\ &&
-\frac{1}{4!}m^{cd}{\cal B}_c\wedge {\cal B}_d\wedge {\cal B}_a\wedge {\cal B}_b\wedge dy \\
\check{\cal F}^{(11)}_{abc}&=&{\cal F}^{(11)}_{abc}+m_{(a}^{\phantom{(a}d}(3{\cal C}^{(10)}_{bc)d} -{\cal B}^{(8)}_{bc)}\wedge {\cal B}_d)\wedge dy
\nonumber \\ 
&& -\frac{1}{5!}m^{de}{\cal B}_d\wedge {\cal B}_e\wedge{\cal B}_a\wedge {\cal B}_b\wedge {\cal B}_c\wedge dy\\
\check{\cal F}^{(11)}_a&=&{\cal F}^{(11)}_a +m_a^{\phantom{a}b}{\cal C}^{(10)}_b\wedge dy .
\end{eqnarray}

Since all the fields are now assumed to be independent of the isometry direction one uses the massless rules when T-dualising the potentials. The massive terms on the IIA side are now represented by the `massive' terms produced by making the substitution (\ref{eq:IIB massive cov deriv}), and it is in this way that the T-duality rules are modified implicitly. In order to T-dualise the field strength equations it is more convenient to work with the non-covariant fields, so we give these in Appendix \ref{sec:IIB massive fields}. Then it is a long-winded but straightforward task to show that the IIB equations map to the equations in $SL(2,\mathbb{R})$ covariant 11-dimensional SUGRA. This acts as a non-trivial check on those equations.

One should also check that the IIB charges given in the previous subsection are still closed under this scheme. We now give the `massive' $SL(2,\mathbb{R})$ covariant differential relations for the bilinears so that this can be checked. These relations are derived from the standard ones given in \cite{Callister:2007jy} by making the above field modifications as well as the substitution (\ref{eq:IIB massive cov deriv}). The results are then given by
\begin{eqnarray}
dK_a&=&i_{K^+}\check{\cal H}_a +m_a^{\phantom{a}b}K_b\wedge dy\\
d\Phi&=&\epsilon^{ab}K_a\wedge \check{\cal H}_b-i_{K^+}\check{\cal F}^{(5)}\\
d\Sigma_a&=&\Phi\wedge \check{\cal H}_a +K_a\wedge \check{\cal F}^{(5)} -i_{K^+}\check{\cal F}^{(7)}_a +m_a^{\phantom{a}b}\Sigma_b\wedge dy\\
d({\cal M}_{ab}\Pi)&=&\Sigma_{(a}\wedge \check{\cal H}_{b)} +K_{(a}\wedge \check{\cal F}^{(7)}_{b)} -i_{K^+}\check{\cal F}^{(9)}_{ab} \nonumber \\
&& +2m_{(a}^{\phantom{(a}c}{\cal M}_{b)c}\Pi\wedge dy \\
d({\cal M}_{(ab}\Omega_{c)})&=&{\cal M}_{(ab}\Pi\wedge \check{\cal H}_{c)} +K_{(a}\wedge \check{\cal F}^{(9)}_{bc)} + i_{K^+}\check{\cal F}^{(11)}_{abc}\nonumber \\
&& +2m_{(a}^{\phantom{a}d}{\cal M}_{|d|b}\Omega_{c)}\wedge dy +m_{(a}^{\phantom{(a}d}{\cal M}_{bc)}\Omega_d\wedge dy\\
d({\cal M}_a^{\phantom{a}b}\Omega_{b})&=&-i_{K^+}\check{\cal F}^{(11)}_a +m_{(a}^{\phantom{(a}b}{\cal M}_b^{\phantom{b}c}\Omega_{c}\wedge dy .
\end{eqnarray}
Here the inclusion of the 11-form field strengths in the last two differential relations are conjectures along the lines of (\ref{eq:11-d dPi doublet}).

Finally, in order for the charges to be closed we require gauge choices for the potentials which satisfy the following conditions
\begin{eqnarray}
{\cal L}_{K^+}{\cal B}_a&=&-K^{+y}m_{a}^{\phantom{a}b}{\cal B}_b - m_{a}^{\phantom{a}b} N^{(1)}_b\wedge dy\\
{\cal L}_{K^+}\tilde{\cal C}^{(4)}&=&\frac{1}{2}m^{ab}N^{(1)}_a\wedge {\cal B}_b\wedge dy\\
{\cal L}_{K^+}{\cal B}_a^{(6)}&=&-K^{+y}m_{a}^{\phantom{a}b}{\cal B}^{(6)}_{b}+m_{a}^{\phantom{a}b}N^{(5)}_{b}\wedge dy \nonumber \\
 & & - \frac{1}{3}m^{bc}N^{(1)}_b\wedge {\cal B}_c\wedge {\cal B}_a\wedge dy\\
{\cal L}_{K^+}{\cal C}^{(8)}_{ab}&=&-2K^{+y}m_{(a}^{\phantom{(a}c}{\cal C}^{(8)}_{b)c} +2m_{(a}^{\phantom{(a}c}N^{(7)}_{b)c}\wedge dy -m_{(a}^{\phantom{(a}c}N^{(5)}_{|c|}\wedge{\cal B}_{b)} \wedge dy \nonumber \\ 
&& +\frac{1}{8}m^{cd}N^{(1)}_c\wedge {\cal B}_d\wedge {\cal B}_a\wedge {\cal B}_b\wedge dy\\
{\cal L}_{K^+}{\cal C}^{(10)}_{abc}&=&-3K^{+y}m_{(a}^{\phantom{(a}d}{\cal C}^{(10)}_{bc)d} -3m_{(a}^{\phantom{(a}d}N^{(9)}_{bc)d}\wedge dy +2m_{(a}^{\phantom{(a}d}N^{(7)}_{|d|b}\wedge {\cal B}_{c)}\wedge dy \nonumber \\
&& -\frac{1}{2}m_{(a}^{\phantom{(a}d}N^{(5)}_{|d|}\wedge {\cal B}_b\wedge {\cal B}_{c)}\wedge dy \nonumber \\
 & & + \frac{1}{30}m^{de}N^{(1)}_d\wedge {\cal B}_e \wedge {\cal B}_a \wedge {\cal B}_b\wedge {\cal B}_c\\
{\cal L}_{K^+}{\cal C}^{(10)}_{a}&=&-K^{+y}m_{a}^{\phantom{a}b}{\cal C}^{(10)}_b +m_{a}^{\phantom{a}b} N^{(9)}_b\wedge dy .
\end{eqnarray}
These are merely the T-duals of the conditions imposed on the IIA potentials. Using these relations one can then check that the IIB charges are closed in this `massive' theory and it is in this sense they can be said to be related through T-duality to the charges in non-covariant IIA.

$K^{+y}$ in the above expressions is related to the IIA F-string charge $M^{(F1)}$ (given in \cite{Callister:2007jy}) via T-duality. Explicitly we have
\begin{eqnarray}
i_{\beta}M^{(F1)} \leftrightarrow K^{+y}
\end{eqnarray}
which is easily shown. $K^{+y}$ must then clearly be closed itself.

\section{Discussion}\label{sec:discussion}

In this paper we have extended the results of \cite{HackettJones:2003vz,Callister:2007jy} by constructing the generalised charges for the BPS states of the $SL(2,\mathbb{R})$ covariant massive 11-dimensional and type IIB SUGRAs. We have also explored the multiplet structure of the 11-dimensional charges by using contractions with the Killing vectors, and confirmed that these map to the known multiplets in IIB and therefore demonstrated that in curved spacetimes the charges transform in the same $SL(2,\mathbb{R})$ representation as the states they correspond to.

In the 11-dimensional theory we found the expressions $\hat{L}^{(7)}_a$ and $\hat{L}^{(9)}_{ab}$ which do not seem to correspond to any (known) BPS states. However, such expressions play a key role in the construction of the various KK-monopole and M9-brane multiplets that map to IIB. An additional doublet $\hat{L}^{(10)}_a$ was also conjectured whose structure was unrelated to that of the other charges, but which was required in order to map to the charges of the IIB 9-brane doublet.

En route to determining the structure of these charges we first investigated the 11-dimensional gauge fields and found their field strength equations. Many of these explicitly involve the Killing vectors present in the theory. Although we focused on the $SL(2,\mathbb{R})$ case, with $n=2$ Killing vectors, the field equations should generalise to arbitrary $n$, and in this way will produce the field equations for the various lower dimensional SUGRAs.

An interesting observation was the existence of the scalar potential $i_{\beta}C^{(1)}$ in IIA, with 1-form field strength $X^{(1)}$. We might therefore expect new scalars to emerge in 11-dimensions for $n=3,6,8$. Specifically we would have $i_{\hat{k}_1\hat{k}_2\hat{k}_3}\hat{A}$, $i_{\hat{k}_1\ldots \hat{k}_6}\hat{C}$ and $i_{\hat{k}_1\ldots \hat{k}_8}\hat{N}^{(8)}$. These would then each have a 9-form dual potential given by $\hat{A}^{(9)}$, $\hat{C}^{(9)}$ and $\hat{N}^{(9)}$ respectively, which would suggest the existence of new families of branes appearing for each of these values of $n$. Schematically the generalised charges for these could be given by
\begin{eqnarray}
i_{\hat{k}_1\hat{k}_2\hat{k}_3}(\hat{\Sigma}\wedge \hat{k}_1\wedge \hat{k}_2\wedge \hat{k}_3) + i_{\hat{K}\hat{k}_1\hat{k}_2\hat{k}_3}\hat{A}^{(9)} \nonumber \\
i_{\hat{k}_1\ldots \hat{k}_6}(\hat{\omega}\wedge \hat{k}_1\wedge \ldots \wedge \hat{k}_6) + i_{\hat{K}\hat{k}_1\ldots \hat{k}_6}\hat{C}^{(9)} \nonumber \\
i_{\hat{k}_1\ldots \hat{k}_8}(i_{\hat{k}_1}\hat{K}\wedge \hat{k}_1\wedge \ldots \hat{k}_8) + i_{\hat{K}\hat{k}_1\ldots \hat{k}_8}\hat{N}^{(9)}
\end{eqnarray}
The expressions for the tensions would include the terms $|\hat{k}_1|^2|\hat{k}_2|^2|\hat{k}_3|^2$, $|\hat{k}_1|^2\ldots |\hat{k}_6|^2$ and $|\hat{k}_1|^3|\hat{k}_2|^2\ldots |\hat{k}_8|^2$ respectively which is in agreement with previous results from considering the full U-duality group \cite{Hull:1997kb,Obers:1997kk}. However from the approach here we would also expect there to be terms involving the contraction of different Killing vectors which are not usually found using those methods. It would be interesting to explore these states further and construct the full generalised charge expressions.

\section*{Acknowledgements}
AK was funded by the Isle of Man Government. The research of DJS is supported by the STFC.

\appendix

\section{Conventions}\label{sec:conventions}

We denote 11-dimensional objects and indices with a hat, whilst the corresponding objects and indices in the 10-dimensional theories are unhatted. Unless specified we use Greek letters to denote spacetime indices, and Roman letters for $SL(2,\mathbb{R})$ indices.

We use metrics with signature $(-,+,\ldots,+)$, and our spacetime antisymmetric symbol is defined (in a $(d+1)$-dimensional spacetime) by
\begin{eqnarray}
\epsilon_{01\ldots d}=+1 .
\end{eqnarray}
Our inner product convention between a vector and a $p$-form $Y$ is defined by contraction with the first index 
\begin{eqnarray}
(i_{k}Y)_{\mu_1\ldots \mu_{p-1}}=k^{\nu}Y_{\nu\mu_1\ldots \mu_{p-1}} .
\end{eqnarray}
Our Hodge dual convention is defined as
\begin{eqnarray}
(\ast F)_{\mu_1\ldots \mu_p}=\frac{\sqrt{|g|}}{(D-p)!}\epsilon_{\mu_1\ldots \mu_p}^{\phantom{\mu_1\ldots \mu_p}\nu_1\ldots \nu_{D-p}}F_{\nu_{1}\ldots \nu_{D-p}} .
\end{eqnarray}
Combinations of $\Gamma$-matrices are assumed antisymmetrised, i.e.
\begin{eqnarray}
\Gamma_{\mu_1\ldots \mu_p}=\Gamma_{[\mu_1\ldots \mu_p]}=\Gamma_{(p)}
\end{eqnarray}
where there is a factor of $\frac{1}{p!}$ in our definition for anti-symmetrisation.

\section{IIA from 11 dimensions} \label{sec:reducs}
Here we present the reduction rules of the 11-dimensional theory to IIA with our definitions and conventions. We perform the dimensional reduction along the isometry defined by $\hat{\alpha}$ and work in a co-ordinate system where $\hat{\alpha}^{\hat{\mu}}=\delta^{\hat{\mu}z}$. The metric then reduces according to  
\begin{eqnarray}
d\hat{s}^2_{(1,10)}&=&e^{-\frac{2}{3}\phi}ds^2_{(1,9)}+e^{\frac{4}{3}\phi}(dz+C^{(1)}
_{\mu}dx^{\mu})^2 .
\end{eqnarray}

The 11-dimensional gauge potentials and Killing vectors (with indices down) defined in the text reduce as
\begin{eqnarray}
\hat{\alpha}&\rightarrow&e^{\frac{4}{3}\phi}C^{(1)}+e^{\frac{4}{3}}dz\\
\hat{\beta}&\rightarrow&e^{-\frac{2}{3}\phi}\beta+e^{\frac{4}{3}\phi}i_{\alpha}C^{(1)}C^{(1)}+e^{\frac{4}{3}\phi}i_{\beta}C^{(1)}dz\\
\hat{A}&\rightarrow&C^{(3)}+B\wedge dz\\
\hat{C}&\rightarrow&B^{(6)}-(C^{(5)}-\frac{1}{2}C^{(3)}\wedge B)\wedge dz\\
\hat{N}^{(8)}&\rightarrow &\frac{4}{3}\phi^{(8)}+(C^{(7)}-\frac{1}{3!}C\wedge B^2)\wedge dz\\
\hat{T}^{(8)}&\rightarrow&N^{(8)}+(N^{(7)}-\frac{2}{3}i_{\beta}\phi^{(8)}-\frac{1}{3}C\wedge i_{\beta}C \wedge B)\wedge dz
\\
\hat{A}^{(10)}&\rightarrow&A^{(10)}+(C^{(9)}-\frac{1}{4!}C^{(3)}\wedge (B)^3)\wedge dz\\
\hat{B}^{(10)}&\rightarrow&B^{(10)}+(B^{(9)}-\frac{1}{16}C^{(3)}\wedge i_{\beta}C^{(3)}\wedge (B)^2)\wedge dz\\
\hat{D}^{(10)}&\rightarrow&D^{(10)}+(D^{(9)}-\frac{1}{8}C^{(3)}\wedge (i_{\beta}C^{(3)})^2\wedge B)\wedge dz .
\end{eqnarray} 
In addition to these we define the double dimensional reduction rules for the doublet of 11-forms as (in covariant notation)
\begin{eqnarray}
i_{\hat{\alpha}}\hat{A}^{(11)}_1&\rightarrow&\overline{A}^{(10)}\\
i_{\hat{\alpha}}\hat{A}^{(11)}_2&\rightarrow&\overline{B}^{(10)} .
\end{eqnarray}
The spinors reduce according to 
\begin{eqnarray}
\hat{\epsilon}&\rightarrow&e^{-\frac{1}{6}\phi}\epsilon
\end{eqnarray}
from which one can infer the reduction rules for the bilinear forms. These are most easily represented in the orthonormal frame defined adapted to $z$ and are given by
\begin{eqnarray}
\hat{K}_{(1)}&\rightarrow&\exp(-2\phi/3)K+\exp(-\phi/3)X\hat{e}^{\underline{z}}\\
\hat{\omega}_{(2)}&\rightarrow&\exp(-\phi)\Omega+\exp(-2\phi/3)\tilde{K}\wedge\hat{e}^{\underline{z}}\\
\hat{\Sigma}_{(5)}&\rightarrow&\exp(-2\phi)\Sigma+\exp(-5\phi/3)Z\wedge\hat{e}^{\underline{z}}\\
\hat{\Lambda}_{(6)}&\rightarrow&\exp(-7\phi/3)\Lambda+\exp(-2\phi)\tilde{\Sigma}\wedge\hat{e}^{\underline{z}}\\
\hat{\Pi}_{(9)}&\rightarrow&\exp(-10\phi/3)\Pi+\exp(-3\phi)\Psi\wedge\hat{e}^{\underline{z}}\\
\hat{\Upsilon}_{(10)}&\rightarrow&\exp(-11\phi/3)\Upsilon+\exp(-10\phi/3)\tilde{\Pi}\wedge\hat{e}^{\underline{z}}
\end{eqnarray}
where $\hat{e}^{\underline{z}}=e^{\frac{2}{3}\phi}(C^{(1)}+dz)$.

\section{$SL(2,\mathbb{R})$ covariant 11-dimensional KK-monopole worldvolume action}\label{sec:KK action} 

We will construct the kinetic term of the worldvolume action of the KK-monopoles in $SL(2,\mathbb{R})$ covariant 11-dimensional massive SUGRA. The purpose here is not to study the worldvolume actions in any great detail but rather just to show that an action can be constructed with two gauged isometries which is invariant under the massive gauge transformations (\ref{eq:11-d massive isom gauge}). We begin by first reviewing the case of a single gauged isometry to demonstrate the general mechanisms at work, before constructing the double isometry case. 

\subsection{Single gauged isometry case}

Our discussion of the single gauged isometry case is based on \cite{Bergshoeff:1998ef} which is an extension of the earlier work \cite{Bergshoeff:1997gy}. Here the action for the KK-monopole in a massive background where the Taub-NUT and massive isometries coincided was considered. In this case we therefore have just a single massive isometry ($\hat{\alpha}$) which is gauged in the worldvolume action by replacing the partial derivatives with a covariant derivative whenever the pullback of a spacetime field to the worldvolume is performed. The action is then that of a gauged sigma model \cite{Hull:1990ms} where one of the embedding co-ordinates is eliminated through the gauging. Specially we make the substitution
\begin{eqnarray}
\partial_i\hat{X}^{\hat{\mu}}\rightarrow D_i\hat{X}^{\hat{\mu}}=\partial_i\hat{X}^{\hat{\mu}}-\hat{C}_i\hat{\alpha}^{\hat{\mu}}
\end{eqnarray}
where
\begin{eqnarray}
\hat{C}_i=\hat{\alpha}^{-2}\partial_i\hat{X}^{\hat{\mu}}\hat{\alpha}_{\hat{\mu}} .
\end{eqnarray}
The kinetic term of the action is then given by\footnote{We set $2\pi\alpha'=1$.}
\begin{eqnarray}\label{eq:single isom action}
\hat{S}_{kin}= -\hat{T}_{KK}\int d^7 \xi \ \hat{\alpha}^2 \sqrt{|\textrm{det}(D_i\hat{X}^{\hat{\mu}}D_j\hat{X}^{\hat{\nu}}\hat{g}_{\hat{\mu}\hat{\nu}}+\hat{\alpha}^{-1}\hat{\cal F}_{ij})|}
\end{eqnarray}
where $\xi^i$ are the worldvolume co-ordinates, $\hat{T}_{KK}$ is the monopole tension and the field strength ${\hat{\cal F}}$ is given by
\begin{eqnarray}\label{eq:11-d BI field strength}
\hat{\cal F}_{ij}=d\hat{\omega}^{(1)}_{ij}+ i_{\hat{\alpha}}\hat{A}_{ij}
\end{eqnarray}
where $\hat{\omega}^{(1)}_i$ is the Born-Infeld 1-form. This field describes intersections of the KK-monopole with an M2-brane where the M2-brane wraps the compact isometry direction.

The action (\ref{eq:single isom action}) is invariant under both local isometry transformations with parameter $\hat{\sigma}^{(0)}(\xi^i)$ and massive gauge transformations which we will now give. Under local isometry transformations the embedding co-ordinates transform as
\begin{eqnarray}
\delta \hat{X}^{\hat{\mu}}&=&\hat{\sigma}^{(0)}\hat{\alpha}^{\hat{\mu}}
\end{eqnarray}
whereas a spacetime field $\hat{T}^{\hat{\mu}_1\ldots \hat{\mu}_i}_{\phantom{\hat{\mu}_1\ldots \hat{\mu}_i}\hat{\nu}_1\ldots \hat{\nu}_j}$ varies according to the general rule
\begin{eqnarray}\label{eq:spacetime sigma rule}
\delta_{\hat{\sigma}}\hat{T}^{\hat{\mu}_1\ldots \hat{\mu}_i}_{\phantom{\hat{\mu}_1\ldots \hat{\mu}_i}\hat{\nu}_1\ldots \hat{\nu}_j}=\hat{\sigma}^{(0)}\hat{\alpha}^{\hat{\lambda}}\partial _{\hat{\lambda}}\hat{T}^{\hat{\mu}_1\ldots \hat{\mu}_i}_{\phantom{\hat{\mu}_1\ldots \hat{\mu}_i}\hat{\nu}_1\ldots \hat{\nu}_j} .
\end{eqnarray}
Using these rules together with the massive rule (\ref{eq:11-d gauge trans}) for $\hat{\alpha}$, one determines that $\hat{C}_i$ transforms as follows:
\begin{eqnarray}
\delta \hat{C}_i&=& d\hat{\sigma}^{(0)}_i+\hat{m}\hat{\lambda}_i
\end{eqnarray}
from which it is a simple matter to calculate
\begin{eqnarray}
\delta D_i\hat{X}^{\hat{\mu}}&=& \hat{\sigma}^{(0)}D_i\hat{X}^{\hat{\nu}}\partial _{\hat{\nu}}\hat{\alpha}^{\hat{\mu}}-\hat{m}\hat{\lambda}_i\hat{\alpha}^{\hat{\mu}} .
\end{eqnarray}
We also have the following transformation rules for the Born-Infeld vector and spacetime metric 
\begin{eqnarray}\label{eq:11-d BI gauge}
\delta \hat{\omega}^{(1)}_i&=&d\hat{\rho}^{(0)}_i+\hat{\lambda}_i\\\label{eq:11-d metric gauge}
\delta \hat{g}_{\hat{\mu}\hat{\nu}}&=&2\hat{m}\hat{\lambda}_{(\hat{\mu}}\hat{\alpha}_{\hat{\nu})} + \hat{\sigma}^{(0)}\hat{\alpha}^{\hat{\rho}}\partial_{\hat{\rho}} \hat{g}_{\hat{\mu}\hat{\nu}} .
\end{eqnarray}
Finally we need the transformation rule for $\hat{A}_{\hat{\mu}_1\hat{\mu}_2\hat{\mu}_3}$ which is given by (\ref{eq:11-d A gauge}) together with (\ref{eq:spacetime sigma rule}).

Using this collection of transformation rules together with the fact that $\hat{\alpha}$ is a Killing isometry it is a straightforward task to show that the action (\ref{eq:single isom action}) is invariant under both local isometry transformations and the massive gauge transformations. This concludes our discussion of the single gauged isometry action.

\subsection{Double gauged isometry case}

We now extend this method to the case where two isometries are gauged in the action. The results here apply to both massive and non-massive isometries. So for example, they can be applied to the case of a KK-monopole in a single M9-brane background where the Taub-NUT and massive isometries do not coincide, as considered in \cite{Eyras:1998kx}. Alternatively, they can also be applied to the case of a KK-monopole in a double M9-brane background where the Taub-NUT isometry does coincide with one of the massive isometries. Which case is described depends on whether the truncation of the mass matrix $\hat{Q}$ (\ref{eq:11-d Q trunc}) or the general mass matrix (\ref{eq:11-d Q}) is used respectively.

In order to write the action in an $SL(2,\mathbb{R})$ covariant manner, we need to make use of the charge matrix $q^{ab}$ as done in \cite{Bergshoeff:2006gs}. The entries of this matrix determines what combination of states within the KK-monopole triplet the action is describing. However, as outlined below we find that we are also required to include the charge vector $q^a$. The only way that this can be meaningfully related to the combination of states being considered is if it is derived from the charge matrix according to $q^{ab}=q^aq^b$, which adds a restriction to $q^{ab}$, in particular we must have $\textrm{det}(q^{ab})=0$. The same restriction was found in \cite{Bergshoeff:2006gs} for the triplet of 7-branes where the Wess-Zumino term of the action was also considered. Mapping this restriction from IIB to 11-dimensions also gives us a restriction on the mass parameters
\begin{eqnarray}\label{eq:11-d mass constraint}
\textrm{det}(\hat{Q}^{ab})=0
\end{eqnarray}
which is equivalent to $\hat{m}_1^2+\hat{m}_-\hat{m}_+=0$. We therefore might also expect this restriction to apply to the action here, however we will show below that this does not necessarily seem to be required for the action (or at least the kinetic term) to be invariant under the massive gauge transformations.

We start constructing the action by generalising the field $\hat{C}_i$ of the previous section. In the present case we have two isometries and so expect to have two such fields. Treating these symmetrically, we propose that each should be defined analogously to $\hat{C}_i$ but each with a `naive' gauging using the other isometry. Explicitly we propose that we have two such fields defined as
\begin{eqnarray}\nonumber
\hat{B}_i&=&\hat{\beta}^{-2}(\partial_i \hat{X}^{\hat{\mu}} - \hat{C}_i\hat{\alpha}^{\hat{\mu}})\hat{\beta}_{\hat{\mu}}\\ \label{eq:isom gauging}
\hat{C}_i&=&\hat{\alpha}^{-2}(\partial_i \hat{X}^{\hat{\mu}} - \hat{B}_i\hat{\beta}^{\hat{\mu}})\hat{\alpha}_{\hat{\mu}}
\end{eqnarray}
where we have used a non-covariant notation with $\hat{\beta}$ denoting the second isometry. These rearrange to give
\begin{eqnarray}
\hat{B}_i&=&\frac{\hat{\alpha}^2}{\hat{\gamma}}\partial_i\hat{X}^{\hat{\mu}}{\hat{\beta}}_{\hat{\mu}}-\frac{\hat{\alpha}\cdot \hat{\beta}}{\hat{\gamma}}\partial_i\hat{X}^{\hat{\mu}}{\hat{\alpha}}_{\hat{\mu}}\\
\hat{C}_i&=&\frac{\hat{\beta}^2}{\hat{\gamma}}\partial_i\hat{X}^{\hat{\mu}}{\hat{\alpha}}_{\hat{\mu}}-\frac{\hat{\alpha}\cdot \hat{\beta}}{\hat{\gamma}}\partial_i\hat{X}^{\hat{\mu}}{\hat{\beta}}_{\hat{\mu}}
\end{eqnarray}
where we have defined
\begin{eqnarray}
\hat{\gamma}=\hat{\alpha}^2\hat{\beta}^2-(\hat{\alpha}\cdot \hat{\beta})^2 .
\end{eqnarray}
These expressions can be written in covariant notation as
\begin{eqnarray}
\hat{C}_{ia}=\frac{2\epsilon^{bc}(\hat{k}_a\cdot \hat{k}_b)\partial_i\hat{X}^{\hat{\mu}}\hat{k}_{\hat{\mu}c}}{\epsilon^{bc}\epsilon^{de}(\hat{k}_b\cdot \hat{k}_d )(\hat{k}_c\cdot \hat{k}_e)}
\end{eqnarray}
with
\begin{eqnarray}
\hat{C}_{ia}=\left(
\begin{array}{c}
\hat{B}_i\\
-\hat{C}_i
\end{array}
\right) .
\end{eqnarray}
The covariant derivative in this case is then defined as
\begin{eqnarray}
D_i\hat{X}^{\hat{\mu}}=\partial_i \hat{X}^{\hat{\mu}} - \epsilon^{ab}\hat{C}_{ia}\hat{k}^{\hat{\mu}}_b .
\end{eqnarray}
It is a simple matter to check that in the limit that the two isometries coincide this definition simplifies to the covariant derivative defined in the previous section.

In order to construct an action that is invariant under the various transformations we consider each subterm individually, beginning with the metric term. The embedding co-ordinates now change under the two local isometry transformations, with parameters $\hat{\sigma}^{(0)}_a(\xi^i)$, according to
\begin{eqnarray}
\delta \hat{X}^{\hat{\mu}}&=&\epsilon^{ab} \hat{\sigma}^{(0)}_a\hat{k}^{\hat{\mu}}_b
\end{eqnarray}
whereas a spacetime field $\hat{T}^{\hat{\mu}_1\ldots \hat{\mu}_i}_{\phantom{\hat{\mu}_1\ldots \hat{\mu}_i}\hat{\nu}_1\ldots \hat{\nu}_j}$ transforms by the general rule
\begin{eqnarray}
\delta_{\hat{\sigma}} \hat{T}^{\hat{\mu}_1\ldots \hat{\mu}_i}_{\phantom{\hat{\mu}_1\ldots \hat{\mu}_i}\hat{\nu}_1\ldots \hat{\nu}_j} =\epsilon^{ab} \hat{\sigma}^{(0)}_a\hat{k}_b^{\hat{\lambda}}\partial_{\hat{\lambda}}\hat{T}^{\hat{\mu}_1\ldots \hat{\mu}_i}_{\phantom{\hat{\mu}_1\ldots \hat{\mu}_i}\hat{\nu}_1\ldots \hat{\nu}_j} .
\end{eqnarray}
Then, using (\ref{eq:11-d massive isom gauge}) one can show through explicit calculation that 
\begin{eqnarray}
\delta \hat{C}_{ia}&=&d\hat{\sigma}^{(0)}_{ia} - \hat{Q}_a^{\phantom{a}b}D_i\hat{X}^{\hat{\mu}}\hat{\lambda}_{\hat{\mu}b}
\end{eqnarray}
and therefore
\begin{eqnarray}
\delta D_i\hat{X}^{\hat{\mu}}=\epsilon^{ab}\hat{\sigma}^{(0)}_aD_i\hat{X}^{\hat{\nu}}\partial _{\hat{\nu}}\hat{k}_b^{\hat{\mu}} - \hat{Q}^{ab}D_i\hat{X}^{\hat{\nu}}\hat{\lambda}_{\hat{\nu}a}\hat{k}_b^{\hat{\mu}} .
\end{eqnarray}
The rule for the metric (\ref{eq:11-d metric gauge}) now becomes
\begin{eqnarray}
\delta \hat{g}_{\hat{\mu}\hat{\nu}}&=&\hat{Q}^{ab}(\hat{\lambda}_{\hat{\mu}a}\hat{k}_{\hat{\nu}b} +\hat{\lambda}_{\hat{\nu}a}\hat{k}_{\hat{\mu}b}) +\epsilon^{ab} \hat{\sigma}^{(0)}_a\hat{k}_b^{\hat{\rho}}\partial_{\hat{\rho}} \hat{g}_{\hat{\mu}\hat{\nu}} .
\end{eqnarray}
It is then a straightforward task to show that the term
\begin{eqnarray}
D_i\hat{X}^{\hat{\mu}}D_j\hat{X}^{\hat{\nu}}\hat{g}_{\hat{\mu}\hat{\nu}}
\end{eqnarray}
is invariant under these covariant transformations using the fact that $\hat{k}_a$ define Killing isometries.

Next we consider the Born-Infeld field strength. Since the Born-Infeld field describes the intersection of the KK-monopole with an M2-brane where the M2-brane wraps a compact isometry direction, we get a doublet of Born-Infeld fields $\hat{\omega}^{(1)}_{ia}$ in the covariant theory since we now have two compact isometry directions. We begin determining the structure of the field strength $\hat{\cal F}_{ija}$ by generalising the transformation rule (\ref{eq:11-d BI gauge}) to
\begin{eqnarray}
\delta \hat{\omega}^{(1)}_{ia}&=&d\hat{\rho}^{(0)}_{ia} +D_i\hat{X}^{\hat{\mu}}\hat{\lambda}_{\hat{\mu}a}
\end{eqnarray}
from which we calculate
\begin{eqnarray}
\delta(d\hat{\omega}^{(1)}_{a})=d\hat{\lambda}_{a}-i_{\hat{k}_2}\hat{\lambda}_1d\hat{C}_a - d(i_{\hat{k}_2}\hat{\lambda}_1)\wedge \hat{C}_a .
\end{eqnarray}

From (\ref{eq:11-d BI field strength}) we would also expect there to be a term $D_i\hat{X}^{\hat{\mu}}D_j\hat{X}^{\hat{\nu}}i_{\hat{k}_a}\hat{A}_{\hat{\mu}\hat{\nu}}$ which, using (\ref{eq:11-d cov 3-form gauge}), we find transforms as
\begin{eqnarray}
\delta (D_i\hat{X}^{\hat{\mu}}D_j\hat{X}^{\hat{\nu}}i_{\hat{k}_a}\hat{A}_{\hat{\mu}\hat{\nu}})& =&-D_i\hat{X}^{\hat{\mu}}D_j\hat{X}^{\hat{\nu}}d\hat{\lambda}_{\hat{\mu}\hat{\nu}a}\nonumber \\ 
&& +i_{\hat{k}_2}\hat{\lambda}_1\hat{Q}_a^{\phantom{a}b}D_i\hat{X}^{\hat{\mu}}D_j\hat{X}^{\hat{\nu}}i_{\hat{k}_b}\hat{A}_{\hat{\mu}\hat{\nu}} .
\end{eqnarray}
Combining these results we see that the naive generalisation of (\ref{eq:11-d BI field strength}) is not invariant under the massive gauge transformations. Following \cite{Eyras:1998kx} we introduce the worldvolume scalar $\hat{\omega}^{(0)}$ which has the following gauge transformation
\begin{eqnarray}
\delta \hat{\omega}^{(0)} = \frac{1}{2}\epsilon^{ab}i_{\hat{k}_a}\hat{\lambda}_b = -i_{\hat{k}_2}\hat{\lambda}_1 .
\end{eqnarray}
We then consider the term $\hat{\omega}^{(0)}d\hat{C}_{ija}$ which transforms as
\begin{eqnarray}
\delta (\hat{\omega}^{(0)}d\hat{C}_{a})=-i_{\hat{k}_2}\hat{\lambda}_1d\hat{C}_{a} - \hat{\omega}^{(0)}\hat{Q}_a^{\phantom{a}b}(d\hat{\lambda}_{b} -i_{\hat{k}_2}\hat{\lambda}_1d\hat{C}_a - d(i_{\hat{k}_2}\hat{\lambda}_1)\wedge \hat{C}_a) .
\end{eqnarray}
Putting these together we see that the following field definition is invariant up to quadratic order of the mass parameters
\begin{eqnarray}
\hat{\cal F}_{ija}&=&d\hat{\omega}^{(1)}_{ija} +(\mathbbm{1}_a^{\phantom{a}b} + \hat{\omega}^{(0)}\hat{Q}_a^{\phantom{a}b})D_i\hat{X}^{\hat{\mu}}D_j\hat{X}^{\hat{\nu}}i_{\hat{k}_b}\hat{A}_{\hat{\mu}\hat{\nu}}\nonumber \\
&& -\hat{\omega}^{(0)}(\mathbbm{1}_a^{\phantom{a}b} + \frac{1}{2}\hat{\omega}^{(0)}\hat{Q}_a^{\phantom{a}b})d\hat{C}_{ijb} .
\end{eqnarray}
Note that we have the identity $\hat{Q}_a^{\phantom{a}b}\hat{Q}_b^{\phantom{b}c}=\textrm{det}(\hat{Q})\mathbbm{1}_a^{\phantom{a}c}$. We can therefore neglect terms that are quadratic or higher order in the mass matrix if we impose the constraint (\ref{eq:11-d mass constraint}). 

Finally we must consider how terms such as $\hat{k}_a\cdot \hat{k}_b$ transform. We find
\begin{eqnarray}
\delta (\hat{k}_a\cdot \hat{k}_b) = 2\hat{Q}_{(a}^{\phantom{(a}c}(\hat{k}_{b)}\cdot \hat{k}_c) i_{\hat{k}_2}\hat{\lambda}_1 .
\end{eqnarray}
We then construct the invariant expression
\begin{eqnarray}
\hat{k}_{ab}=(\mathbbm{1}_{(a}^{\phantom{(a}c} +2\hat{\omega}^{(0)}\hat{Q}_{(a}^{\phantom{(a}c})(\hat{k}_{b)}\cdot \hat{k}_c)
\end{eqnarray}
where once again we restrict ourselves to the case where the terms quadratic in the mass parameters vanish.

At this point we realise that there are no appropriate fields with which we can contract the $SL(2,\mathbb{R})$ free index of the Born-Infeld field strength. The Killing vectors $\hat{k}_a$ are not suitable since this would leave a free spacetime index. We are therefore forced to include the charge vector $q^a$ discussed at the start of this section. Putting these expressions together we find that an $SL(2,\mathbb{R})$ covariant action with two gauged isometries can be written as
\begin{eqnarray}\label{eq:11-d cov KK action}
\hat{S}_{kin}&=& -\hat{T}_{KK}\int d^7 \xi \ q^{ab}\hat{k}_{ab}\sqrt{|\textrm{det}(D_i\hat{X}^{\hat{\mu}}D_j\hat{X}^{\hat{\nu}}\hat{g}_{\hat{\mu}\hat{\nu}}+\frac{q^a\hat{\cal F}_{ija}}{(q^{ab}\hat{k}_{ab})^{\frac{1}{2}}})|}\nonumber \\ &&
\end{eqnarray} 
which is invariant under both the massive gauge transformations and local isometry transformations. Note that after performing the truncation (\ref{eq:11-d Q trunc}) there are differences between the action presented here and the action given in \cite{Eyras:1998kx}. This originates from the differences between the massive gauge transformation rules (\ref{eq:11-d massive isom gauge}) and the symmetric manner in which we gauged both isometries (\ref{eq:isom gauging}).

In \cite{Bergshoeff:2006gs} the IIB 7-brane action was constructed for the $\textrm{det}(q)=0$ case and the result has the same general structure as (\ref{eq:11-d cov KK action}). This action was then generalised to positive $\textrm{det}(q)$ in \cite{Bergshoeff:2007aa} to second order in the Born-Infeld term.  In the current situation, while it is not obvious how to generalise the action fully, we note that the restriction (\ref{eq:11-d mass constraint}) can be lifted without spoiling the gauge invariance if we define the Born-Infeld field strength as
\begin{eqnarray}
\hat{\cal F}_{ija}&=&d\hat{\omega}^{(1)}_{ija} +\textrm{exp}(\hat{\omega}^{(0)}\hat{Q})_{a}^{\phantom{a}b}D_i\hat{X}^{\hat{\mu}}D_j\hat{X}^{\hat{\nu}}i_{\hat{k}_b}\hat{A}_{\hat{\mu}\hat{\nu}}\nonumber \\
&& -(\hat{Q}^{-1})_{a}^{\phantom{a}b}(\textrm{exp}(\hat{\omega}^{(0)}\hat{Q})_{b}^{\phantom{b}c}-\mathbbm{1}_{b}^{\phantom{b}c})d\hat{C}_{ijc}
\end{eqnarray}
and the Killing vector term as
\begin{eqnarray}
\hat{k}_{ab}=\textrm{exp}(2\hat{\omega}^{(0)}\hat{Q})_{(a} {\phantom{}^c}(\hat{k}_{b)}\cdot \hat{k}_c)
\end{eqnarray}
The action is then exponentially dependent on the mass parameters.

\section{Mapping 11-dimensional 10-forms to IIB}\label{sec:mapping 10-forms} 

We now map the 11-dimensional 10-form potentials to IIB and determine the resulting quadruplet of field equations for the IIB 10-form potentials. These equations were calculated in \cite{Bergshoeff:2005ac} in an $SU(1,1)$ covariant form. The reason we re-derive them here is to determine the T-duality rules for the IIA potentials which map to these potentials. These are required to T-dualise the IIA 8/9-form charges. Furthermore, we reveal the existence of a pair of IIB 9-form potentials which form part of a doublet which we conjecture should couple to a doublet of KK-type 9-branes. 

We begin by considering (\ref{eq:11-d 11-form constraint}) which shows that there are in fact only two independent 11-form field strengths in the 11-dimensional theory. We make this fact explicit upon performing the dimensional reduction to IIA. Specifically we have (reducing along $z$)
\begin{eqnarray}
\hat{F}^{(11)}_{11}&\rightarrow & G^{(11)} + F^{(10)}\wedge (C^{(1)} + dz) \\
\hat{F}^{(11)}_{12}&\rightarrow & i_{\beta}C^{(1)}G^{(11)} + H^{(11)} + (i_{\beta}C^{(1)}F^{(10)} + H^{(10)})\wedge (C^{(1)} + dz) \\
\hat{F}^{(11)}_{22}&\rightarrow & ((i_{\beta}C^{(1)})^2 - e^{-2\phi}R_{\beta}^2)G^{(11)} + 2i_{\beta}C^{(1)}H^{(11)} \nonumber \\
&& +\biggl(((i_{\beta}C^{(1)})^2 - e^{-2\phi}R_{\beta}^2)F^{(10)} + 2i_{\beta}C^{(1)}H^{(10)}\biggr)\wedge (C^{(1)} + dz)\nonumber \\ &&
\end{eqnarray}
where $F^{(10)}$ is the Ramond-Ramond 10-form and is related to the mass parameters via 
\begin{eqnarray}
F^{(10)}&=& \ast (-m_+ +((i_{\beta}C^{(1)})^2 - e^{-2\phi}R_{\beta}^2)m_- - 2i_{\beta}C^{(1)}m_1) \nonumber \\
&=& -\ast F^{(0)} 
\end{eqnarray}
which is a generalisation of the usual relation found in Romans' theory. $H^{(10)}$ is a second 10-form which also has a relation with the mass parameters given by
\begin{eqnarray}
H^{(10)} &=& e^{-2\phi}R_{\beta}^2\ast (2i_{\beta}C^{(1)}m_- - 2m_1)\nonumber \\
&=& e^{-2\phi}R_{\beta}^2\ast H^{(0)} .
\end{eqnarray}
Obviously the two 11-form fields $G^{(11)}$ and $H^{(11)}$ have no Hodge dual interpretation and in reality are identically zero due to their rank. 

In this case it is not possible to deduce the T-duality rules for the field strengths by a Hodge dual argument as done previously. However, we find that in order to produce well formed IIB field equations the IIA field strengths must T-dualise as follows
\begin{eqnarray}
e^{-2\phi}R_{\beta}^2F^{(10)}_{\mu_1\ldots \mu_{10}}&\rightarrow & \frac{3}{4}({\cal S}^{(10)}+{\cal R}_{\beta}^{-2}i_{\beta}{\cal S}^{(9)}\wedge \beta \nonumber \\
&& +e^{-2\varphi}({\cal F}^{(9)}+{\cal R}_{\beta}^{-2}i_{\beta}{\cal F}^{(9)}\wedge \beta)\wedge i_{\beta}{\cal B})_{\mu_1\ldots \mu_{10}} \\
H^{(10)}_{\mu_1\ldots \mu_{10}}&\rightarrow & ( i_{\beta}{\cal H}^{(11)} +({\cal H}^{(9)}-{\cal R}^{-2}_{\beta}i_{\beta}{\cal H}^{(9)}\wedge \beta)\wedge i_{\beta}{\cal B} \nonumber \\
&& -\frac{1}{2}{\cal G}^{(10)} -\frac{1}{2}{\cal R}_{\beta}^{-2}i_{\beta}{\cal G}^{(10)}\wedge \beta )_{\mu_1\ldots \mu_{10}}\\
e^{-2\phi}R_{\beta}^2i_{\beta}G^{(11)}_{\mu_1\ldots \mu_{10}}&\rightarrow & i_{\beta}{\cal G}^{(11)}_{\mu_1\ldots \mu_{10}}\\
i_{\beta}G^{(11)}_{\mu_1\ldots \mu_{10}}&\rightarrow & ({\cal G}^{(10)} +{\cal R}_{\beta}^{-2}i_{\beta}{\cal G}^{(10)}\wedge \beta)_{\mu_1\ldots \mu_{10}}\\
i_{\beta}H^{(11)}_{\mu_1\ldots \mu_{10}}&\rightarrow & (i_{\beta}{\cal X}^{(11)} +\frac{3}{8}{\cal S}^{(10)} +\frac{3}{8}{\cal R}_{\beta}^{-2}i_{\beta}{\cal S}^{(10)}\wedge \beta )_{\mu_1\ldots \mu_{10}}
\end{eqnarray}
along with the usual rule for the Ramond-Ramond 10-form which is just the inversion of (\ref{eq:RR T-dual}). It may seem strange that $F^{(10)}$ and $i_{\beta}G^{(11)}$ obey different rules depending on whether a factor of $e^{-2\phi}R_{\beta}^2$ is present or not. However, essentially the same situation occurs when T-dualising $i_{\beta}{\cal F}^{(9)}$ from IIB to IIA which maps to either $F^{(8)}$ or $i_{\beta}X^{(9)}$ (\ref{eq:X9 T-dual}). In that case the reason for it can be traced back to the fact that both the IIA field strengths are related to $F^{(2)}$ through Hodge duality. Obviously there is no such interpretation in the current case.

The field strengths ${\cal S}^{(10)}$ and ${\cal G}^{(10)}$ are the T-duals of $X^{(9)}$ and $H^{(9)}$ respectively and have the associated 9-form potentials ${\cal T}^{(9)}$ and ${\cal N}^{(9)}$ respectively. These fields are present in a non-covariant (in the spacetime sense) massive IIB SUGRA which we will discuss further below. Due to a cancellation they do not appear in the field equations for the quadruplet of 10-forms which is obtained by mapping $i_{\hat{k}_{(a}}\hat{A}^{(10)}_{bc)}$ from 11-dimensions, as we now see.

We find that in order to produce well-formed IIB field strength equations the relevant IIA potentials T-dualise as 

\begin{eqnarray}
i_{\beta}A^{(10)}_{\mu_1\ldots \mu_9}&\rightarrow & \biggl[-{\cal N}^{(9)} + {\cal R}_{\beta}^{-2}i_{\beta}{\cal N}^{(9)}\wedge \beta +\frac{1}{4}i_{\beta}{\cal C}^{(8)}\wedge {\cal C}^{(2)} \nonumber \\
&& +\frac{1}{4}{\cal R}_{\beta}^{-2}i_{\beta}{\cal C}^{(8)}\wedge i_{\beta}{\cal C}^{(2)}\wedge \beta \biggr]_{\mu_1\ldots \mu_9}
\end{eqnarray}

\begin{eqnarray}
B^{(9)}_{\mu_1\ldots \mu_9}&\rightarrow & \biggl[ -\frac{3}{2}i_{\beta}{\cal B}^{(10)} -\frac{1}{2}{\cal N}^{(9)} +\frac{1}{2}{\cal R}_{\beta}^{-2}i_{\beta}{\cal N}^{(9)}\wedge \beta -\frac{3}{8}i_{\beta}{\cal C}^{(8)}\wedge {\cal C}^{(2)}\nonumber \\
&& -\frac{3}{8}{\cal R}_{\beta}^{-2}i_{\beta}{\cal C}^{(8)}\wedge i_{\beta}{\cal C}^{(2)}\wedge \beta +(\varphi^{(8)} +{\cal R}_{\beta}^{-2}i_{\beta}\varphi^{(8)}\wedge \beta +\frac{1}{2}{\cal C}^{(6)}\wedge {\cal C}^{(2)} \nonumber \\
&& +\frac{1}{2}{\cal R}_{\beta}^{-2}i_{\beta}{\cal C}^{(6)}\wedge {\cal C}^{(2)}\wedge \beta   +\frac{1}{2}{\cal R}_{\beta}^{-2}{\cal C}^{(6)}\wedge i_{\beta}{\cal C}^{(2)}\wedge \beta)\wedge i_{\beta}{\cal B} \biggr]_{\mu_1\ldots \mu_9}
\end{eqnarray}

\begin{eqnarray}
i_{\beta}{\cal B}^{(10)}_{\mu_1\ldots \mu_9}&\rightarrow & \biggl[ \frac{3}{2}i_{\beta}{\cal A}^{(10)}  -\frac{3}{8}{\cal T}^{(9)} +\frac{3}{8}{\cal R}_{\beta}^{-2}i_{\beta}{\cal T}^{(9)}\wedge \beta +\frac{3}{4}i_{\beta}\varphi^{(8)}\wedge {\cal C}^{(2)} \nonumber \\
&& +\frac{3}{4}{\cal R}_{\beta}^{-2}i_{\beta}\varphi^{(8)}\wedge i_{\beta}{\cal C}^{(2)} \wedge \beta -\frac{1}{4}{\cal N}^{(7)}\wedge {\cal C}^{(2)} +\frac{1}{4}{\cal R}_{\beta}^{-2} i_{\beta}{\cal N}^{(7)}\wedge {\cal C}^{(2)}\wedge \beta \nonumber \\
&& -\frac{1}{4}{\cal R}_{\beta}^{-2}{\cal N}^{(7)}\wedge i_{\beta}{\cal C}^{(2)}\wedge \beta +\frac{1}{8}i_{\beta}{\cal C}^{(6)}\wedge ({\cal C}^{(2)})^2 \nonumber \\
&& +\frac{1}{4}{\cal R}_{\beta}^{-2} i_{\beta}{\cal C}^{(6)}\wedge i_{\beta}{\cal C}^{(2)}\wedge {\cal C}^{(2)}\wedge \beta \biggr]_{\mu_1\ldots \mu_9} 
\end{eqnarray}

\begin{eqnarray}
D^{(9)}_{\mu_1\ldots \mu_9}&\rightarrow & \biggl[-\frac{3}{4}{\cal T}^{(9)} +\frac{3}{4}{\cal R}_{\beta}^{-2}i_{\beta}{\cal T}^{(9)}\wedge \beta -\frac{1}{2}i_{\beta}\varphi^{(8)}\wedge {\cal C}^{(2)} \nonumber \\
&& -\frac{1}{2}{\cal R}_{\beta}^{-2}i_{\beta}\varphi^{(8)}\wedge i_{\beta}{\cal C}^{(2)} \wedge \beta -\frac{1}{2}{\cal N}^{(7)}\wedge {\cal C}^{(2)} +\frac{1}{2}{\cal R}_{\beta}^{-2} i_{\beta}{\cal N}^{(7)}\wedge {\cal C}^{(2)}\wedge \beta \nonumber \\
&& -\frac{1}{2}{\cal R}_{\beta}^{-2}{\cal N}^{(7)}\wedge i_{\beta}{\cal C}^{(2)}\wedge \beta -\frac{1}{4}i_{\beta}{\cal C}^{(6)}\wedge ({\cal C}^{(2)})^2 -\frac{1}{2}{\cal R}_{\beta}^{-2}i_{\beta}{\cal C}^{(6)}\wedge i_{\beta}{\cal C}^{(2)}\wedge {\cal C}^{(2)} \wedge \beta\nonumber \\
&& +(-{\cal N}^{(8)} -{\cal R}_{\beta}^{-2} i_{\beta}{\cal N}^{(8)}\wedge \beta +\frac{1}{2}{\cal C}^{(4)}\wedge ({\cal C}^{(2)})^2 +\frac{1}{2}{\cal R}_{\beta}^{-2} i_{\beta}{\cal C}^{(4)}\wedge ({\cal C}^{(2)})^2\wedge \beta \nonumber \\
&& +{\cal R}_{\beta}^{-2} {\cal C}^{(4)}\wedge i_{\beta}{\cal C}^{(2)}\wedge {\cal C}^{(2)}\wedge \beta)\wedge i_{\beta}{\cal B} \biggr]_{\mu_1\ldots \mu_9} 
\end{eqnarray}

\begin{eqnarray}
i_{\beta}D^{(10)}_{\mu_1\ldots \mu_9}&\rightarrow& \biggl[-i_{\beta}{\cal D}^{(10)} +{\cal N}^{(8)}\wedge i_{\beta}{\cal C}^{(2)} -{\cal R}^{-2}_{\beta}i_{\beta}{\cal N}^{(8)}\wedge i_{\beta}{\cal C}^{(2)}\wedge \beta \nonumber \\
&& +\frac{1}{8}i_{\beta}{\cal C}^{(4)}\wedge ({\cal C}^{(2)})^3+\frac{3}{8}{\cal R}^{-2}_{\beta}i_{\beta}{\cal C}^{(4)}\wedge i_{\beta}{\cal C}^{(2)} \wedge ({\cal C}^{(2)})^2\wedge \beta \biggr]_{\mu_1\ldots \mu_9}
\end{eqnarray}
along with the standard rule for the Ramond-Ramond potential $C^{(9)}$.

We then find the field equations to be given by
\begin{eqnarray}\label{eq:IIB 11-form 1}
\frac{2}{3}{\cal X}^{(11)}&=&d{\cal A}^{(10)}-l^2{\cal F}^{(11)} -\frac{4}{3}l{\cal H}^{(11)} +\frac{2}{3}{\cal H}^{(9)}\wedge {\cal C}^{(2)} +\frac{2}{3}l{\cal F}^{(9)}\wedge {\cal C}^{(2)}\nonumber \\
&& +\frac{1}{3}{\cal H}\wedge {\cal N}^{(8)} -\frac{1}{6}{\cal F}^{(7)}\wedge ({\cal C}^{(2)})^2 -\frac{1}{3}{\cal H}\wedge {\cal B}^{(6)}\wedge {\cal C}^{(2)} \\
\frac{2}{3}{\cal H}^{(11)}&=&d{\cal B}^{(10)} -l{\cal F}^{(11)} +\frac{1}{3}{\cal F}^{(9)}\wedge {\cal C}^{(2)} -\frac{2}{3}{\cal H}\wedge \varphi^{(8)}\\
{\cal F}^{(11)}&=&d{\cal C}^{(10)} - {\cal C}^{(8)}\wedge {\cal H}\\ 
{\cal G}^{(11)}&=& d{\cal D}^{(10)} + l^3{\cal F}^{(11)} +2l^2{\cal H}^{(11)} +2l{\cal X}^{(11)} \nonumber \\ \label{eq:IIB 11-form 2}
&& -{\cal N}^{(8)}\wedge d{\cal C}^{(2)} -\frac{1}{6}d{\cal C}^{(4)}\wedge ({\cal C}^{(2)})^3 +\frac{1}{24}{\cal H}\wedge ({\cal C}^{(2)})^4
\end{eqnarray}
where we have included the Ramond-Ramond field equation which takes the usual form. These can be written as a quadruplet which is shown in Section \ref{sec:IIB cov 9-charge quad}.

We now briefly discuss the IIB 9-form potentials ${\cal T}^{(9)}$ and ${\cal N}^{(9)}$. The field strength equations for these potentials can be determined by mapping the 11-dimensional doublet $\epsilon^{bc}i_{\hat{k}_b}\hat{F}^{(11)}_{ca}$ to IIB. Alternatively they are the direct T-duals of the IIA potentials $N^{(8)}$ and $\phi^{(8)}$. Doing this T-duality is equivalent to carrying out a Scherk-Schwarz dimensional reduction of IIA using the global $SO(1,1)$ symmetry \cite{Bergshoeff:2002mb,Bergshoeff:2002nv}. Using the above rules we find that mapping the charge doublet (\ref{eq:M9 charge doublet}) to IIB would produce the charges for a doublet of KK-type 9-branes which minimally couple to the combination of potentials $\frac{3}{2}i_{\beta}{\cal B}^{(10)}+\frac{3}{2}{\cal N}^{(9)}$ and $\frac{3}{2}i_{\beta}{\cal A}^{(10)} -\frac{9}{8}{\cal T}^{(9)}$ which form a doublet. Therefore even though the states will be 9-branes, due to the presence of an isometry direction and the fact that they do minimally couple to ${\cal T}^{(9)}$ and ${\cal N}^{(9)}$, these branes should act as the sources of a doublet of masses. These branes would therefore appear to be the IIB origin of the doublet $(m_4,\tilde{m}_4)$ discussed in \cite{Bergshoeff:2002nv}.

\section{Non-covariant massive IIB fields}\label{sec:IIB massive fields}

The `massive' IIB field strength equations, which T-dualise to the non-covariant IIA equations, are given by
\begin{eqnarray}
\check{\cal F}^{(1)}&=&{\cal F}^{(1)}+\biggl[m_+ + m_-(e^{-2\varphi}-l^2) +2m_1l\biggr]\wedge dy\\
\check{\cal H}^{(1)}&=&{\cal H}^{(1)}+\biggl[2m_-l -2m_1\biggr]\wedge dy\\
\check{\cal F}^{(3)}&=&{\cal F}^{(3)}+\biggl[m_+{\cal B}-m_-l{\cal C}^{(2)}+m_1({\cal C}^{(2)} +l{\cal B})\biggr]\wedge dy\\
\check{\cal H}&=&{\cal H}+\biggl[m_-{\cal C}^{(2)}-m_1{\cal B}\biggr]\wedge dy\\
\check{\cal F}^{(5)}&=&{\cal F}^{(5)} +\biggl[\frac{1}{2}m_+({\cal B})^2-\frac{1}{2}m_-({\cal C}^{(2)})^2+m_1{\cal B}\wedge {\cal C}^{(2)}\biggr]\wedge dy\\
\check{\cal F}^{(7)}&=&{\cal F}^{(7)} +\biggl[\frac{1}{3!}m_+({\cal B})^3 -m_-{\cal B}^{(6)} \nonumber \\
&& +m_1(-{\cal C}^{(6)} +{\cal C}^{(4)}\wedge {\cal B})\biggr]\wedge dy\\
\check{\cal H}^{(7)}&=&{\cal H}^{(7)} + m_+\biggl[ -{\cal C}^{(6)} +{\cal C}^{(4)}\wedge {\cal B} -\frac{1}{2}{\cal C}^{(2)}\wedge ({\cal B})^2 +\frac{1}{3!}l({\cal B})^3 \biggr]\wedge dy \nonumber \\
&& +m_-\biggl[ \frac{1}{3!}({\cal C}^{(2)})^3 -l({\cal B})^3 \biggr]\wedge dy +m_1\biggl[ {\cal B}^{(6)} -\frac{1}{2}({\cal C}^{(2)})^2\wedge {\cal B} \nonumber \\ 
&&-l{\cal C}^{(6)} +l{\cal C}^{(4)}\wedge {\cal B}\biggr]\wedge dy\\
\check{\cal F}^{(9)}&=&{\cal F}^{(9)} +\biggl[\frac{1}{4!}m_+({\cal B})^4 +2m_-\varphi^{(8)} +m_1(-2{\cal C}^{(8)} +{\cal C}^{(6)}\wedge {\cal B})\biggr]\wedge dy\\
\check{\cal H}^{(9)}&=&{\cal H}^{(9)} +m_+\biggl[{\cal C}^{(8)} -\frac{1}{2}{\cal C}^{(6)}\wedge {\cal B} +\frac{1}{12}{\cal C}^{(2)}\wedge ({\cal B})^3 -\frac{1}{4!}l({\cal B})^4\biggr]\wedge dy\nonumber \\
&& m_-\biggl[-{\cal N}^{(8)} +\frac{1}{2}{\cal B}^{(6)}\wedge {\cal C}^{(2)} -2l\varphi^{(8)}\biggr]\wedge dy +m_1\biggl[-\frac{1}{2}{\cal C}^{(6)}\wedge {\cal C}^{(2)} \nonumber \\
&& -\frac{1}{2}{\cal B}^{(6)}\wedge {\cal B} +\frac{1}{2}{\cal C}^{(4)}\wedge {\cal C}^{(2)}\wedge {\cal B} +2l{\cal C}^{(8)} -l{\cal C}^{(6)}\wedge {\cal B}\biggr]\wedge dy\\
\check{(e^{-2\varphi}{\cal F}^{(9)})}&=& e^{-2\varphi}{\cal F}^{(9)} +m_+\biggl[-2\varphi^{(8)}- {\cal B}^{(6)}\wedge {\cal B} -{\cal C}^{(6)}\wedge {\cal C}^{(2)} +{\cal C}^{(4)}\wedge {\cal C}^{(2)}\wedge {\cal B} \nonumber \\ 
&& -\frac{1}{4}({\cal C}^{(2)})^2\wedge ({\cal B})^2 +2l{\cal C}^{(8)} -l{\cal C}^{(6)}\wedge {\cal B} +\frac{1}{3!}l{\cal C}^{(2)}\wedge ({\cal B})^3 -\frac{1}{4!}l^2({\cal B})^4\biggr]\wedge dy \nonumber \\
&& +m_-\biggl[\frac{1}{4!}({\cal C}^{(2)})^4 -2l{\cal N}^{(8)} +l{\cal B}^{(6)}\wedge {\cal C}^{(2)} -2l^2\varphi ^{(8)} \biggr]\wedge dy \nonumber \\
&& +m_1\biggl[ 2{\cal N}^{(8)} -{\cal B}^{(6)}\wedge {\cal C}^{(2)} -\frac{1}{3!}({\cal C}^{(2)})^3\wedge {\cal B} +2l^2{\cal C}^{(8)} -l^2{\cal C}^{(6)}\wedge {\cal B} \nonumber \\ 
&& -l{\cal C}^{(6)}\wedge {\cal C}^{(2)} -l{\cal B}^{(6)}\wedge {\cal B} +l{\cal C}^{(4)}\wedge {\cal C}^{(2)}\wedge {\cal B}\biggr]\wedge dy .
\end{eqnarray}

\end{document}